\documentclass[twocolumn,10pt,twoside]{article}

\usepackage{graphics}

\begin{document}

\hyphenation{palaeonto-logi-cal}

\def\etal{{\it{}et~al.}}        
\def\eref#1{(\protect\ref{#1})}
\def\fref#1{\protect\ref{#1}}
\def\sref#1{\protect\ref{#1}}
\def\tref#1{\protect\ref{#1}}
\def\av#1{\langle#1\rangle}
\def\th#1{$#1^{\rm th}$}
\def\d{{\rm d}}
\def\e{{\rm e}}
\def\half{\mbox{$\frac12$}}
\def\xx{}
\def\NK{{\it NK}}
\def\NKCS{{\it NKCS}}

{\newif\ifnotend
\notendtrue
\def\veclist{ABCDEFGHIJKLMNOPQRSTUVWXYZabcdefghijklmnopqrstuvwxyz.}
\def\top#1#2.{#1}
\def\tail#1#2.{#2.}
\loop\expandafter\xdef\csname v\expandafter\top\veclist\endcsname%
{{\noexpand\bf\expandafter\top\veclist}}
\edef\veclist{\expandafter\tail\veclist}
\if\veclist.\notendfalse\fi\ifnotend\repeat}

%
\def\tcapt#1{\refstepcounter{table}\bigskip\hbox to \textwidth{%
       \hfil\vbox{\hsize=\captwidth\renewcommand{\baselinestretch}{1}\small
       {\sc Table \thetable}\quad#1}\hfil}\bigskip}

%
\newdimen\captwidth
\captwidth=12cm                                
\newdimen\sidecaptwidth
\sidecaptwidth=3.5cm                             
\newdimen\widefigwidth
\widefigwidth=13cm                             
\newdimen\normalfigwidth
\normalfigwidth=10cm                           
\newdimen\tinyfigwidth
\tinyfigwidth=6cm                              
\newdimen\sidefigwidth
\sidefigwidth=4cm                              

\def\widecapt#1{\refstepcounter{figure}\bigskip\hbox to \linewidth{%
       \hfil\vbox{\hsize=\captwidth\renewcommand{\baselinestretch}{1}\small
       {\sc Figure \thefigure}\quad#1}\hfil}\bigskip}

\def\capt#1{\refstepcounter{figure}
       \vbox{\null\vspace{10pt}\hsize=\linewidth
        \renewcommand{\baselinestretch}{1}\small
       {\sc Figure \thefigure}\quad#1}}

\def\normalfigure#1{\hbox to\textwidth{%
       \hfil\resizebox{\normalfigwidth}{!}{\includegraphics{#1}}\hfil}}

\def\columnfigure#1{\resizebox{\linewidth}{!}{\includegraphics{#1}}}

\def\sidefigure#1#2{\hbox to \linewidth{%
       \resizebox{\sidefigwidth}{!}{\includegraphics{#1}}\hfill
       \refstepcounter{figure}
       \vbox{\hsize=\sidecaptwidth\renewcommand{\baselinestretch}{1}
       \small\raggedright{\sc Figure \thefigure}\quad#2}}}

\def\sixfigure#1#2#3#4#5#6{\hbox to \textwidth{%
       \hfil\hbox to\widefigwidth{%
       \resizebox{\tinyfigwidth}{!}{\includegraphics{#1}}\hfil
       \resizebox{\tinyfigwidth}{!}{\includegraphics{#2}}}\hfil}
       \bigskip
       \hbox to \textwidth{%
       \hfil\hbox to\widefigwidth{%
       \resizebox{\tinyfigwidth}{!}{\includegraphics{#3}}\hfil
       \resizebox{\tinyfigwidth}{!}{\includegraphics{#4}}}\hfil}
       \bigskip
       \hbox to \textwidth{%
       \hfil\hbox to\widefigwidth{%
       \resizebox{\tinyfigwidth}{!}{\includegraphics{#5}}\hfil
       \resizebox{\tinyfigwidth}{!}{\includegraphics{#6}}}\hfil}}

\begin{titlepage}
\null
\vspace{2cm}
\begin{center}
\huge
Models of Extinction\\
\medskip\LARGE
A Review\\
\vspace{1cm}\Large
M. E. J. Newman and R. G. Palmer\\
\medskip\normalsize\it
Santa Fe Institute, 1399 Hyde Park Road,\\
Santa Fe, New Mexico 87501.  U.S.A.
\end{center}

\bigskip\rm
\begin{center}
{\bf Abstract}
\end{center}
\begin{quote}
  We review recent work aimed at modelling species extinction over
  geological time.  We discuss a number of models which, rather than
  dealing with the direct causes of particular extinction events, attempt
  to predict overall statistical trends, such as the relative frequencies
  of large and small extinctions, or the distribution of the lifetimes of
  species, genera or higher taxa.  We also describe the available fossil
  and other data, and compare the trends visible in these data with the
  predictions of the models.
\end{quote}
\end{titlepage}

\onecolumn
\thispagestyle{empty}
\tableofcontents

\newpage
\pagestyle{headings}
\renewcommand{\sectionmark}[1]{\markboth{\thesection\hspace{10pt}#1}{\thesection\hspace{10pt}#1}}
\twocolumn
\noindent Of the estimated one to four billion species which have existed
on the Earth since life first appeared here (Simpson~1952), less than 50
million are still alive today (May~1990).  All the others became extinct,
typically within about ten million years~(My) of their first appearance.
It is clearly a question of some interest what the causes are of this high
turnover, and much research has been devoted to the topic (see for example
Raup~(1991a) and Glen~(1994) and references therein).  Most of this work
has focussed on the causes of extinction of individual species, or on the
causes of identifiable mass extinction events, such as the end-Cretaceous
event.  However, a recent body of work has examined instead the statistical
features of the history of extinction, using mathematical models of
extinction processes and comparing their predictions with global properties
of the fossil record.  In this paper we review a number of these models,
describing their mathematical basis, the extinction mechanisms which they
incorporate, and their predictions.  We also discuss the trends in fossil
and other data which they attempt to predict and ask how well they achieve
that goal.  As we will see, a number of them give results which are in
reasonable agreement with the general features of the data.

The outline of the paper is as follows.  In Section~\sref{causes} we give a
brief synopsis of the current debate over the causes of extinction.  In
Section~\sref{data} we describe the fossil record as it pertains to the
models we will be discussing, as well as a number of other types of data
which have been cited in support of these models.  In Sections~\sref{early}
to~\sref{reset} we describe in detail the modelling work which is the
principal topic of this review, starting with early work such as that of
Willis~(1922) and van~Valen~(1973), but concentrating mainly on new results
from the last five years or so.  In Section~\sref{conclusions} we give our
conclusions.

\section{Causes of extinction}
\label{causes}
There are two primary colleges of thought about the causes of extinction.
The traditional view, still held by most palaeontologists as well as many
in other disciplines, is that extinction is the result of external stresses
imposed on the ecosystem by the environment~(Benton~1991, Hoffmann and
Parsons~1991, Parsons~1993).  There are indeed excellent arguments in
favour of this viewpoint, since we have good evidence for particular
exogenous causes for a number of major extinction events in the Earth's
history, such as marine regression (sea-level drop) for the late-Permian
event (Jablonski~1985, Hallam~1989), and bolide impact for the
end-Cretaceous (Alvarez~\etal~1980, Alvarez~1983, 1987).  These
explanations are by no means universally accepted (Glen~1994), but almost
all of the alternatives are also exogenous in nature, ranging from the
mundane (climate change (Stanley~1984, 1988), ocean anoxia (Wilde and
Berry~1984)) to the exotic (volcanism (Duncan and Pyle~1988,
Courtillot~\etal~1988), tidal waves (Bourgeois~\etal~1988), magnetic field
reversal (Raup~1985, Loper~\etal~1988), supernovae (Ellis and
Schramm~1995)).  There seems to be little disagreement that, whatever the
causes of these mass extinction events, they are the result of some change
in the environment.  However, the mass extinction events account for only
about 35\% of the total extinction evident in the fossil record at the
family level, and for the remaining 65\% we have no firm evidence favouring
one cause over another.  Many believe, nonetheless, that all extinction can
be accounted for by environmental stress on the ecosystem.  The extreme
point of view has been put forward (though not entirely seriously) by
Raup~(1992), who used statistical analyses of fossil extinction and of the
effects of asteroid impact to show that, within the accuracy of our present
data, it is conceivable that {\em all\/} terrestrial extinction has been
caused by meteors and comets.  This however is more a demonstration of the
uncertainty in our present knowledge of the frequency of impacts and their
biotic effects than a realistic theory.

At the other end of the scale, an increasing number of biologists and
ecologists are supporting the idea that extinction has biotic causes---that
extinction is a natural part of the dynamics of ecosystems and would take
place regardless of any stresses arising from the environment.  There is
evidence in favour of this viewpoint also, although it is to a large extent
anecdotal.  Maynard Smith~(1989) has given a variety of different examples
of modern-day extinctions caused entirely by species interactions, such as
the effects of overzealous predators, or the introduction of new
competitors into formerly stable systems.  The problem is that extinction
events of this nature usually involve no more than a handful of species at
the most, and are therefore too small to be picked out over the
``background'' level of extinction in the fossil data, making it difficult
to say with any certainty whether they constitute an important part of this
background extinction.  (The distinction between mass and background
extinction events is discussed in more detail in Section~\sref{rates}.)
The recent modelling work which is the primary focus of this review
attempts to address this question by looking instead at statistical trends
in the extinction record, such as the relative frequencies of large and
small extinction events.  Using models which make predictions about these
trends and comparing the results against fossil and other data, we can
judge whether the assumptions which go into the models are plausible.  Some
of the models which we discuss are based on purely biotic extinction
mechanisms, others on abiotic ones, and still others on some mixture of the
two.  Whilst the results of this work are by no means conclusive
yet---there are a number of models based on different extinction mechanisms
which agree moderately well with the data---there has been some encouraging
progress, and it seems a promising line of research.

\section{The data}
\label{data}
In this section we review the palaeontological data on extinction.  We also
discuss a number of other types of data which may have bearing on the
models we will be discussing.

\subsection{Fossil data}
\label{fossils}
The discovery and cataloguing of fossils is a painstaking business, and the
identification of a single new species is frequently the sole subject of a
published article in the literature.  The models with which we are here
concerned, however, predict statistical trends in species extinction,
origination, diversification and so on.  In order to study such statistical
trends, a number of authors have therefore compiled databases of the
origination and extinction times of species described in the literature.
The two most widely used such databases are those of Sepkoski~(1992) and of
Benton~(1993).  Sepkoski's data are labelled by both genus and family,
although the genus-level data are, at the time of writing, unpublished.
The database contains entries for approximately forty thousand marine
genera, primarily invertebrates, from about five thousand families.  Marine
invertebrates account for the largest part of the known fossil record, and
if one is to focus one's attention in any single area, this is the obvious
area to choose.  Benton's database by contrast covers both marine and
terrestrial biotas, though it does so only at the family level, containing
data on some seven thousand families.  The choice of taxonomic level in a
compilation such as this is inevitably a compromise.  Certainly we would
like data at the finest level possible, and a few studies have even been
attempted at the species level (e.g.,~Patterson and Fowler~1996).  However,
the accuracy with which we can determine the origination and extinction
dates of a particular taxon depend on the number of fossil representatives
of that taxon.  In a taxon for which we have very few specimens, the
chances of one of those specimens lying close to the taxon's extinction
date are slim, so that our estimate of this date will tend to be early.
This bias is known as the Signor--Lipps effect~(Signor and Lipps~1982).
The reverse phenomenon, sometimes humorously referred to as the
``Lipps--Signor'' effect, is seen in the origination times of taxa, which
in general err on the late side in poorly represented taxa.  By grouping
fossil species into higher taxa, we can work with denser data sets which
give more accurate estimates of origination and extinction dates, at the
expense of throwing out any information which is specific to the lower
taxonomic levels (Raup and Boyajian~1988).  (Higher taxa do, however,
suffer from a greater tendency to paraphyly---see the discussion of
pseudoextinction in Section~\sref{pseudoextinction}.)

\subsubsection{Biases in the fossil data}
\label{biases}
The times of origination and extinction of species are usually recorded to
the nearest geological stage.  Stages are intervals of geological time
determined by stratigraphic methods, or in some cases by examination of the
fossil species present.  Whilst this is a convenient and widely accepted
method of dating, it presents a number of problems.  First, the dates of
the standard geological stages are not known accurately.  They are
determined mostly by interpolation between a few widely-spaced calibration
points, and even the timings of the major boundaries are still contested.
In the widely-used timescale of Harland~\etal~(1990), for example, the
Vendian--Cambrian boundary, which approximately marks the beginning of the
explosion of multi-cellular life, is set at around 625 million years
ago~(Ma).  However, more recent results indicate that its date may be
nearer 545~Ma, a fairly significant correction (Bowring~\etal~1993).

Another problem, which is particularly annoying where studies of extinction
are concerned, is that the stages are not of even lengths.  There are 77
stages in the Phanerozoic (the interval from the start of the Cambrian till
the present, from which virtually all the data are drawn) with a mean
length of 7.3~My, but they range in length from about 1~My to 20~My.  If
one is interested in calculating extinction rates, i.e.,~the number of
species becoming extinct per unit time, then clearly one should divide the
number dying out in each stage by the length of the stage.  However, if, as
many suppose, extinction occurs not in a gradual fashion, but in intense
bursts, this can give erroneous results.  A single large burst of
extinction which happens to fall in a short stage, would give an
anomalously high extinction rate, regardless of whether the average
extinction rate was actually any higher than in surrounding times.
Benton~(1995) for example has calculated familial extinction rates in this
way and finds that the apparent largest mass extinction event in the
Earth's history was the late Triassic event, which is measured to be 20
times the size of the end-Cretaceous one.  This result is entirely an
artifact of the short duration (1~to 2~My) of the Rhaetian stage at the end
of the Triassic.  In actual fact the late Triassic event killed only about
half as many families as the end-Cretaceous.  In order to minimize effects
such as these, it has become common in studies of extinction to examine not
only extinction rates (taxa becoming extinction per unit time) but also
total extinction (taxa becoming extinct in each stage).  While the total
extinction does not suffer from large fluctuations in short stages as
described above, it obviously gives a higher extinction figure in longer
stages in a way which rate measures do not.  However, some features of the
extinction record are found to be independent of the measure used, and in
this case it is probably safe to assume that they are real effects rather
than artifacts of the variation in stage lengths.

The use of the stages as a time scale has other problems associated with it
as well.  For example, it appears to be quite common to assign a different
name to specimens of the same species found before and after a major stage
boundary~(Raup and Boyajian~1988), with the result that stage boundaries
``generate'' extinctions---even species which did not become extinct during
a mass extinction event may appear to do so, by virtue of being assigned a
new name after the event.

There are many other shortcomings in the fossil record.  Good discussions
have been given by Raup~(1979a), Raup and Boyajian~(1988) and
Sepkoski~(1996).  Here we just mention briefly a few of the most glaring
problems.  The ``pull of the recent'' is a name which refers to the fact
that species diversity appears to increase towards recent times because
recent fossils tend to be better preserved and easier to dig up.  Whether
this in fact accounts for all of the observed increase in diversity is an
open question, one which we discuss further in Section~\sref{origination}.
A related phenomenon affecting recent species (or higher taxa) is that some
of them are still alive today.  Since our sampling of living species is
much more complete than our sampling of fossil ones, this biases the recent
record heavily in favour of living species.  This bias can be corrected for
by removing living species from our fossil data.

The ``monograph'' effect is a source of significant bias in studies of
taxon origination.  The name refers to the apparent burst of speciation
seen as the result of the work of one particularly zealous researcher or
group of researchers investigating a particular period; the record will
show a peak of speciation over a short period of geological time, but this
is only because that period has been so extensively researched.  A closely
related phenomenon is the so-called ``Lagerst\"atten'' effect, which refers
to the burst of speciation seen when the fruits of a particularly
fossil-rich site are added to the database.  These and other fluctuations
in the number of taxa---the standing diversity---over geologic time can be
partly corrected for by measuring extinction as a fraction of diversity.
Such ``per taxon'' measures of extinction may however miss real effects
such as the slow increase in overall diversity over time discussed in
Section~\sref{origination}.  For this reason it is common in fact to
calculate both per taxon and actual extinction when looking for trends in
fossil data.  Along with the two ways of treating time described above,
this gives us four different extinction ``metrics'': total number of taxa
becoming extinct per stage, percentage of taxa becoming extinct per stage,
number per unit time, and percentage per unit time.

A source of bias in measures of the sizes of mass extinction events is poor
preservation of fossils after a large event because of environmental
disturbance.  It is believed that many large extinction events are caused
by environmental changes, and that these same changes may upset the
depositional regime under which organisms are fossilized.  In some cases
this results in the poor representation of species which actually survived
the extinction event perfectly well, thereby exaggerating the measured size
of the event.  There are a number of examples of so-called Lazarus taxa
(Flessa and Jablonski~1983) which appear to become extinct for exactly this
reason, only to reappear a few stages later.  On the other hand, the
Signor--Lipps effect discussed above tends to bias results in the opposite
direction.  Since it is unlikely that the last representative of a
poorly-represented taxon will be found very close to the actual date of a
mass-extinction event, it sometimes appears that species are dying out for
a number of stages before the event itself, even if this is not in fact the
case.  Thus extinction events tend to get ``smeared'' backwards in time.
In fact, the existence of Lazarus taxa can help us to estimate the
magnitude of this problem, since the Signor--Lipps effect should apply to
these taxa also, even though we know that they existed right up until the
extinction event (and indeed beyond).

With all these biases present in the fossil data, one may well wonder
whether it is possible to extract any information at all from the fossil
record about the kinds of statistical trends with which our models are
concerned.  However, many studies have been performed which attempt to
eliminate one or more of these biases, and some results are common to all
studies.  This has been taken as an indication that at least some of the
trends visible in the fossil record transcend the rather large error bars
on their measurement.  In the next section we discuss some of these trends,
particularly those which have been used as the basis for models of
extinction, or cited as data in favour of such models.

\begin{figure}[t]
\columnfigure{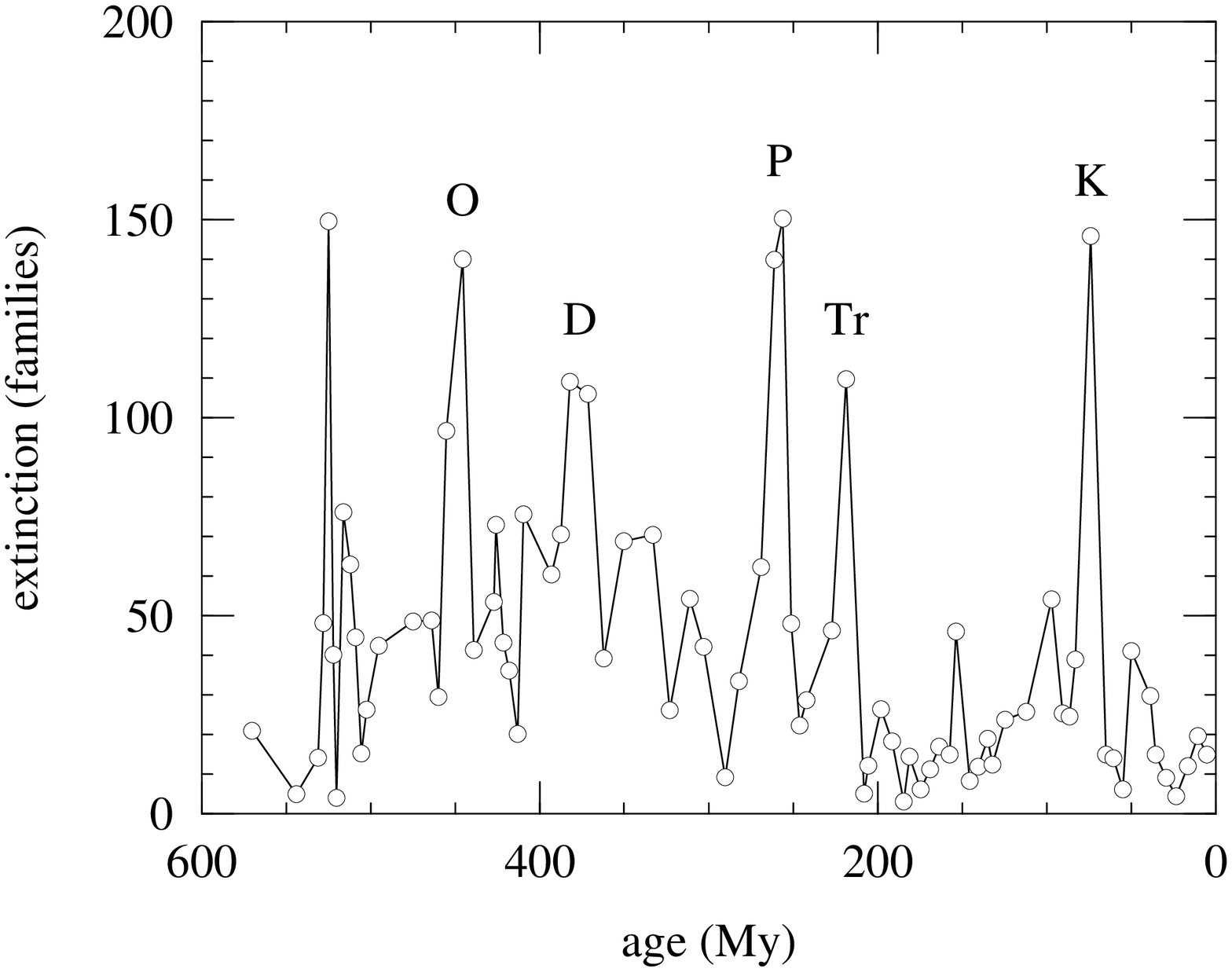}
\capt{The number of families of known marine organisms becoming extinct per
  stratigraphic stage as a function of time during the Phanerozoic.  The
  positions of the ``big five'' mass extinctions discussed in the text are
  marked with letters.  The data are from the compilation by
  Sepkoski~(1992).}
\label{extrates}
\end{figure}

\subsection{Trends in the fossil data}
\label{trends}
There are a number of general trends visible in the fossil data.  Good
discussions have been given by Raup~(1986) and by Benton~(1995).  Here we
discuss some of the most important points, as they relate to the models
with which this review is concerned.

\begin{figure}[t]
\columnfigure{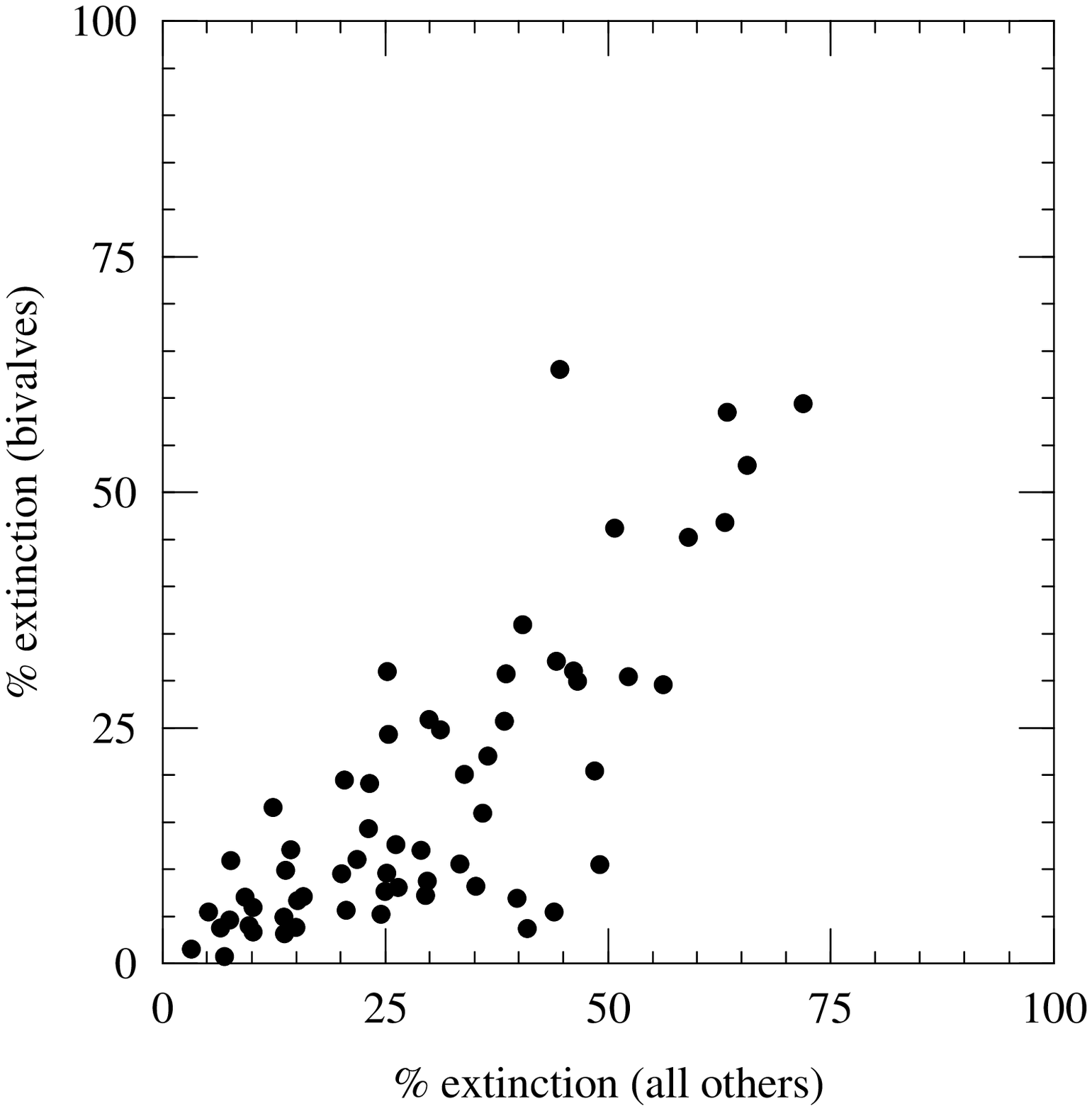}
\capt{The percentage of genera of bivalves becoming extinct in each stage
  plotted against the percentage extinction of all other genera.  The
  positive correlation ($r=0.78$) indicates of a common cause of
  extinction.  After Raup and Boyajian~(1988).}
\label{bivalves}
\end{figure}

\subsubsection{Extinction rates}
\label{rates}
In Figure~\fref{extrates} we show a plot of the number of families of
marine organisms becoming extinct in each geological stage since the start
of the Phanerozoic.  The data are taken from an updated version of the
compilation by Sepkoski~(1992).  It is clear from this plot that, even
allowing for the irregular sizes of the stages discussed above, there is
more variation in the extinction rate than could be accounted for by simple
Poissonian fluctuations.  In particular, a number of mass extinction events
can be seen in the data, in which a significant fraction of the known
families were wiped out simultaneously.  Palaeontology traditionally
recognizes five large extinction events in terrestrial history, along with
quite a number of smaller ones (Raup and Sepkoski~1982).  The ``big five''
are led by the late Permian event (indicated by the letter P in the figure)
which may have wiped out more than 90\% of the species on the planet
(Raup~1979b).  The others are the events which ended the Ordovician (O),
the Devonian (D), the Triassic (Tr) and the Cretaceous (K).  A sixth
extinction peak at about 525~Ma is also visible in the figure (the leftmost
large peak), but it is still a matter of debate whether this peak
represents a genuine historical event or just a sampling error.

As discussed in Section~\sref{causes}, the cause of mass extinction events
is a topic of much debate.  However, it seems to be widely accepted that
those causes, whatever they are, are abiotic, which lends strength to the
view, held by many palaeontologists, that {\em all\/} extinction may have
been caused by abiotic effects.  The opposing view is that large extinction
events may be abiotic in origin, but that smaller events, perhaps even at
the level of single species, have biotic causes.  Raup and Boyajian~(1988)
have investigated this question by comparing the extinction profiles of the
nine major invertebrate groups throughout the Phanerozoic.  While the
similarities between these profiles is not as strong as between the
extinction profiles of different subsets of the same group, they
nonetheless find strong correlations between groups in the timing of
extinction events.  This may be taken as evidence that there is
comparatively little taxonomic selectivity in the processes giving rise to
mass extinction, which in turn favours abiotic rather than biotic causes.
In Figure~\fref{bivalves}, for example, reproduced from data given in their
paper, we show the percentage extinction of bivalve families against
percentage extinction of all other families, for each stage of the
Phanerozoic.  The positive correlation ($r^2=0.78$) of these data suggest a
common cause for the extinction of bivalves and other species.

\begin{figure}[t]
\columnfigure{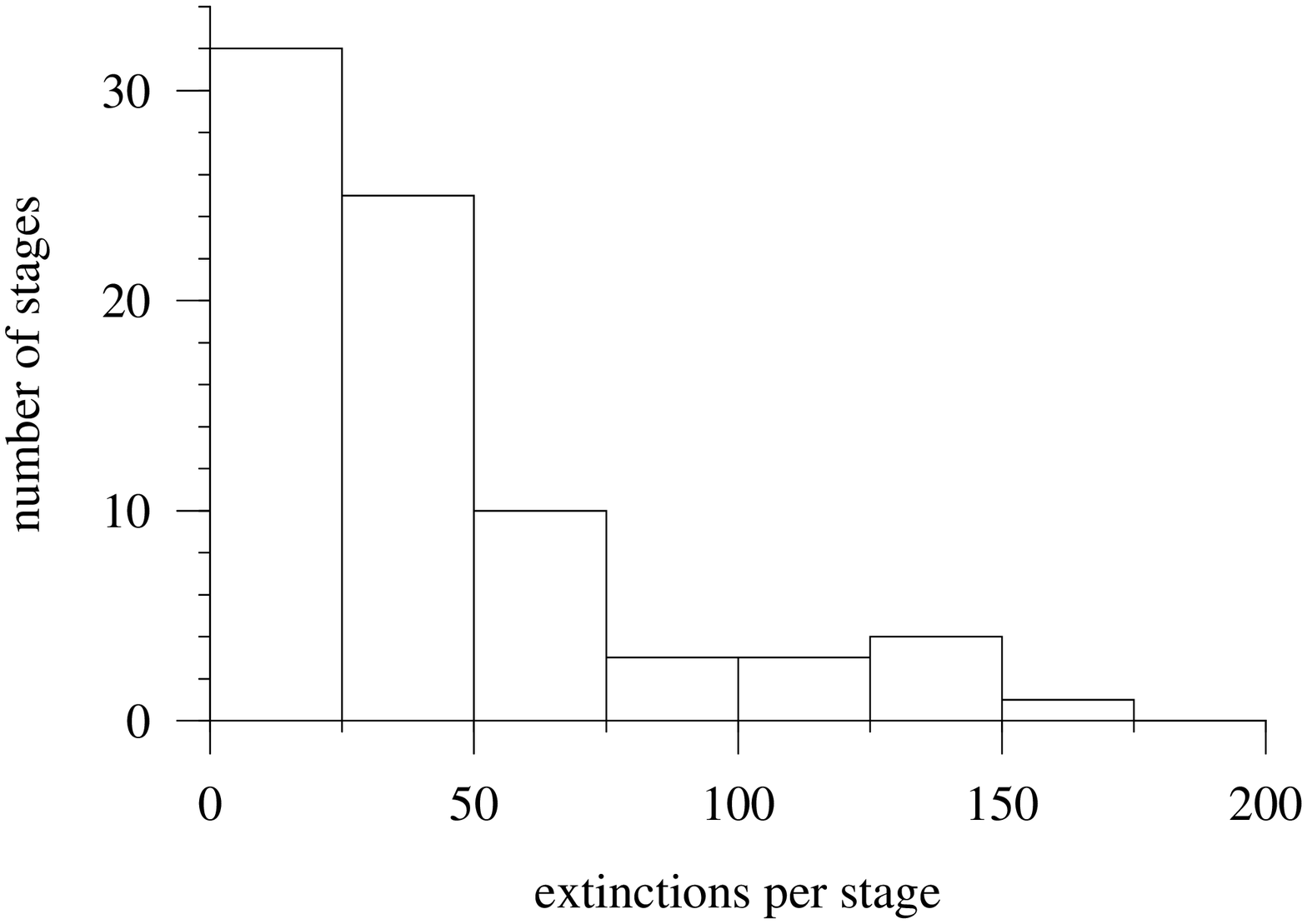}
\capt{Histogram of the number of families of marine organisms becoming
  extinct per stratigraphic stage during the Phanerozoic.  The data are
  drawn from Sepkoski~(1992).}
\label{extdist}
\end{figure}

The shortcoming of these studies is that they can still only yield
conclusions about correlations between extinction events large enough to be
visible above the noise level in the data.  It is perfectly reasonable to
adopt the position that the large extinction events have exogenous causes,
but that there is a certain level of ``background'' events which are
endogenous in origin.  In order to address this issue a number of
researchers have constructed plots of the distribution of the sizes of
extinction events; non-uniformity in such a distribution might offer
support for distinct mass and background extinction mechanisms (Raup~1986,
Kauffman~1993, Sol\'e and Bascompte~1996).  One such distribution is shown
in Figure~\fref{extdist}, which is a histogram of the number of families
dying out per stage.  This is not strictly the same thing as the sizes of
extinction events, since several distinct events may contribute to the
total in a given stage.  However, since most extinction dates are only
accurate to the nearest stage it is the best we can do.  If many
independent extinction events were to occur in each stage, then one would
expect, from Poisson statistics (see, for instance, Grimmett and
Stirzaker~1992), that the histogram would be approximately normally
distributed.  In actual fact, as the figure makes clear, the distribution
is highly skewed and very far from a normal distribution (Raup~1996).  This
may indicate that extinction at different times is correlated, with a
characteristic correlation time of the same order of magnitude as or larger
than the typical stage length so that the extinctions within a single stage
are not independent events (Newman and Eble~1999a).

The histogram in Figure~\fref{extdist} shows no visible discontinuities,
within the sampling errors, and therefore gives no evidence for any
distinction between mass and background extinction events.  An equivalent
result has been derived by Raup~(1991b) who calculated a ``kill curve'' for
marine extinctions in the Phanerozoic by comparing Monte Carlo calculations
of genus survivorship with survivorship curves drawn from the fossil data.
The kill curve is a cumulative frequency distribution of extinctions which
measures the frequency with which one can expect extinction events of a
certain magnitude.  Clearly this curve contains the same information as the
distribution of extinction sizes, and it can be shown that the conversion
from one to the other involves only a simple integral transform
(Newman~1996).  On the basis of Raup's calculations, there is again no
evidence for a separation between mass and background extinction events in
the fossil record.

This result is not necessarily a stroke against extinction models which are
based on biotic causes.  First, it has been suggested (Jablonski~1986,
1991) that although there may be no quantitative distinction between mass
and background events, there could be a qualitative one; it appears that
the traits which confer survival advantages during periods of background
extinction may be different from those which allow species to survive a
mass extinction, so that the selection of species becoming extinction under
the two regimes is different.

\begin{figure}[t]
\columnfigure{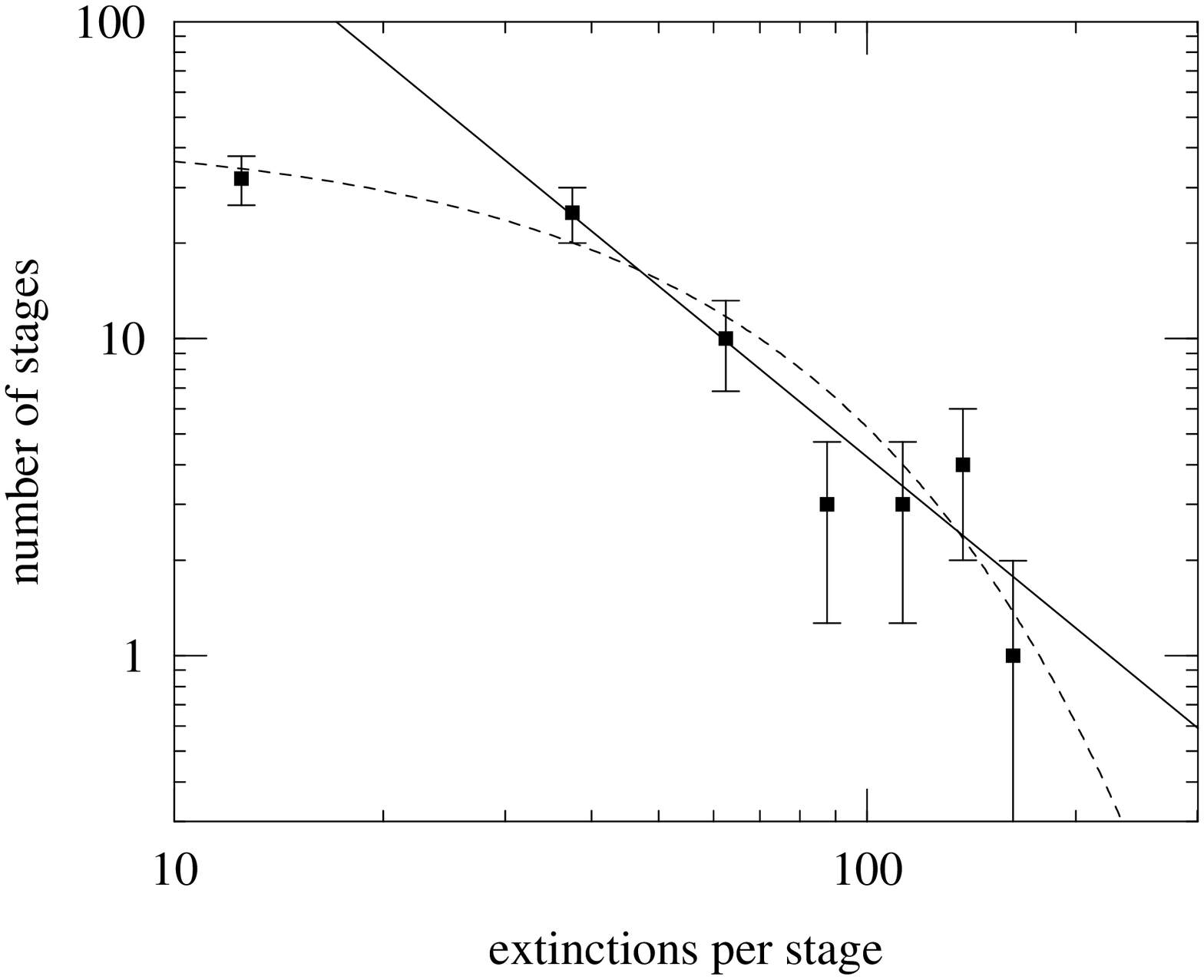}
\capt{The data from Figure~\fref{extdist} replotted on logarithmic scales,
  with Poissonian error bars.  The solid line is the best power-law fit to
  the data.  The dashed line is the best exponential fit.}
\label{extlog}
\end{figure}

Second, there are a number of models which predict a smooth distribution of
the sizes of extinction events all the way from the single species level up
to the size of the entire ecosystem simply as a result of biotic
interactions.  In fact, the distribution of extinction sizes is one of the
fundamental predictions of most of the models discussed in this review.
Although the details vary, one of the most striking features which these
models have in common is their prediction that the extinction distribution
should follow a power law, at least for large extinction events.  In other
words, the probability $p(s)$ that a certain fraction $s$ of the extant
species/genera/families will become extinct in a certain time interval (or
stage) should go like
\begin{equation}
p(s) \propto s^{-\tau},
\label{powerlaw}
\end{equation}
for large $s$, where $\tau$ is an exponent whose value is determined by the
details of the model.  This is a conjecture which we can test against the
fossil record.  In Figure~\fref{extlog} we have replotted the data from
Figure~\fref{extdist} using logarithmic scales, on which a power-law form
should appear as a straight line with slope $-\tau$.  As pointed out by
Sol\'e and Bascompte~(1996), and as we can see from the figure, the data
are indeed compatible with the power-law form,\footnote{In this case we
  have excluded the first point on the graph from our fit, which is
  justifiable since the power law is only expected for large values of
  $s$.} but the error bars are large enough that they are compatible with
other forms as well, including the exponential shown in the figure.  

In cases such as this, where the quality of the data makes it difficult to
distinguish between competing forms for the distribution, a useful tool is
the {\em rank/frequency plot.}  A rank/frequency plot for extinction is
constructed by taking the stratigraphic stages and numbering them in
decreasing order of number of taxa becoming extinct.  Thus the stage in
which the largest number of taxa become extinct is given rank~1, the stage
with the second largest number is given rank~2, and so forth.  Then we plot
the number of taxa becoming extinct as a function of rank.  It is
straightforward to show (Zipf~1949) that distributions which appear as
power laws or exponentials in a histogram such as Figure~\fref{extlog} will
appear as power laws and exponentials on a rank/frequency plot also.
However, the rank frequency plot has the significant advantage that the
data points need not be grouped into bins as in the histogram.  Binning the
data reduces the number of independent points on the plot and throws away
much of the information contained in our already sparse data set.  Thus the
rank/frequency plot often gives a better guide to the real form of a
distribution.

\begin{figure}[t]
\columnfigure{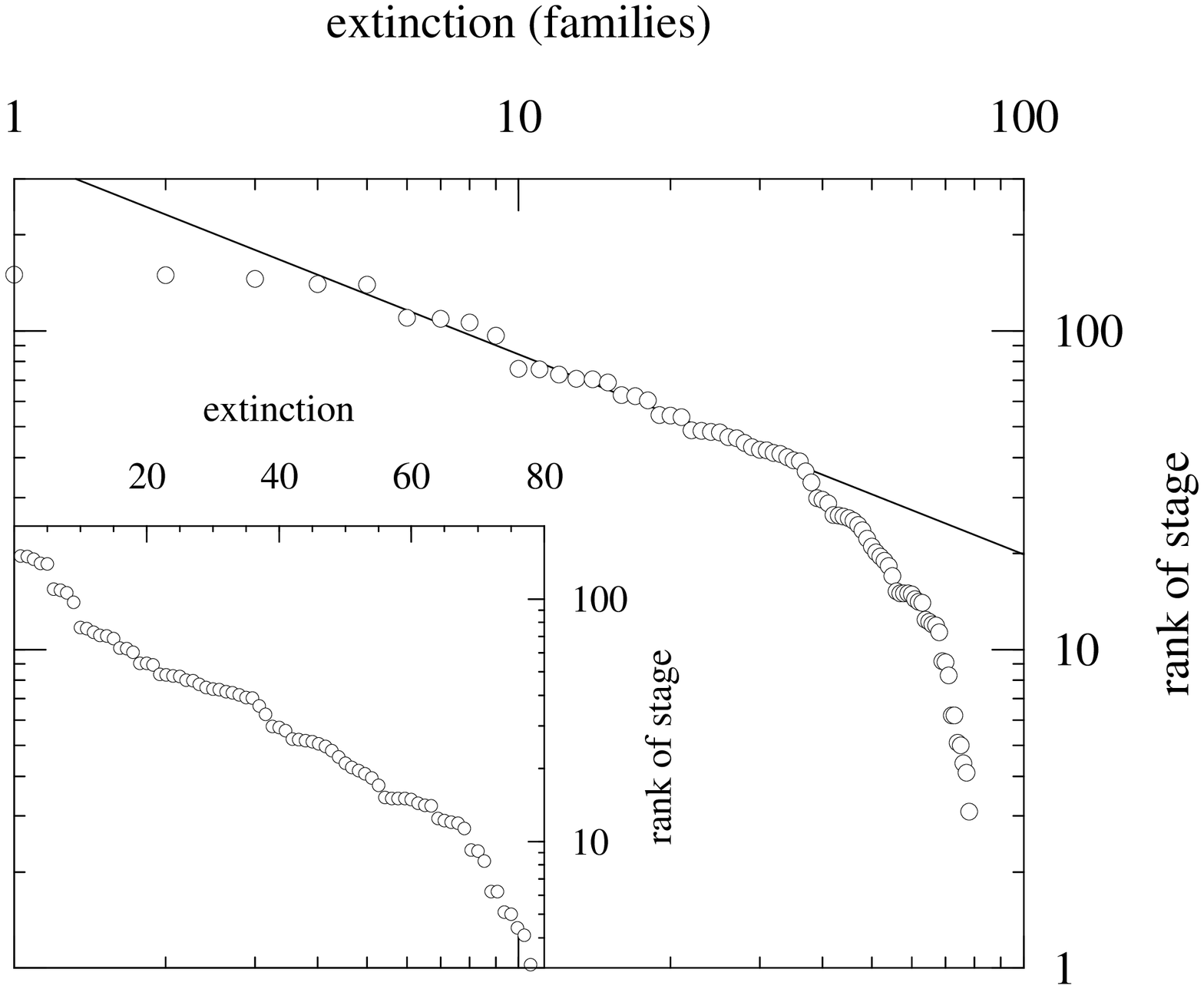}
\capt{Main figure: a rank/frequency plot of the numbers of families of
marine organisms becoming extinct in each stratigraphic stage of the
Phanerozoic.  The straight line is a power-law fit to the points.  Inset:
the same data replotted on semi-logarithmic axes.}
\label{rfplot}
\end{figure}

In Figure~\fref{rfplot} we show a rank/frequency plot of extinctions of
marine families in each stage of the Phanerozoic on logarithmic scales.  As
we can see, this plot does indeed provide a clearer picture of the
behaviour of the data, although ultimately the conclusions are rather
similar.  The points follow a power law quite well over the initial portion
of the plot, up to extinctions on the order of 40 families or so, but
deviate markedly from power law beyond this point.  The inset shows the
same data on semi-logarithmic scales, and it appears that they may fall on
quite a good straight line, although there are deviations in this case as
well.  Thus it appears again that the fossil data could indicate either a
power-law or an exponential form (and possibly other forms as well).

More sophisticated analysis (Newman~1996) has not settled this question,
although it does indicate that the Monte Carlo results of Raup~(1991b) are
in favour of the power-law form, rather than the exponential one, and also
allows for a reasonably accurate measurement of the exponent of the power
law, giving $\tau=2.0\pm0.2$.  This value can be compared against the
predictions of the models.

\begin{figure}[t]
\columnfigure{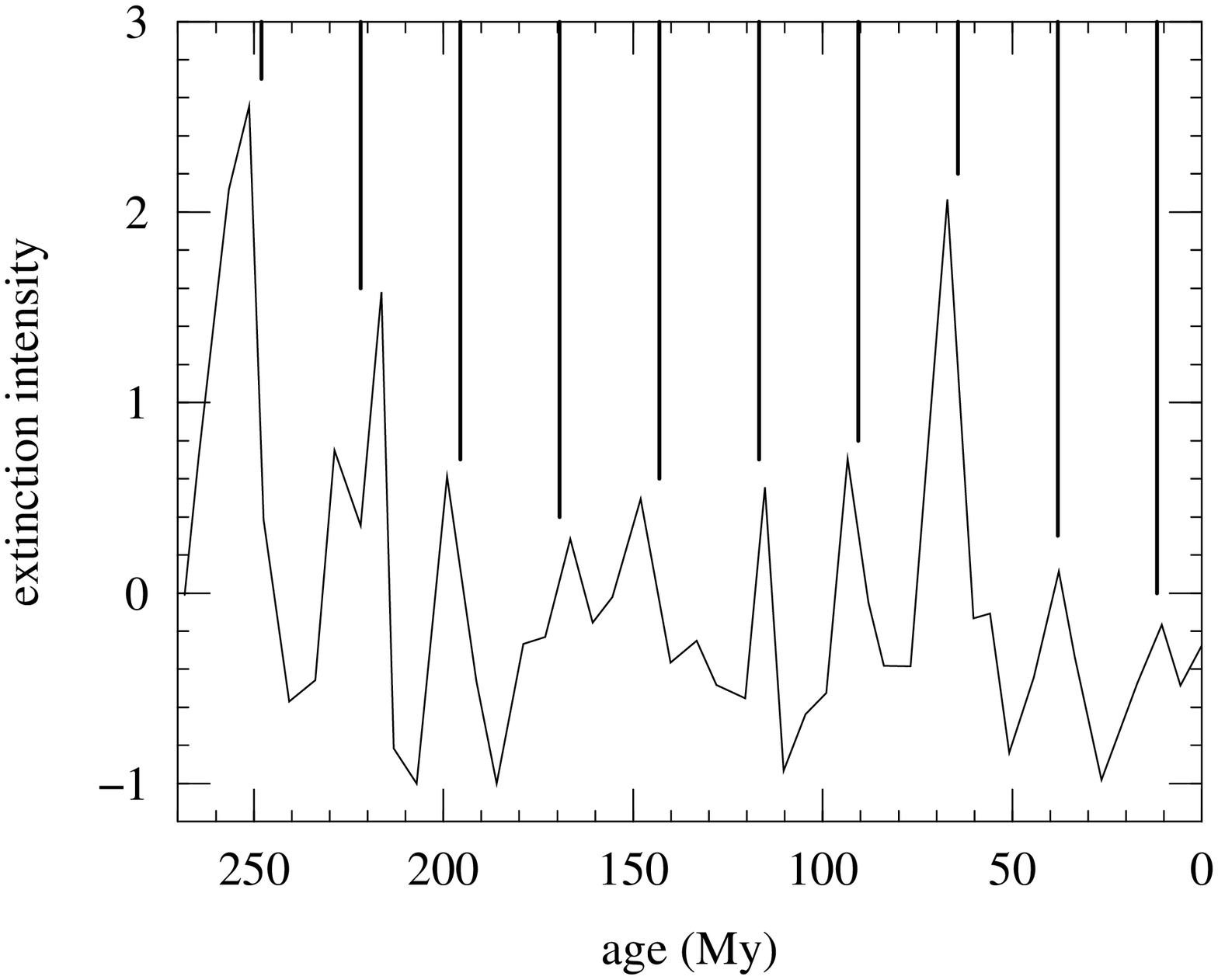}
\capt{The number of genera of marine invertebrates becoming extinct per
  stratigraphic stage over the last 270~My.  The vertical scale is in units
  of standard deviations from the mean extinction rate.  The vertical bars
  indicate the positions of the periodic extinctions postulated by Raup and
  Sepkoski~(1984).  After Sepkoski~(1990).}
\label{period}
\end{figure}

\begin{figure}[t]
\columnfigure{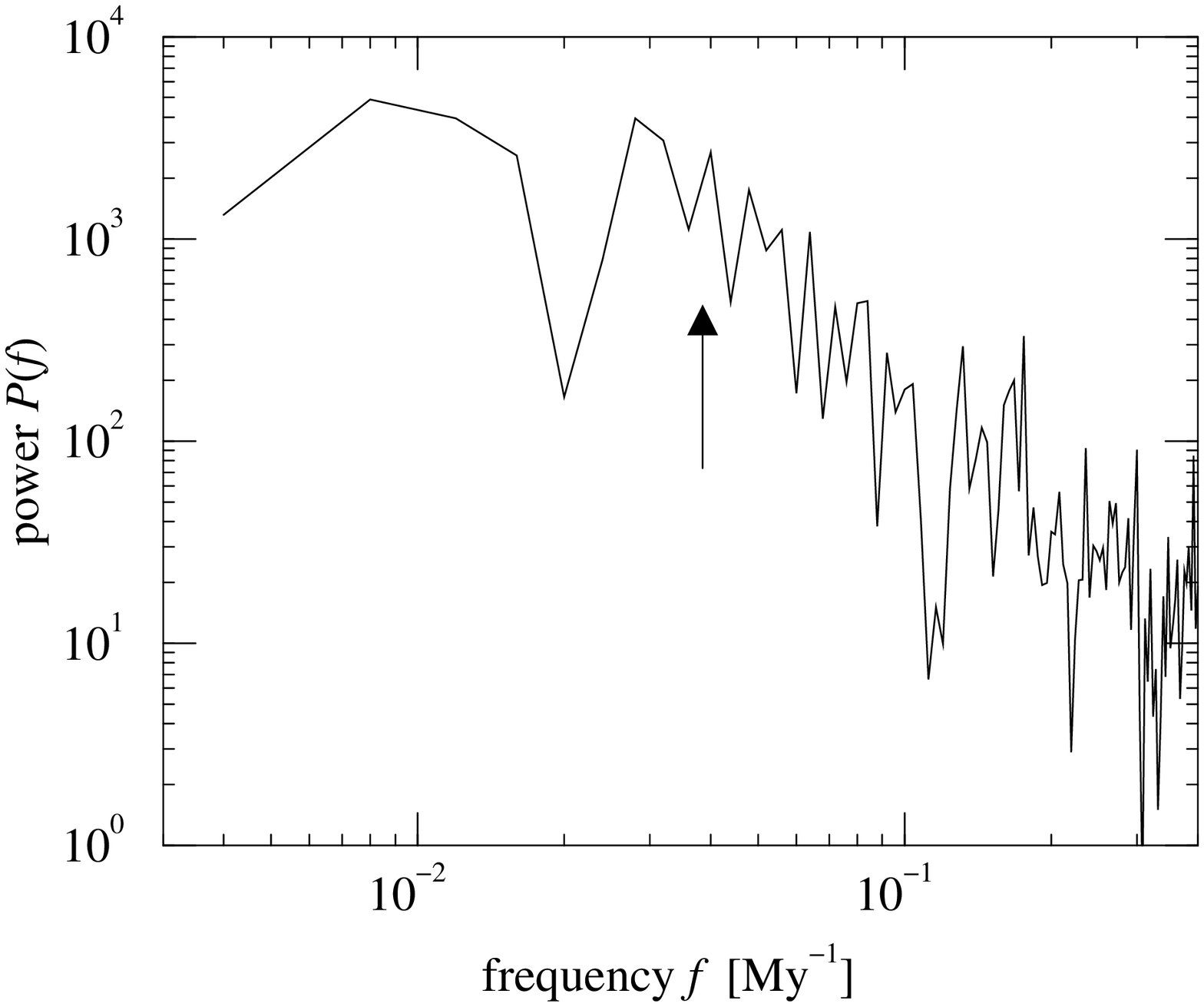}
\capt{The power spectrum of familial extinction for marine animals over the
  last 250~My, calculated using data from the database compiled by
  Sepkoski~(1992).  The arrow marks the frequency corresponding to the
  conjectured 26~My periodicity of extinctions.  Note that the scales on
  both axes are logarithmic.}
\label{ppspec}
\end{figure}

\subsubsection{Extinction periodicity}
\label{periodicity}
In an intriguing paper published in 1984, Raup and Sepkoski have suggested
that the mass extinction events seen in the most recent 250~My or so of the
fossil record occur in a periodic fashion, with a period of about 26~My
(Raup and Sepkoski~1984, 1986, 1988, Sepkoski~1989, 1990).
Figure~\fref{period} shows the curve of extinction intensity for marine
invertebrate genera from the middle Permian to the Recent from Sepkoski's
data, with the postulated periodicity indicated by the vertical lines.  A
number of theories have been put forward, mostly based on astronomical
causes, to explain how such a periodicity might arise (Davis~\etal~1984,
Rampino and Stothers~1984, Whitmire and Jackson~1984, Hut~\etal~1987).
More recently however, it has been suggested that the periodicity has more
mundane origins.  Patterson and Smith~(1987, 1989), for instance, have
theorized that it may be an artifact of noise introduced into the data by
poor taxonomic classification (Sepkoski and Kendrick~(1993) argue
otherwise), while Stanley~(1990) has suggested that it may be a result of
delayed recovery following large extinction events.

A quantitative test for periodicity of extinction is to calculate the power
spectrum of extinction over the appropriate period and look for a peak at
the frequency corresponding to 26~My.  We have done this in
Figure~\fref{ppspec} using data for marine families from the Sepkoski
compilation.  As the figure shows, there is a small peak in the spectrum
around the relevant frequency (marked with an arrow), but it is not
significant given the level of noise in the data.  On the other hand,
similar analyses by Raup and Sepkoski~(1984) and by Fox~(1987) using
smaller databases do appear to produce a significant peak.  The debate on
this question is still in progress.

The power spectrum of fossil extinction is interesting for other reasons.
Sol\'e~\etal~(1997) have suggested on the basis of calculations using
fossil data from the compilation by Benton~(1993) that the spectrum has a
$1/f$ form, i.e.,~it follows a power law with exponent $-1$.  This result
would be intriguing if true, since it would indicate that extinction at
different times in the fossil record was correlated on arbitrarily long
time-scales.  However, it now appears likely that the form found by
Sol\'e~\etal\ is an artifact of the method of analysis, rather than a real
result (Kirchner and Weil~1998).  Spectra calculated using other methods do
not show the $1/f$ form and can be explained without assuming any long-time
correlations: they are consistent with an exponential form at low
frequencies crossing over to a $1/f^2$ behaviour at high frequencies
(Newman and Eble~1999a).

\begin{figure}[t]
\columnfigure{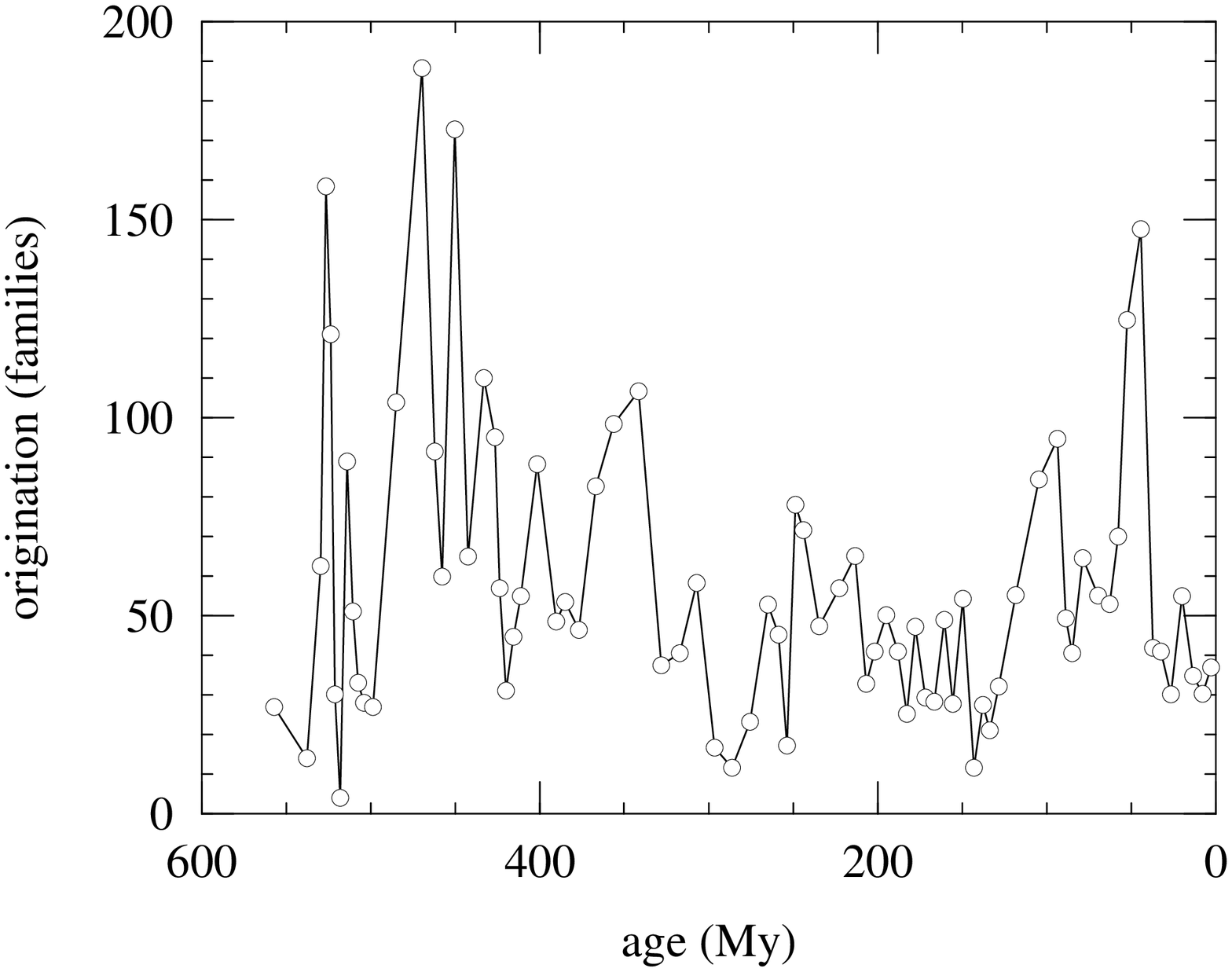}
\capt{The number of families of known marine organisms appearing for the
  first time in each stratigraphic stage as a function of time throughout
  the Phanerozoic.  The data come from the compilation by Sepkoski~(1992).}
\label{origin}
\end{figure}

\subsubsection{Origination and diversity}
\label{origination}
The issue of origination rates of species in the fossil record is in a
sense complementary to that of extinction rates, but has been investigated
in somewhat less depth.  Interesting studies have been carried out by, for
example, Gilinsky and Bambach~(1987), Jablonski and Bottjer~(1990a, 1990b,
1990c), Jablonski~(1993), Sepkoski~(1998) and Eble~(1998, 1999).  One clear
trend is that peaks in the origination rate appear in the immediate
aftermath of large extinction events.  In Figure~\fref{origin} we show the
number of families of marine organisms appearing per stage.  Comparison
with Figure~\fref{extrates} shows that there are peaks of origination
corresponding to all of the prominent extinction peaks, although the
correspondence between the two curves is by no means exact.

The usual explanation for these bursts of origination is that new species
find it easier to get a toe-hold in the taxonomically under-populated world
which exists after a large extinction event.  As the available niches in
the ecosystem fill up, this is no longer the case, and origination slows.
Many researchers have interpreted this to mean that there is a saturation
level above which a given ecosystem can support no more new species, so
that, apart from fluctuations in the immediate vicinity of the large
extinction events, the number of species is approximately constant.  This
principle has been incorporated into most of the models considered in this
review; the models assume a constant number of species and replace any
which become extinct by an equal number of newly-appearing ones.  (The
``reset'' model considered in Section~\sref{reset} is an important
exception.)

\begin{figure}[t]
\columnfigure{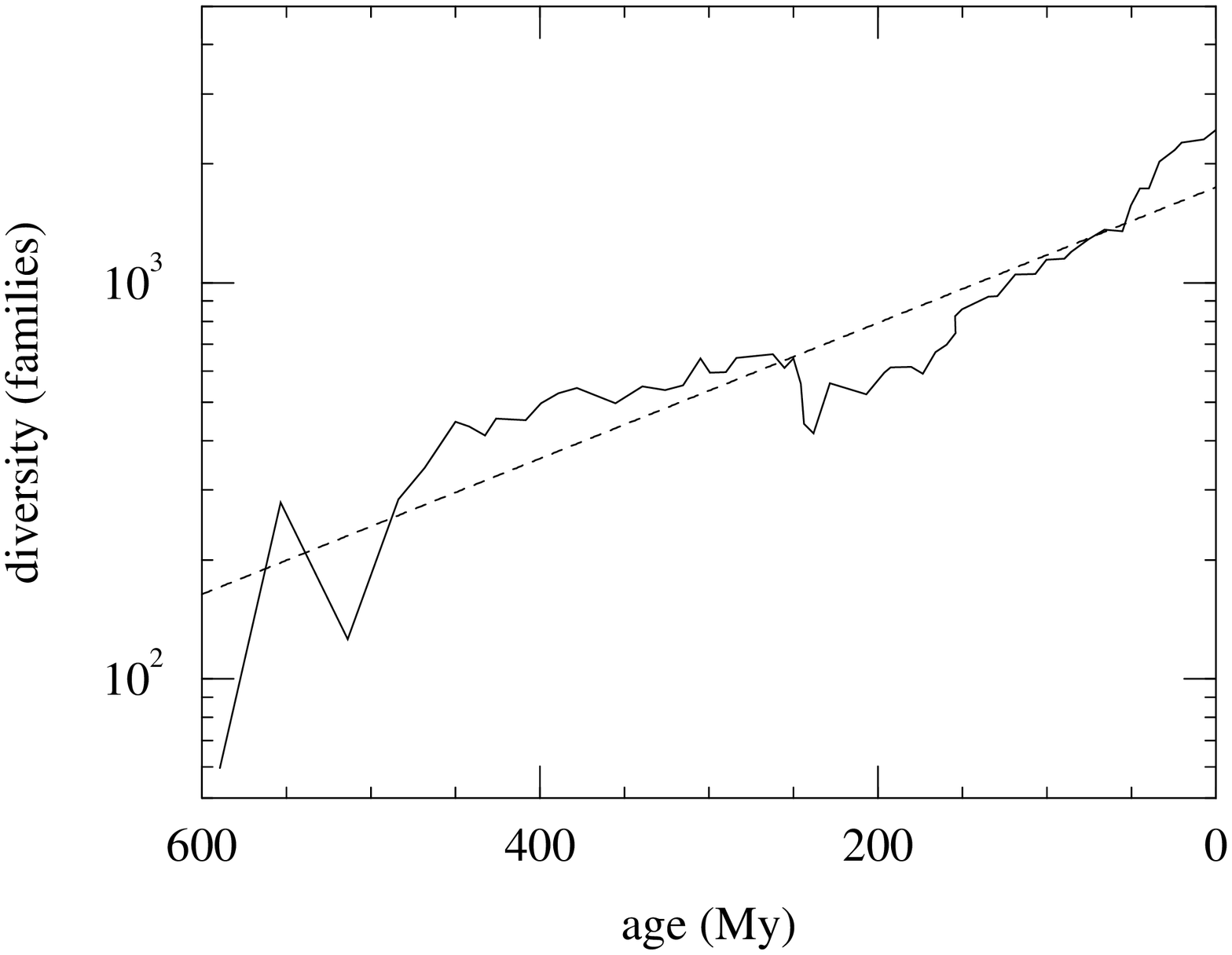}
\capt{The total number of families of known fossil organisms as a function
  of time during the Phanerozoic.  The vertical axis is logarithmic, and
  the dashed line is an exponential fit to the data.  After Benton~(1995).}
\label{diversity}
\end{figure}

However, the hypothesis of constant species number is not universally
accepted.  In the short term, it appears to be approximately correct to say
that a certain ecosystem can support a certain number of species.
Modern-day ecological data on island biogeography support this view (see
for example Rosenzweig~(1995)).  However, on longer timescales, the
diversity of species on the planet appears to have been increasing, as
organisms discover for the first time ways to exploit new habitats or
resources.  In Figure~\fref{diversity} we show the total number of known
fossil families as a function of geological time.  The vertical axis is
logarithmic, and the approximately straight-line form indicates that the
increase in diversity is roughly exponential, although logistic and linear
growth forms have been suggested as well (Sepkoski~1991, Newman and
Sibani~1999).  As discussed in Section~\sref{biases}, one must be careful
about the conclusions we draw from such figures, because of the apparent
diversity increase caused by the ``pull of the recent''.  However, current
thinking mostly reflects the view that there is a genuine diversity
increase towards recent times associated with the expansion of life into
new domains.  As Benton~(1995) has put it: ``There is no evidence in the
fossil record of a limit to the ultimate diversity of life on Earth''.

\begin{figure}[t]
\columnfigure{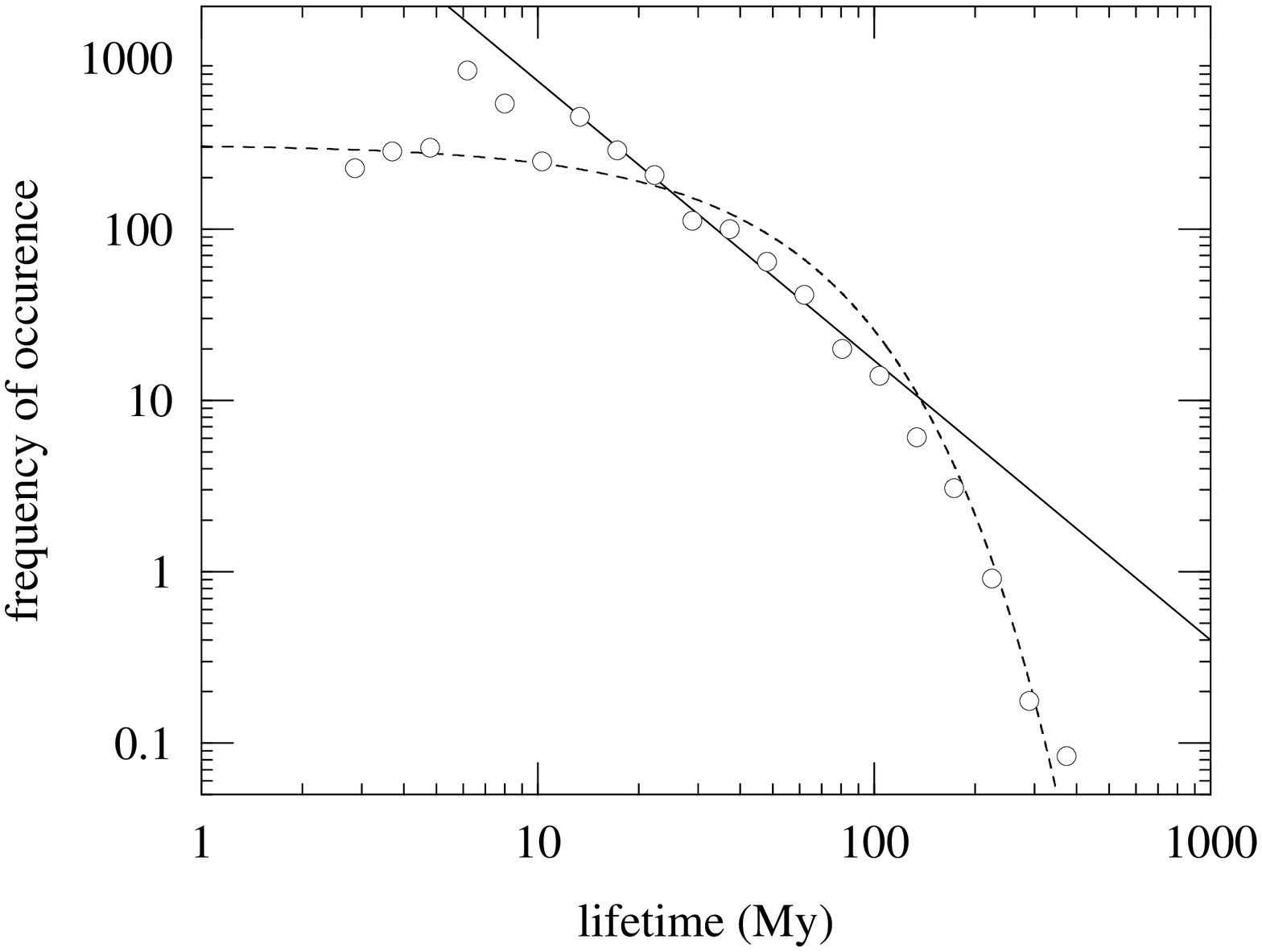}
\capt{Frequency distribution of marine genus lifetimes in the fossil
  record.  The solid line is the best power-law fit to the data between 10
  and 100~My, while the dotted line is the best exponential fit to all the
  data.  After Newman and Sibani~(1999).}
\label{lifedist}
\end{figure}

\subsubsection{Taxon lifetimes}
\label{lifetimes}
Another quantity which has been compared with the predictions of a variety
of extinction models is the distribution of the lifetimes of taxa.  In
Figure~\fref{lifedist} we show a histogram of the lifetimes of marine
genera in the Sepkoski database.  The axes of the figure are logarithmic
and the solid and dotted lines represent respectively power-law and
exponential fits to the data.

At first glance it appears from this figure that the lifetime distribution
is better fitted by the exponential form.  This exponential has a time
constant of $40.1$~My, which is of the same order of magnitude as the mean
genus lifetime of $30.1$~My.  An exponential distribution of this type is
precisely what one would expect to see if taxa are becoming extinct at
random with a constant average rate (a Poisson process).  A number of
authors have however argued in favour of the power-law fit
(Sneppen~\etal~1995, Bak~1996).  The power-law fit in the figure is a fit
only to the data between 10 and 100~My.  In this interval it actually
matches the data quite well, but for longer or shorter lifetimes the
agreement is poor.  Why then should we take this suggestion seriously?  The
answer is that both very long and very short lifetimes are probably
under-represented in the database because of systematic biases.  First,
since the appearance and disappearance of genera are recorded only to the
nearest stage, lifetimes of less than the length of the corresponding stage
are registered as being zero and do not appear on the histogram.  This
means that lifetimes shorter than the average stage length of about 7~My
are under-represented.  Second, as mentioned briefly in
Section~\sref{biases}, a taxon is sometimes given a different name before
and after a major stage boundary, even though little or nothing about that
taxon may have changed.  This means that the number of species with
lifetimes longer than the typical separation of these major boundaries is
also underestimated in our histogram.  This affects species with lifetimes
greater than about 100~My.  Thus there are plausible reasons for performing
a fit only in the central region of Figure~\fref{lifedist} and in this case
the power-law form is quite a sensible conjecture.

The exponent of the power law for the central region of the figure is
measured to be $\alpha=1.6\pm0.1$.  This value is questionable however,
since it depends on which data we choose to exclude at long and short
times.  In fact, a case can be made for any value between about
$\alpha=1.2$ and $2.2$.  In this review we take a working figure of
$\alpha=1.7\pm0.3$ for comparison with theoretical models.  Several of
these models provide explanations for a power-law distribution of taxon
lifetimes, with figures for $\alpha$ in reasonable agreement with this
value.

We should point out that there is a very simple possible explanation for a
power-law distribution of taxon lifetimes which does not rely on any
detailed assumptions about the nature of evolution.  If the addition and
removal of species from a genus (or any sub-taxa from a taxon) are
stochastically constant and take place at roughly the same rate, then the
number of species in the genus will perform an ordinary random walk.  When
this random walk reaches zero---the so-called first return time---the genus
becomes extinct.  Thus the distribution of the lifetimes of genera is also
the distribution of first return times of a one-dimensional random walk.
As is easily demonstrated (see Grimmett and Stirzaker~(1992), for example),
the distribution of first return times follows a power law with exponent
$\frac32$, in reasonable agreement with the figure extracted from the
fossil record above.  An alternative theory is that speciation and
extinction should be multiplicative, i.e.,~proportional to the number of
species in the genus.  In this case the {\em logarithm\/} of the size of
the genus performs a random walk, but the end result is the same: the
distribution of lifetimes is a power law with exponent $\frac32$.

\subsubsection{Pseudoextinction and paraphyly}
\label{pseudoextinction}
One possible source of discrepancy between the models considered in this
paper and the fossil data is the way in which an extinction is defined.  In
the palaeontological literature a distinction is usually drawn between
``true extinction'' and ``pseudoextinction''.  The term pseudoextinction
refers to the evolution of a species into a new form, with the resultant
disappearance of the ancestral form.  The classic example is that of the
dinosaurs.  If, as is supposed by some, modern birds are the descendants of
the dinosaurs (Gauthier~1986, Chiappe~1995), then the dinosaurs did not
truly become extinct, but only pseudoextinct.  Pseudoextinction is of
course a normal part of the evolution process; Darwin's explanation of the
origin of species is precisely the replacement of strains by their own
fitter mutant offspring.  And certainly this is a form of extinction, in
that the ancestral strain will no longer appear in the fossil record.
However, palaeontology makes a distinction between this process and true
extinction---the disappearance of an entire branch of the phylogenetic tree
without issue---presumably because the causes of the two are expected to be
different.  Pseudoextinction is undoubtedly a biotic process (although the
evolution of a species and subsequent extinction of the parent strain may
well be brought on by exogenous pressures---see Roy~(1996), for example).
On the other hand, many believe that we must look to environmental effects
to find the causes of true extinction (Benton~1991).

Some of the models discussed in this review are models of true extinction,
with species becoming extinct and being replaced by speciation from other,
unrelated species.  Others however deal primarily with pseudoextinction,
predicting varying rates of evolution over the course of time, with mass
extinction arising as the result of periods of intense evolutionary
activity in which many species evolve to new forms, causing the
pseudoextinction of their parent forms.  It may not be strictly fair to
compare models such as these to the fossil data on extinction presented
above.  To be sure, the data on extinction dates from which the statistics
are drawn do not distinguish between true extinction and pseudoextinction;
all that is recorded is the last date at which a specimen of a certain
species is found.  However, the grouping of the data into higher taxa, as
discussed in Section~\sref{fossils}, does introduce such a distinction.
When a species evolves to a new form, causing the pseudoextinction of the
ancestral form, the new species is normally assigned to the same higher
taxa---genus and family---as the ancestor.  Thus a compilation of data at
the genus or family level will not register the pseudoextinction of a
species at this point.  The extinction of a genus or family can normally
only occur when its very last constituent species becomes (truly) extinct,
and therefore the data on the extinction times of higher taxa reflect
primarily true extinctions.

However, the situation is not entirely straightforward.  The assignment of
species to genera and genera to families is, to a large extent, an
arbitrary process, with the result that whilst the argument above may apply
to a large portion of the data, there are many anomalies of taxonomy which
give rise to exceptions.  Strictly, the correct way to construct a
taxonomic tree is to use cladistic principles.  A clade is a group of
species which all claim descendence from one ancestral species.  In theory
one can construct a tree in which each taxon is monophyletic, i.e.,~is
composed only of members of one clade.  Such a tree is not unique; there is
still a degree of arbitrariness introduced by differences of opinion over
when a species should be considered the founding member of a new taxon.
However, to the extent that such species are a small fraction of the total,
the arguments given above for the absence of pseudoextinction from the
fossil statistics, at the genus level and above, are valid.  In practice,
however, cladistic principles are hard to apply to fossil species, whose
taxonomic classification is based on morphology rather than on a direct
knowledge of their lines of descent.  In addition, a large part of our
present classification scheme has been handed down to us by a tradition
which predates the introduction of cladism.  The distinction between
dinosaurs and birds, for example, constitutes exactly such a traditional
division.  As a result, many---indeed most---taxonomic groups, particularly
higher ones, tend to be paraphyletic: the members of the taxa are descended
from more than one distinct ancestral species, whose own common ancestor
belonged to another taxon.  Not only does this failing upset our arguments
concerning pseudoextinction above, but also, by virtue of the resulting
unpredictable nature of the taxonomic hierarchy, introduces errors into our
statistical measures of extinction which are hard to quantify (Sepkoski and
Kendrick~1993).  As Raup and Boyajian~(1988) put it: ``If all paraphyletic
groups were eliminated from taxonomy, extinction patterns would certainly
change''.

\subsection{Other forms of data}
\label{otherdata}
There are a few other forms of data which are of interest in connection
with the models we will be discussing.  Chief amongst these are taxonomic
data on modern species, and simulation data from so-called artificial life
experiments.

\subsubsection{Taxonomic data}
\label{taxonomy}
\begin{figure}[t]
\columnfigure{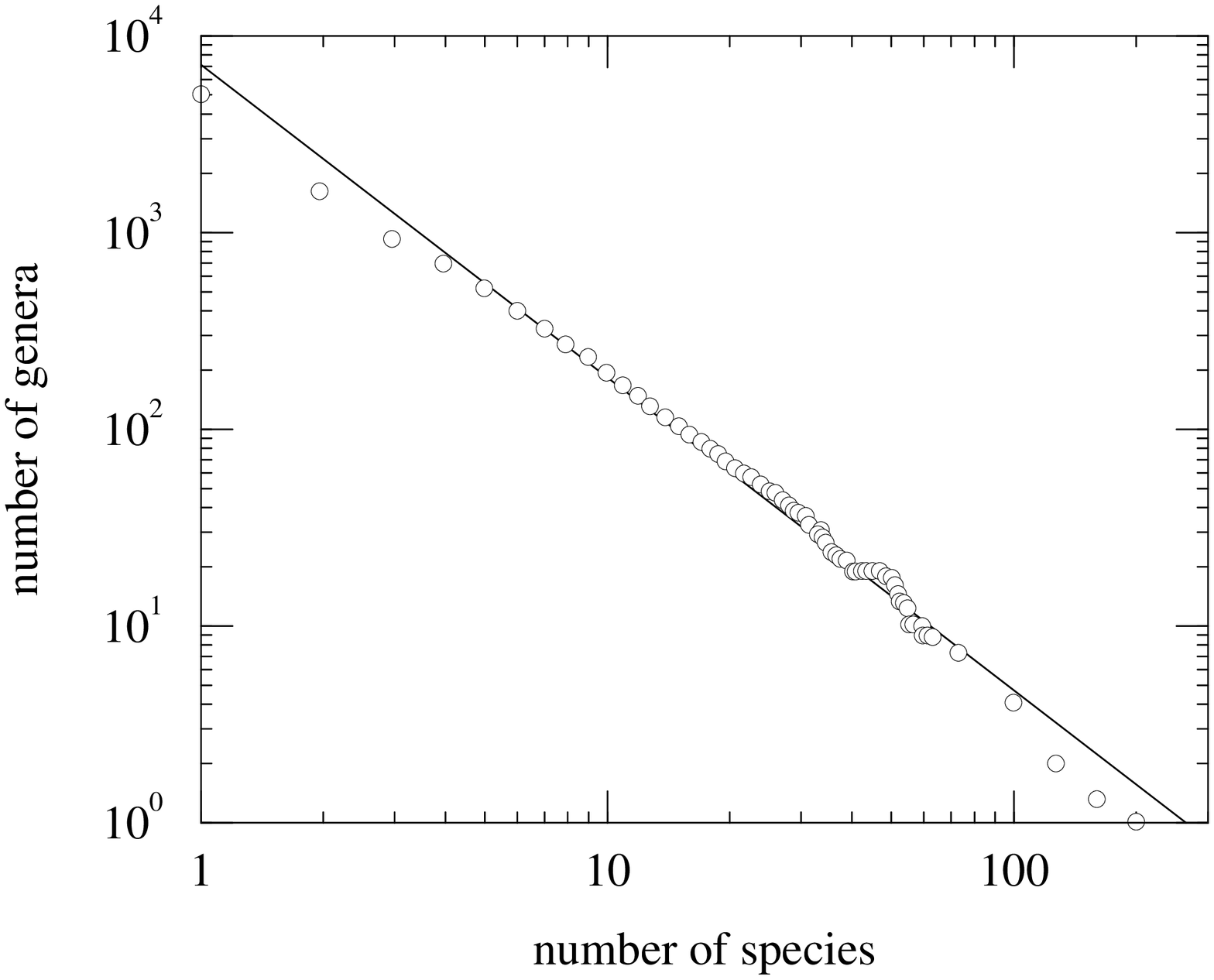}
\capt{Histogram of the number of species per genus of flowering plants.
  The solid line is the best power-law fit to the data.  After
  Willis~(1922).}
\label{willis}
\end{figure}

As long ago as 1922, it was noted that if one takes the taxonomic hierarchy
of current organisms, counts the number of species $n_s$ in each genus, and
makes a histogram of the number of genera $n_g$ for each value of $n_s$,
then the resulting graph has a form which closely follows a power law
(Willis~1922, Williams~1944):
\begin{equation}
n_g \sim n_s^{-\beta}.
\label{willislaw}
\end{equation}
In Figure~\fref{willis}, for example, we reproduce the results of Willis
for the number of species per genus of flowering plants.  The measured
exponent in this case is $\beta=1.5\pm0.1$.  Recently, Burlando~(1990,
1993) has extended these results to higher taxa, showing that the number of
genera per family, families per order, and so forth, also follow power
laws, suggesting that the taxonomic tree has a fractal structure, a result
of some interest to those working on ``critical'' models of extinction (see
Section~\sref{bsspec}).

In certain cases, for example if one makes the assumption that speciation
and extinction rates are stochastically constant, it can be shown that the
average number of species in a genus bears a power-law relation to the
lifetime of the genus, in which case Willis's data are merely another
consequence of the genus lifetime distribution discussed in
Section~\sref{lifetimes}.  Even if this is true however, these data are
nonetheless important, since they are derived from a source entirely
different from the ones we have so far considered, namely from living
species rather than fossil ones.

Note that we need to be careful about the way these distributions are
calculated.  A histogram of genus sizes constructed using fossil data drawn
from a long period of geologic time is not the same thing as one
constructed from a snapshot of genera at a single point in time.  A
snapshot tends to favour longer lived genera which also tend to be larger,
and this produces a histogram with a lower exponent than if the data are
drawn from a long time period.  Most of the models discussed in this review
deal with long periods of geologic time and therefore mimic data of the
latter kind better than those of the former.  Willis's data, which are
taken from living species, are inherently of the ``snapshot'' variety, and
hence may have a lower value of $\beta$ than that seen in fossil data and
in models of extinction.

\subsubsection{Artificial life}
\label{alife}
Artificial life~(Langton~1995) is the name given to a class of evolutionary
simulations which attempt to mimic the processes of natural selection,
without imposing a particular selection regime from outside.  (By contrast,
most other computation techniques employing ideas drawn from evolutionary
biology call upon the programmer to impose fitness functions or
reproductive selection on the evolving population.  Genetic algorithms
(Mitchell~1996) are a good example of such techniques.)  Probably the work
of most relevance to the evolutionary biologist is that of Ray and
collaborators (Ray~1994a, 1994b), who created a simulation environment
known as Tierra, in which computer programs reproduce and compete for the
computational resources of CPU time and memory.  The basic idea behind
Tierra is to create an initial ``ancestor'' program which makes copies of
itself.  The sole function of the program is to copy the instructions which
comprise it into a new area of the computer's memory, so that, after some
time has gone by, there will be a large number of copies of the same
program running at once.  However, the trick is that the system is set up
so that the copies are made in an unreliable fashion.  Sometimes a perfect
copy is made, but sometimes a mistake occurs, so that the copy differs from
the ancestor.  Usually such mistakes result in a program which is not able
to reproduce itself any further.  However, occasionally they result in a
program which reproduces more efficiently than its ancestor, and hence
dominates over the ancestor after a number of generations.  In systems such
as this, many of the features of evolving biological systems have been
observed, such as programs which cooperate in order to aid one another's
efficient reproduction and parasitic programs which steal resources from
others in order to reproduce more efficiently.

\begin{figure}[t]
\columnfigure{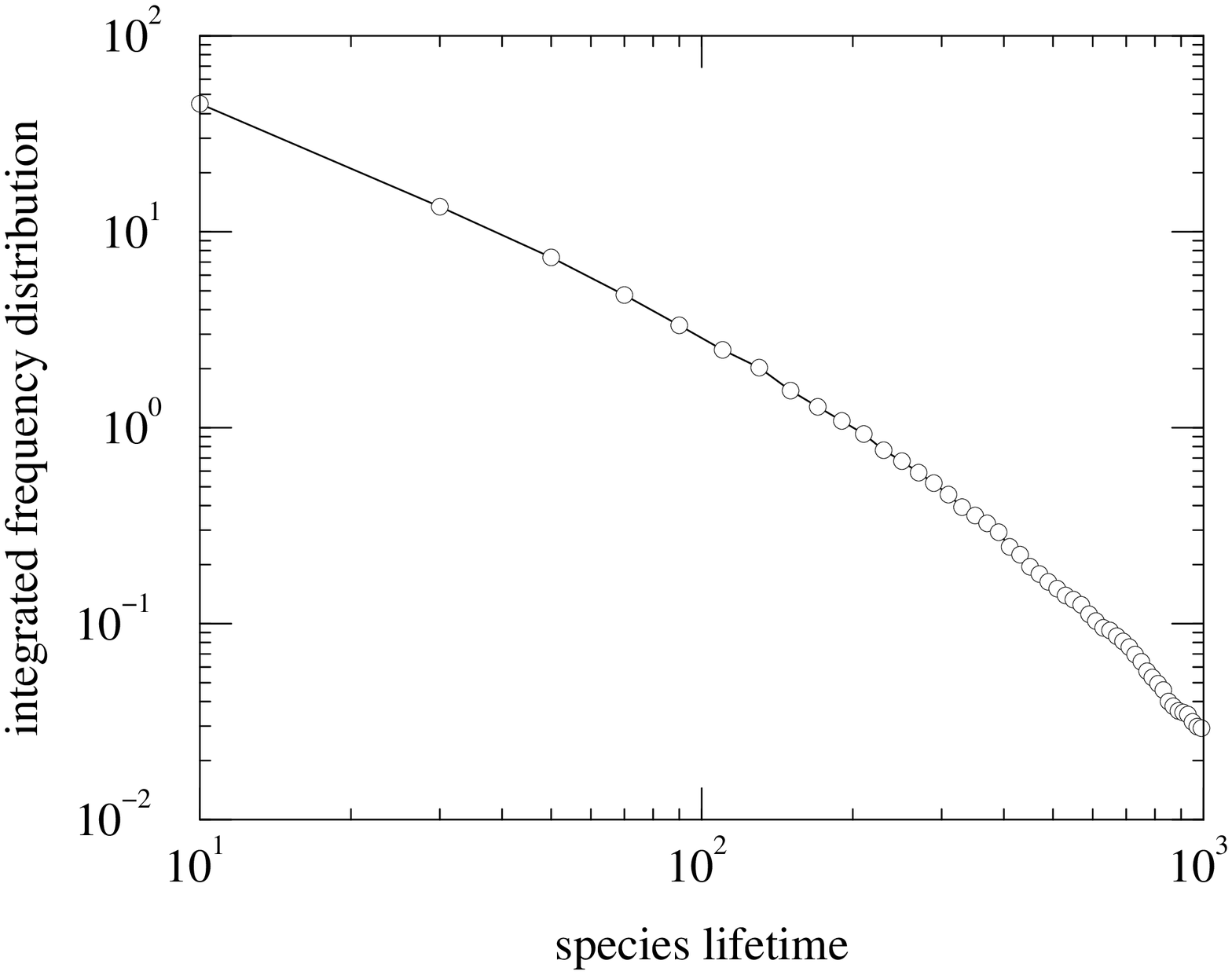}
\capt{Plot of the integrated distribution of ``species'' lifetimes in runs
  of the Tierra artificial life simulation.  The plot is approximately
  power-law in form except for a fall-off at large times due to the finite
  length of the runs.  After Adami~(1995).}
\label{adami}
\end{figure}

In the context of the kinds of models we will be studying here, the recent
work of Adami~(1995) using the Tierra system has attracted attention.  In
his work, Adami performed a number of lengthy runs of the Tierra simulation
and observed the lifetimes of the species appearing throughout the course
of the simulations.  In Figure~\fref{adami} we show some of his results.
The distribution of lifetimes appears again to follow a power law, except
for a fall-off at long lifetimes, which may be accounted for by the finite
length of the simulations.\footnote{Although the integrated distribution in
  Figure~\fref{adami} does not appear to follow a straight line very
  closely, Adami~(1995) shows that in fact it has precisely the form
  expected if the lifetimes are cut off exponentially.} This result appears
to agree with the fossil evidence discussed in Section~\sref{lifetimes},
where the lifetimes of taxa were also found to follow a distribution
approximately power-law in form.  Possible explanations of this result have
been discussed by Newman~\etal~(1997).

\section{Early models of extinction}
\label{early}
Most discussion of extinction has taken place at the species level, which
is natural since extinction is intrinsically a species-level effect---by
extinction we mean precisely the disappearance of a species, although the
concept is frequently extended to cover higher taxa as well.  Our
discussion will also take place mostly at the species and higher taxonomic
levels, but we should bear in mind that the processes underlying extinction
occur at the level of the individual.  McLaren~(1988), for instance, has
argued that it would be better to use the term ``mass killing'', rather
than ``mass extinction'', since it is the death of individuals rather than
species which is the fundamental process taking place.

Although many fossils of extinct species were unearthed during the
eighteenth and early nineteenth centuries, it was not until the theory of
evolution gained currency in the latter half of the nineteenth century that
extinction became an accepted feature of the history of life on Earth.

One of the earliest serious attempts to model extinction was that of
Lyell~(1832) whose ideas, in some respects, still stand up even today.  He
proposed that when species first appear (he did not tackle the then vexed
question of exactly how they appear) they possess varying fitnesses, and
that those with the lowest fitness ultimately become extinct as a result of
selection pressure from other species, and are then replaced by new
species.  While this model does not explain many of the most interesting
features of the fossil record, it does already take a stand on a lot of the
crucial issues in today's extinction debates: it is an equilibrium model
with (perhaps) a roughly constant number of species and it has an explicit
mechanism for extinction (species competition) which is still seriously
considered as one of the causes of extinction.  It also hints at of a way
of quantifying the model by using a numerical fitness measure.

A few years after Lyell put forward his ideas about extinction, Darwin
extended them by emphasizing the appearance of new species through
speciation from existing ones.  In his view, extinction arose as a result
of competition between species and their descendants, and was therefore
dominated by the process which we referred to as ``pseudoextinction'' in
Section~\sref{pseudoextinction}.  The Darwin--Lyell viewpoint is
essentially a gradualist one.  Species change gradually, and become extinct
one by one as they are superseded by new fitter variants.  As Darwin wrote
in the {\sl Origin of Species\/} (Darwin~1859): ``Species and groups of
species gradually disappear, one after another, first from one spot, then
from another, and finally from the world.''  The obvious problem with this
theory is the regular occurrence of mass extinctions in the fossil record.
Although the existence of mass extinctions was well-known in Darwin's time,
Darwin and Lyell both argued strongly that they were probably a sampling
artifact generated by the inaccuracy of dating techniques rather than a
real effect.  Today we know this not to be the case, and a purely
gradualist picture no longer offers an adequate explanation of the facts.
Any serious model of extinction must take mass extinction into account.

\begin{figure}[t]
\columnfigure{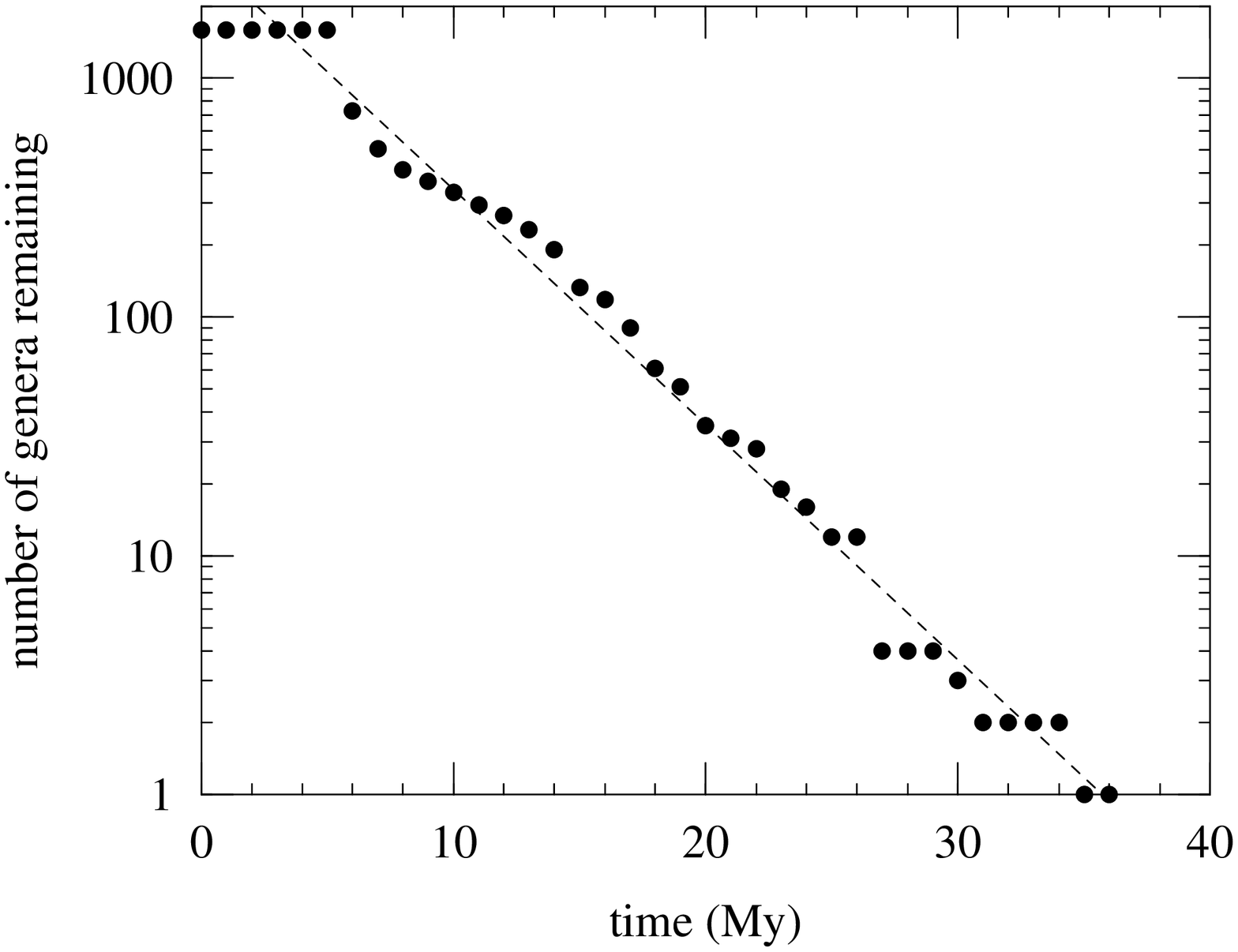}
\capt{The number of genera of mammals surviving out of an initial
  group of 1585, over a period of 36~My.  The dotted line is the best
  fit exponential, and has a time constant of $4.41\pm0.08$~My.  After
  van~Valen~(1973).}
\label{vanvalen}
\end{figure}

With the advent of reasonably comprehensive databases of fossil species, as
well as computers to aid in their analysis, a number of simple models
designed to help interpret and understand extinction data were put forward
in the 1970s and 1980s.  In 1973, van~Valen proposed what he called the
``Red Queen hypothesis'': the hypothesis that the probability per unit time
of a particular species becoming extinct is independent of time.  This
``stochastically constant'' extinction is equivalent to saying that the
probability of a species surviving for a certain length of time $t$ decays
exponentially with $t$.  This is easy to see, since if $p$ is the constant
probability per unit time of the species becoming extinct, then $1-p$ is
the probability that it does not become extinct in any unit time interval,
and
\begin{equation}
P(t) = (1-p)^t = \e^{-t/\tau}
\end{equation}
is the probability that it survives $t$ consecutive time intervals, where
\begin{equation}
\tau = -{1\over\log(1-p)} \simeq {1\over p},
\end{equation}
where the second relation applies for small $p$.  Van~Valen used this
argument to validate his hypothesis, by plotting ``survivorship curves''
for many different groups of species (van~Valen~1973).  A survivorship
curve is a plot of the number of species surviving out of an initial group
as a function of time starting from some arbitrary origin.  In other words,
one takes a group of species and counts how many of them are still present
in the fossil record after time $t$.  It appears that the time constant
$\tau$ is different for the different groups of organisms examined by van
Valen but roughly constant within groups, and in this case the survivorship
curves should fall off exponentially.  In Figure~\fref{vanvalen} we
reproduce van~Valen's results for extinct genera of mammals.  The
approximately straight-line form of the survivorship curve on
semi-logarithmic scales indicates that the curve is indeed exponential, a
result now known as ``van~Valen's law''.  Van~Valen constructed similar
plots for many other groups of genera and families and found similar
stochastically constant extinction there as well.

Van Valen's result, that extinction is uniform in time has been used as the
basis for a number of other simple extinction models, some of which are
discussed in this paper.  However, for a number of reasons, it must
certainly be incorrect.  First, it is not mathematically possible for
van~Valen's law to be obeyed at more than one taxonomic level.  As
Raup~(1991b) has demonstrated, if species become extinct at a
stochastically constant rate $p$, the survivorship curve $S$ for {\em
  genera\/} will not in general be exponential, because it depends not only
on the extinction rate but also on the speciation rate.  The general form
for the genus survivorship curve is
\begin{equation}
S = 1 - {p[\e^{(q-p)t} - 1]\over q\e^{(q-p)t} - p},
\label{rauplaw}
\end{equation}
where $q$ is the average rate of speciation within the genus.  A similar
form applies for higher taxa as well.

Second, van~Valen's law clearly cannot tell the whole story since, just
like the theories of Lyell and Darwin, it is a gradualist model and takes
no account of known mass extinction events in the fossil record.
Raup~(1991b, 1996) gives the appropriate generalization of van~Valen's work
to the case in which extinction is not stochastically constant.  In this
case, the {\em mean\/} survivorship curve follows van~Valen's law (or
Equation~\eref{rauplaw} for higher taxa), but individual curves show a
dispersion around this mean whose width is a measure of the distribution of
the sizes of extinction events.  It was in this way that Raup extracted the
kill curve discussed in Section~\sref{rates} for Phanerozoic marine
invertebrates.

These models however, are all fundamentally just different ways of looking
at empirical data.  None of them offer actual explanations of the observed
distributions of extinction events, or explain the various forms discussed
in Section~\sref{data}.  In the remainder of this review we discuss a
variety of quantitative models which have been proposed in the last ten
years to address these questions.

\section{Fitness landscape models}
\label{flmodels}
Kauffman~(1993, 1995, Kauffman and Levin~1987, Kauffman and Johnsen~1991)
has proposed and studied in depth a class of models referred to as \NK\
models, which are models of random fitness landscapes on which one can
implement a variety of types of evolutionary dynamics and study the
development and interaction of species.  (The letters $N$ and $K$ do not
stand for anything, they are the names of parameters in the model.)  Based
on the results of extensive simulations of \NK\ models Kauffman and
co-workers have suggested a number of possible connections between the
dynamics of evolution and the extinction rate.  To a large extent it is
this work which has sparked recent interest in biotic mechanisms for mass
extinction.  In this section we review Kauffman's work in detail.

\subsection{The NK model}
\label{nk}
An \NK\ model is a model of a single rugged landscape, which is similar in
construction to the spin-glass models of statistical physics (Fischer and
Hertz~1991), particularly $p$-spin models (Derrida~1980) and random energy
models (Derrida~1981).  Used as a model of species fitness\footnote{\NK\
  models have been used as models of a number of other things as
  well---see, for instance, Kauffman and Weinberger~(1989) and Kauffman and
  Perelson~(1990).} the \NK\ model maps the states of a model genome onto a
scalar fitness $W$.  This is a simplification of what happens in real life,
where the genotype is first mapped onto phenotype and only then onto
fitness.  However, it is a useful simplification which makes simulation of
the model for large systems tractable.  As long as we bear in mind that
this simplification has been made, the model can still teach us many useful
things.
\begin{figure*}
\hbox to \textwidth{\hfil
\hbox to \captwidth{%
\begin{tabular}[b]{c|ccc|c}
genotype & $w_1$ & $w_2$ & $w_3$ & $W$ \\
\hline
\hline
000 & 0.487 & 0.076 & 0.964 & 0.509 \\
001 & 0.851 & 0.372 & 0.398 & 0.540 \\
010 & 0.487 & 0.097 & 0.162 & 0.249 \\
011 & 0.851 & 0.566 & 0.062 & 0.493 \\
100 & 0.235 & 0.076 & 0.964 & 0.425 \\
101 & 0.311 & 0.372 & 0.398 & 0.360 \\
110 & 0.235 & 0.097 & 0.162 & 0.165 \\
111 & 0.311 & 0.566 & 0.062 & 0.313 \\
\end{tabular}
\hfil
\resizebox{3.5cm}{!}{\includegraphics{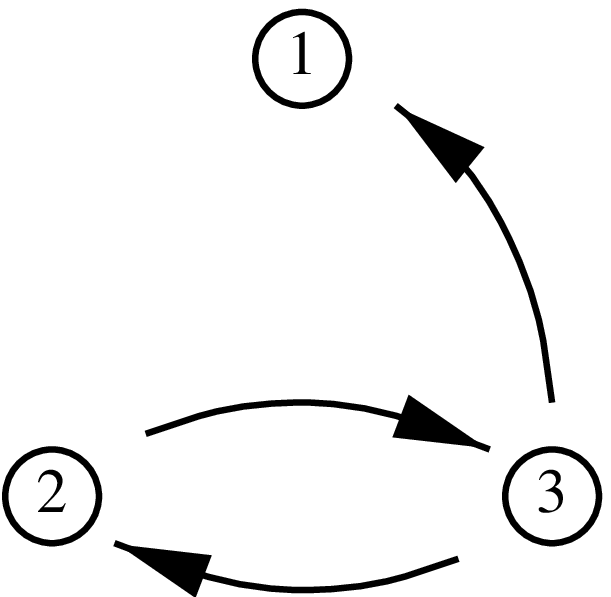}}
}\hfil}
\medskip
\widecapt{Calculation of the fitnesses for an \NK\ model with three binary
  genes.  In this case $K=1$ with the epistatic interactions as indicated
  in the figure on the right.}
\label{nkf}
\end{figure*}

The \NK\ model is a model of a genome with $N$ genes.  Each gene has $A$
alleles.  In most of Kauffman's studies of the model he used $A=2$, a
binary genetic code, but his results are not limited to this case.  The
model also includes epistatic interactions between genes---interactions
whereby the state of one gene affects the contribution of another to the
overall fitness of the species.  In fact, it is these epistatic
interactions which are responsible for the ruggedness of the fitness
landscape.  Without any interaction between genes it is possible (as we
will see) to optimize individually the fitness contribution of each single
gene, and hence to demonstrate that the landscape has the so-called
Fujiyama form, with only a single global fitness peak.

In the simplest form of the \NK\ model, each gene interacts epistatically
with $K$ others, which are chosen at random.  The fitness contribution
$w_j$ of gene $j$ is a function of the state of the gene itself and each of
the $K$ others with which it interacts.  For each of the $A^{K+1}$ possible
states of these $K+1$ genes, a value for $w_j$ is chosen randomly from a
uniform distribution between zero and one.  The total fitness is then the
average over all genes of their individual fitness contributions:
\begin{equation}
W = {1\over N} \sum_{j=1}^N w_j.
\label{nkfitness}
\end{equation}
This procedure is illustrated in Figure~\fref{nkf} for a simple three-gene
genome with $A=2$ and $K=1$.

Some points to notice about the \NK\ model are:
\begin{enumerate}
\item The choices of the random numbers $w_j$ are ``quenched'', which is to
  say that once they have been chosen they do not change again.  The
  choices of the $K$ other genes with which a certain gene interacts are
  also quenched.  Thus the fitness attributed to a particular genotype is
  the same every time we look at it.
\item There is no correlation between the contribution $w_j$ of gene $j$ to
  the total fitness for different alleles of the gene, or for different
  alleles of any of the genes with which it interacts.  If any single one
  of these $K+1$ genes is changed to a different state, the new value of
  $w_j$ is completely unrelated to its value before the change.  This is an
  extreme case.  In reality, epistatic interactions may have only a small
  effect on the fitness contribution of a gene.  Again, however, this is a
  simplifying assumption which makes the model tractable.
\item In order to think of the \NK\ model as generating a fitness
  ``landscape'' with peaks and valleys, we have to say which genotypes are
  close together and which far apart.  In biological evolution, where the
  most common mutations are mutations of single genes, it makes sense to
  define the distance between two genotypes to be the number of genes by
  which they differ.  This definition of distance, or ``metric'', is used
  in all the studies discussed here.  A (local) peak is then a genotype
  that has higher fitness than all $N(A-1)$ of its nearest neighbours,
  those at distance 1 away.
\item The fact of taking an {\em average\/} over the fitness contributions
  of all the genes in Equation~\eref{nkfitness} is crucial to the behaviour
  of the model.  Taking the average has the effect that the typical height
  of fitness peaks diminishes with increasing $N$.  In fact, one can
  imagine defining the model in a number of other ways.  One could simply
  take the total fitness to be the sum of the contributions from all the
  genes---organisms with many genes therefore tending to be fitter than
  ones with fewer.  In this case one would expect to see the reverse of the
  effect described above, with the average height of adaptive peaks
  increasing with increasing $N$.  One might also note that since $W$ is
  the sum of a number of independent random variables, its values should,
  by the central limit theorem, be approximately normally distributed with
  a standard deviation increasing as $\sqrt{N}$ with the number of genes.
  Therefore, it might make sense to normalize the sum with a factor of
  $N^{-1/2}$, so that the standard deviation remains constant as $N$ is
  varied.  Either of these choices would change some of the details of the
  model's behaviour.  For the moment however, we stick with the model as
  defined above.
\end{enumerate}

What kind of landscapes does the \NK\ model generate?  Let us begin by
considering two extreme cases.  First, consider the case $K=0$, in which
all of the genes are entirely non-interacting.  In this case, each gene
contributes to the total fitness an amount $w_j$, which may take any of $A$
values depending on the allele of the gene.  The maximum fitness in this
case is achieved by simply maximizing the contribution of each gene in
turn, since their contributions are independent.  Even if we assume an
evolutionary dynamics of the most restrictive kind, in which we can only
change the state of one gene at a time, we can reach the state of maximum
fitness of the $K=0$ model starting from any point on the landscape and
only making changes which increase the fitness.  Landscapes of this type
are known as Fujiyama landscapes, after Japan's Mount Fuji: they are smooth
and have a single global optimum.

Now consider the other extreme, in which $K$ takes the largest possible
value, $K=N-1$.  In this case each gene's contribution to the overall
fitness $W$ depends on itself and all $N-1$ other genes in the genome.
Thus if any single gene changes allele, the fitness contribution of every
gene changes to a new random number, uncorrelated with its previous value.
Thus the total fitness $W$ is entirely uncorrelated between different
states of the genome.  This gives us the most rugged possible fitness
landscape with many fitness peaks and valleys.  The $K=N-1$ model is
identical to the random energy spin-glass model of Derrida~(1981) and has
been studied in some detail~(Kauffman and Levin~1987, Macken and
Perelson~1989).  The fitness $W$ in this case is the average of $N$
independent uniform random variables between zero and one, which means that
for large $N$ it will be normally distributed about $W=\frac12$ with
standard deviation $1/\sqrt{12N}$.  This means that the typical height of
the fitness peaks on the landscape decreases as $N^{-1/2}$ with increasing
size of the genome.  It also decreases with increasing $K$, since for
larger $K$ it is not possible to achieve the optimum fitness contribution
of every gene, so that the average over all genes has a lower value than
$K=0$ case, even at the global optimum.

For values of $K$ intermediate between the two extremes considered here,
the landscapes generated by the \NK\ model possess intermediate degrees of
ruggedness.  Small values of $K$ produce highly correlated, smooth
landscapes with a small number of high fitness peaks.  High values of $K$
produce more rugged landscapes with a larger number of lower peaks and less
correlation between the fitnesses of similar genotypes.

\subsection{Evolution on NK landscapes}
\label{nkevo}
In order to study the evolution of species using his \NK\ landscapes,
Kauffman made a number of simplifying assumptions.  First, he assumed that
evolution takes place entirely by the mutation of single genes, or small
numbers of genes in an individual.  That is, he neglected recombination.
(This is a reasonable first approximation since, as we mentioned above,
single gene mutations are the most common in biological evolution.) He also
assumed that the mutation of different genes are {\it a priori\/}
uncorrelated, that the rate at which genes mutate is the same for all
genes, and that that rate is low compared to the time-scale on which
selection acts on the population.  This last assumption means that the
population can be approximated by a single genotype, and population
dynamical effects can be ignored.  (This may be valid for some populations,
but is certainly not true in general.)

In addition to these assumptions it is also necessary to state how the
selection process takes place, and Kauffman examined three specific
possibilities, which he called the ``random'', ``fitter'' and ``greedy''
dynamics.  If, as discussed above, evolution proceeds by the mutations of
single genes, these three possibilities are as follows.  In the random
dynamics, single-gene mutations occur at random and, if the mutant genotype
possesses a higher value of $W$ than its ancestral strain, the mutant
replaces the ancestor and the species ``moves'' on the landscape to the new
genotype.  A slight variation on this scheme is the fitter dynamics, in
which a species examines all the genotypes which differ from the current
genotype by the mutation of a single gene, its ``neighbours'', and then
chooses a new genotype from these, either in proportion to fitness, or
randomly amongst those which have higher fitness than the current genotype.
(This last variation differs from the previous scheme only in a matter of
time-scale.)  In the greedy dynamics, a species examines each of its
neighbours in turn and chooses the one with the highest fitness $W$.
Notice that whilst the random and fitter schemes are stochastic processes,
the greedy one is deterministic; this gives rise to qualitative differences
in the behaviour of the model.

The generic behaviour of the \NK\ model of a single species is for the
fitness of the species to increase until it reaches a local fitness
peak---a genotype with higher fitness than all of the neighbouring
genotypes on the landscape---at which point it stops evolving.  For the
$K=0$ case considered above (the Fujiyama landscape), it will typically
take on the order of $N$ mutations to find the single fitness peak (or
$N\log N$ for the random dynamics).  For instance, in the $A=2$ case, half
of the alleles in a random initial genotype will on average be favourable
and half unfavourable.  Thus if evolution proceeds by the mutation of
single genes, $\frac12 N$ mutations are necessary to reach the fitness
maximum.  In the other extreme, when $K=N-1$, one can show that, starting
from a random initial genotype, the number of directions which lead to
higher fitness decreases by a constant factor at each step, so that the
number of steps needed to reach one of the local maxima of fitness goes as
$\log N$.  For landscapes possessing intermediate values of $K$, the number
of mutations needed to reach a local maximum lies somewhere between these
limits.  In other words, as $N$ becomes large, the length of an adaptive
walk to a fitness peak decreases sharply with increasing $K$.  In fact, it
appears to go approximately as $1/K$.  This point will be important in our
consideration of the many-species case.  Recall also that the height of the
typical fitness peak goes down with increasing $K$.  Thus when $K$ is high,
a species does not have to evolve far to find a local fitness optimum, but
in general that optimum is not very good.

\subsection{Coevolving fitness landscapes}
\label{nkcoevo}
The real interest in \NK\ landscapes arises when we consider the behaviour
of a number of coevolving species.  Coevolution arises as a result of
interactions between different species.  The most common such interactions
are predation, parasitism, competition for resources, and symbiosis.  As a
result of interactions such as these, the evolutionary adaptation of one
species can prompt the adaptation of another (Vermeij~1987).  Many
examples are familiar to us, especially ones involving predatory or
parasitic interactions.  Plotnick and McKinney~(1993) have given a number
of examples of coevolution in fossil species, including predator-prey
interactions between echinoids and gastropods (McNamara~1990) and
mutualistic interactions between algae and foraminifera~(Hallock~1985).

How is coevolution introduced into the \NK\ model?  Consider $S$ species,
each evolving on a different \NK\ landscape.  For the moment, let us take
the simplest case in which each species has the same values of $N$ and $K$,
but the random fitnesses $w_j$ defining the landscapes are different.
Interaction between species is achieved by coupling their landscapes so
that the genotype of one species affects the fitness of another.  Following
Kauffman and Johnsen~(1991), we introduce two new quantities: $S_i$ which
is the number of neighbouring species with which species $i$
interacts,\footnote{Although this quantity is denoted $S_i$, it is in fact
  a constant over all species in most of Kauffman's studies; the subscript
  $i$ serves only to distinguish it from $S$, which is the total number of
  species.  Of course, there is no reason why one cannot study a
  generalized model in which $S_i$ (or indeed any of the other variables in
  the model, such as $N$ or $K$) is varied from species to species, and
  Kauffman and Johnsen~(1991) give some discussion and results for models
  of this type, although this is not their main focus.} and $C$ which is
the number of genes in each of those neighbouring species which affect the
fitness contribution of each gene in species $i$.  On account of these two
variables this variation of the model is sometimes referred to as the
\NKCS\ model.

Each gene in species $i$ is ``coupled'' to $C$ randomly chosen genes in
each of the $S_i$ neighbouring species, so that, for example, if $C=1$ and
$S_i=4$, each of $i$'s genes is coupled to four other genes, one randomly
chosen from each of four neighbouring species.  The coupling works in
exactly the same way as the epistatic interactions of the last
section---the fitness contribution $w_j$ which a particular gene $j$ makes
to the total fitness of its host is now a function of the allele of that
gene, of each of the $K$ genes to which it is coupled {\em and\/} of the
alleles of the $CS_i$ genes in other species with which it interacts.  As
before, the values $w_j$ are chosen randomly for each of the possible
states of these genes.

The result is that when a species evolves so as to improve its own fitness,
it may in the process change the allele of one of its genes which affects
the fitness contribution of a gene in another species.  As a result, the
fitness of the other species will change.  Clearly the further a species
must evolve to find a fitness peak, the more alleles it changes, and the
more likely it is to affect the fitness of its neighbours.  Since the
distance to a fitness peak depends on the value of $K$, so also does the
chance of one species affecting another, and this is the root cause of the
novel behaviour seen in Kauffman's coevolution models.

The $S_i$ neighbouring species of species $i$ can be chosen in a variety of
different ways.  The most common are either to chose them at random (but in
a ``quenched'' fashion---once chosen, they remain fixed) or to place the
species on a regular lattice, such as a square lattice in two dimensions,
and then make the nearest neighbours of a species on the lattice its
neighbours in the evolutionary sense.

In their original work on coevolving \NK\ systems, Kauffman and
Johnsen~(1991) examined a number of different variations on the basic model
outlined above.  Here we consider the simplest case of relevance to
extinction, the case of uniform $K$ and $S_i$.

\subsection{Coevolutionary avalanches}
\label{nkaval}
Consider the case of binary genes ($A=2$), with single-gene mutations.
Starting from an initial random state, species take turns in strict
rotation, and attempt by mutation to increase their own fitness
irrespective of anyone else's.  It is clear that if at any time all species
in the system simultaneously find themselves at local fitness optima then
all evolution will stop, since there will be no further mutations of any
species which can increase fitness.  This state is known as a Nash
equilibrium, a name taken from game theoretic models in which similar
situations arise.\footnote{A related concept is that of the
  ``evolutionarily stable strategy'' (Maynard Smith and Price~1973), which
  is similar to a Nash equilibrium but also implies non-invadability at the
  individual level.  The simulations of Kauffman and Johnsen considered
  here take place entirely at the species level, so ``Nash equilibrium'' is
  the appropriate nomenclature in this case.} The fundamental question is
whether such an equilibrium is ever reached.  This, it turns out, depends
on the value of $K$.

For large values of $K$, individual species landscapes are very rugged, and
the distance that a species needs to go to reach a local fitness maximum is
short.  This means that the chance of it affecting its neighbours' fitness
is rather small, and hence the chance of all species simultaneously finding
a fitness maximum is quite good.  On the other hand, if $K$ is small,
species must change many genes to reach a fitness maximum, and so the
chances are high that they will affect the fitnesses of their neighbours.
This in turn will force those neighbours to evolve, by moving the position
of the maxima in their landscapes.  They in turn may have to evolve a long
way to find a new maximum, and this will affect still other species,
resulting in an avalanche of coevolution which for small enough $K$ never
stops.  Thus as $K$ is decreased from large values to small, the typical
size of the coevolutionary avalanche resulting from a random initial state
increases until at some critical value $K_c$ it becomes infinite.

What is this critical value?  The product $CS_i$ is the number of genes in
other species on which the fitness contribution of a particular gene in
species $i$ depends.  A rough estimate of the chance that at least one of
these genes mutates during an avalanche is $CS_iL$, where $L$ is the
typical length of an adaptive walk of an isolated species (i.e.,~the number
of genes which change in the process of evolving to a fitness peak).
Assuming, as discussed in Section~\sref{nkevo}, that $L$ varies inversely
with $K$, the critical value $K_c$ at which the avalanche size diverges
should vary as $K_c\sim CS_i$.  This seems to be supported by numerical
evidence: Kauffman and Johnsen found that $K_c\simeq CS_i$ in the
particular case where every species is connected to every other ($S_i=S$).

The transition from the high-$K$ ``frozen'' regime in which avalanches are
finite to the low-$K$ ``chaotic'' regime in which they run forever appears
to be a continuous phase transition of the kind much studied in statistical
physics (Binney~\etal~1992).  Bak~\etal~(1992) have analysed this
transition in some detail, showing that it does indeed possess genuine
critical properties.  Precisely at $K_c$, the distribution of the sizes $s$
of the avalanches appears to be scale free and takes the form of a power
law, Equation~\eref{powerlaw}, which is typical of the ``critical
behaviour'' associated with such a phase transition.  Kauffman and Johnson
also pointed out that there are large fluctuations in the fitness of
individual species near $K_c$, another characteristic of continuous phase
transitions.

\begin{figure}[t]
\columnfigure{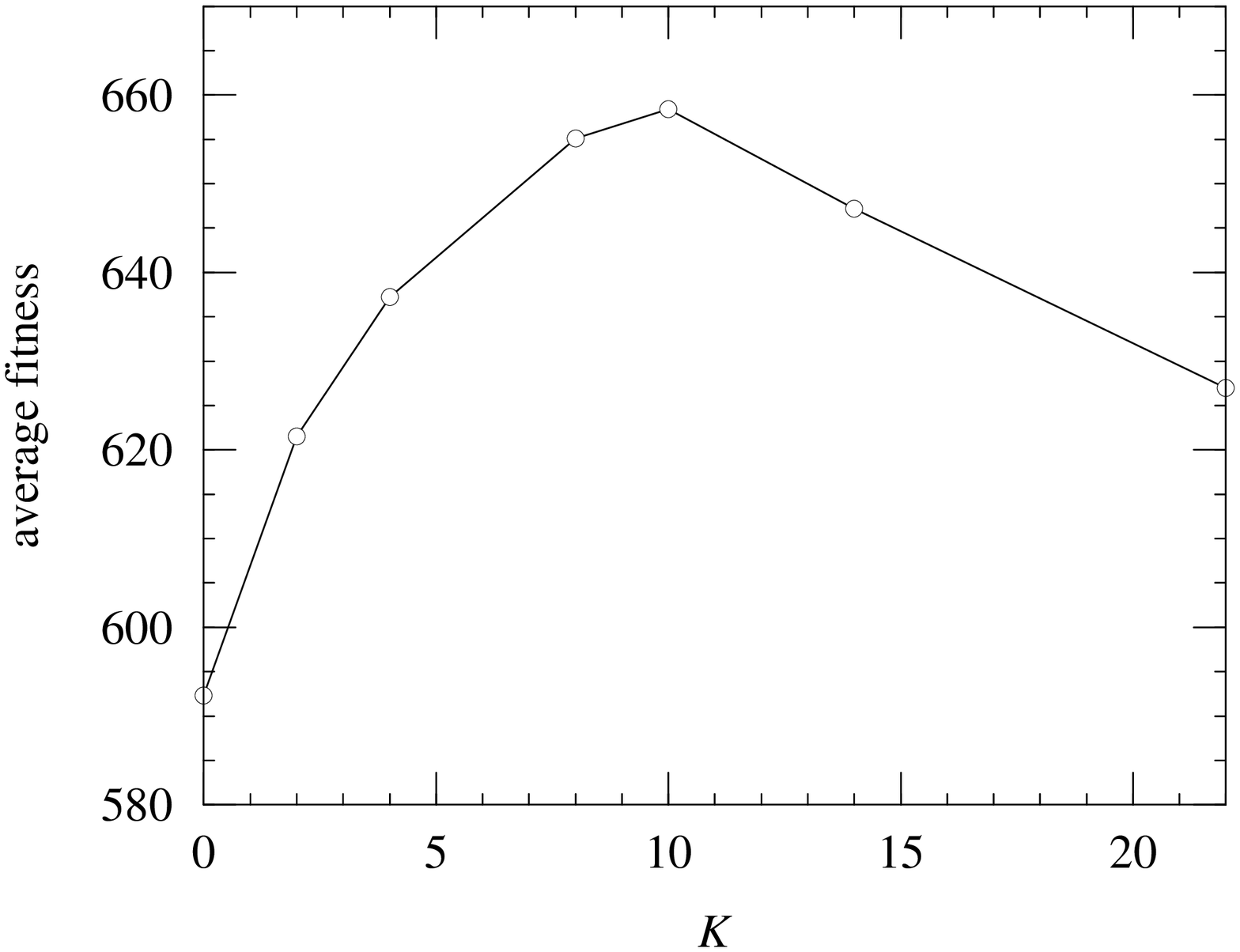}
\capt{The average fitness of species in an \NKCS\ model as a function of
  $K$.  Twenty-five species were arranged in a $5\times5$ array so that
  each one interacted with $S_i=4$ neighbours (except for those on the
  edges, for which $S_i=3$, and those at the corners, for which $S_i=2$).
  Each species had $N=24$ and $C=1$.  The fitness plotted is that of the
  Nash equilibrium if reached, or the time average after transients if not.
  After Kauffman and Johnsen~(1991).}
\label{nkfitter}
\end{figure}

Figure~\fref{nkfitter} shows the average fitness of the coevolving species
as a function of $K$ for one particular case investigated by Kauffman and
Johnsen.  For ecosystems in the frozen $K>K_c$ regime the average fitness
of the coevolving species increases from the initial random state until a
Nash equilibrium is reached, at which point the fitness stops changing.  As
we pointed out earlier, the typical fitness of local optima increases with
decreasing $K$, and this is reflected in the average fitness at Nash
equilibria in the frozen phase: the average fitness of species at
equilibrium increases as $K$ approaches $K_c$ from above.

In the chaotic $K<K_c$ regime a Nash equilibrium is never reached, but
Kauffman and Johnsen measured the ``mean sustained fitness'', which is the
average fitness of species over time, after an initial transient period in
which the system settles down from its starting state.  They found that
this fitness measure {\em decreased\/} with decreasing $K$ in the chaotic
regime, presumably because species spend less and less time close to local
fitness optima.  Thus, there should be a maximum of the average fitness at
the point $K=K_c$.  This behaviour is visible in Figure~\fref{nkfitter},
which shows a clear maximum around $K=10$.  The boundary between frozen and
chaotic regimes was separately observed to occur at around $K_c=10$ for
this system.

On the basis of these observations, Kauffman and Johnsen then argued as
follows.  If the level of epistatic interactions in the genome is an
evolvable property, just as the functions of individual genes are, and our
species are able to ``tune'' the value of their own $K$ parameter to
achieve maximum fitness, then Figure~\fref{nkfitter} suggests that they
will tune it to the point $K=K_c$, which is precisely the critical point at
which we expect to see a power-law distribution of coevolutionary
avalanches.  As we suggested in Section~\sref{pseudoextinction}, mass
extinction could be caused by pseudoextinction processes in which a large
number of species evolve to new forms nearly simultaneously.  The
coevolutionary avalanches of the \NKCS\ model would presumably give rise to
just such large-scale pseudoextinction.  Another possibility, also noted by
Kauffman and Johnson is that the large fluctuations in species fitness in
the vicinity of $K_c$ might be a cause of true extinction, low fitness
species being more susceptible to extinction than high fitness ones.

These ideas are intriguing, since they suggest that by tuning itself to the
point at which average fitness is maximized, the ecosystem also tunes
itself precisely to the point at which species turnover is maximized, and
indeed this species turnover is a large part of the reason why $K=K_c$ is a
fit place to be in first place.  Although extinction and pseudoextinction
can certainly be caused by exogenous effects, even without these effects we
should still see mass extinction.

Some efforts have been made to determine from the fossil evidence whether
real evolution has a dynamics similar to the one proposed by Kauffman and
co-workers.  For example, Patterson and Fowler~(1996) analysed fossil data
for planktic foraminifera using a variety of time-series techniques and
concluded that the results were at least compatible with critical theories
such as Kauffman's, and Sol\'e~\etal~(1997) argued that the form of the
extinction power spectrum may indicate an underlying critical
macroevolutionary dynamics, although this latter suggestion has been
questioned (Kirchner and Weil~1998, Newman and Eble~1999a).

\subsection{Competitive replacement}
\label{nkcompete}
There is however a problem with the picture presented above.  Numerically,
it appears to be true that the average fitness of species in the model
ecosystem is maximized when they all have $K$ close to the critical value
$K_c$.  However, it would be a mistake to conclude that the system
therefore must evolve to the critical point under the influence of
selection pressure.  Natural selection does not directly act to maximize
the average fitness of species in the ecosystem, but rather it acts to
increase individual fitnesses in a selfish fashion.  Kauffman and Johnsen
in fact performed simulations in which only two species coevolved, and they
found that the fitness of both species was greater if the two had different
values of $K$ than if both had the value of $K$ which maximized mean
fitness.  Thus, in a system in which many species could freely vary their
own $K$ under the influence of selection pressure, we would expect to find
a range of $K$ values, rather than all $K$ taking the value $K_c$.

There are also some other problems with the original \NKCS\ model.  For
instance, the values of $K$ in the model were not actually allowed to vary
during the simulations, but one would like to include this possibility.  In
addition, the mechanism by which extinction arises is rather vague; the
model really only mimics evolution and the idea of extinction is tacked on
somewhat as an afterthought.

To tackle all of these problems Kauffman and Neumann~(unpublished) proposed
a refinement of the \NKCS\ model in which $K$ can change and an explicit
extinction mechanism is included, that of competitive replacement.  (An
account of this work can be found in Kauffman~(1995).)  In this variation
of the model, a number $S$ of species coevolve on \NK\ fitness landscapes
just as before.  Now however, at each turn in the simulation, each species
may change the state of one of its genes, change the value of its $K$ by
$\pm1$, it may be invaded by another species (see below), or it can do
nothing.  In their calculations, Kauffman and Neumann used the ``greedy''
dynamics described above and choose the change which most improves the
fitness, but ``fitter'' and ``random'' variants are also possible.
Allowing $K$ to vary gives species the ability to evolve the ruggedness of
their own landscapes in order to optimize their fitness.

Extinction takes place in the model when a species invades the niche
occupied by another.  If the invading species is better at exploiting the
particular set of resources in the niche, it drives the niche's original
occupant to extinction.  In this model, a species' niche is determined by
its neighbouring species---there is no environmental component to the
niche, such as climate, terrain, or food supply.  Extinction by competitive
replacement is actually not a very well-documented mode of extinction
(Benton~1987).  Maynard Smith~(1989) has discussed the question at some
length, but concludes that it is far more common for a species to adapt to
the invasion of a new competitor than for it to become extinct.
Nonetheless, there are examples of extinction by competitive replacement,
and to the extent that it occurs, Kauffman and Neumann's work provides a
model of the process.  In the model, they add an extra ``move'' which can
take place when a species' turn comes to evolve: it can be invaded by
another species.  A randomly chosen species can create a copy of itself
(i.e.,~of its genome) which is then placed in the same niche as the first
species and its fitness is calculated with respect to the genotypes of the
neighbours in that niche.  If this fitness exceeds the fitness of the
original species in that niche, the invader supersedes the original
occupant, which becomes extinct.  In this way, fit species spread through
the ecosystem making the average fitness over all species higher, but at
the same time making the species more uniform, since over time the
ecosystem will come to contain many copies of a small number of fit
species, rather than a wide diversity of less fit ones.

\begin{figure}[t]
\columnfigure{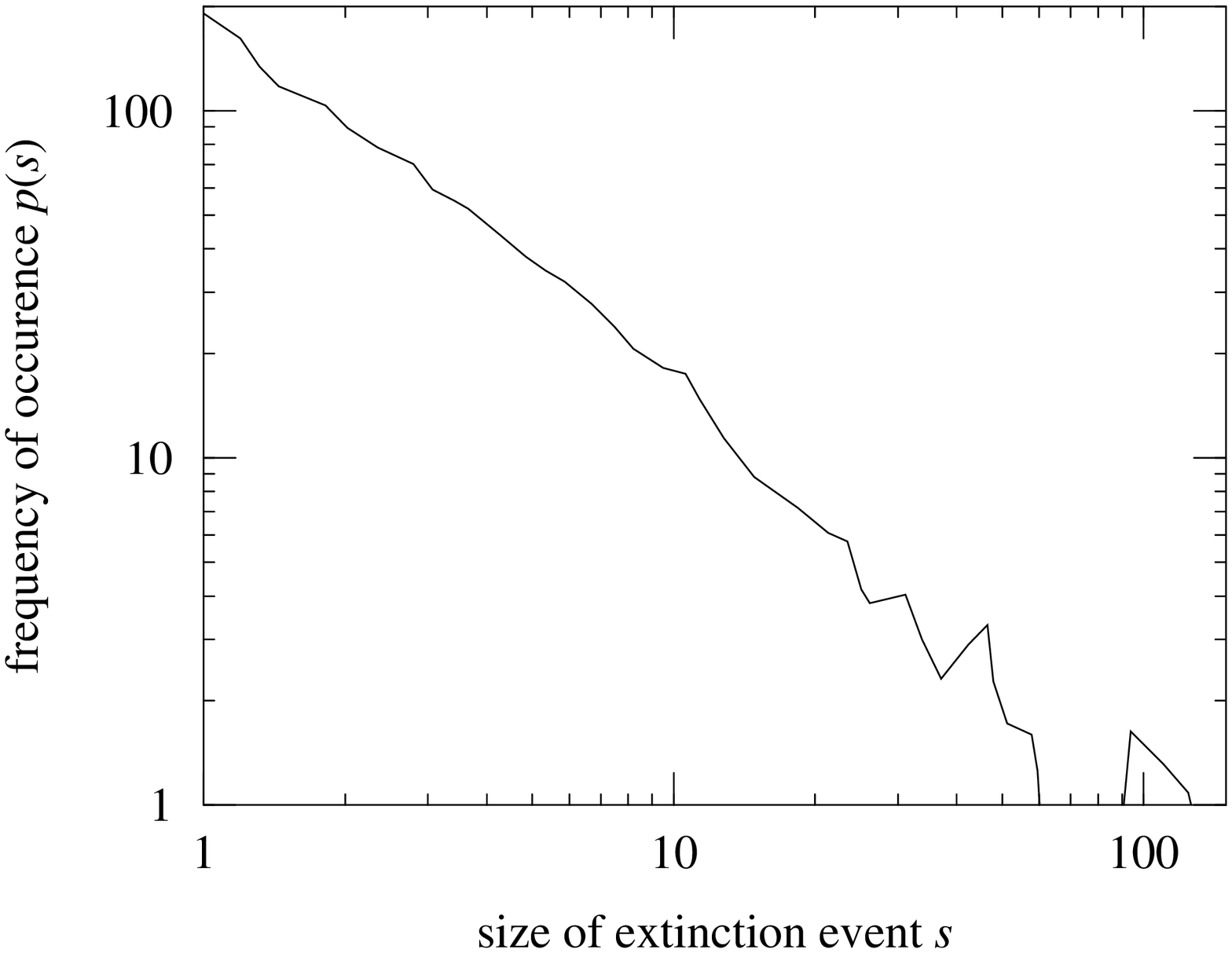}
\capt{The distribution of the sizes of extinction events measured in a
  simulation of the model described in the text.  The distribution is
  approximately power-law in form with an exponent measured to be
  $\tau=1.18\pm0.03$.  After Kauffman~(1995).}
\label{nkresults}
\end{figure}

In numerical simulations this model shows a number of interesting features.
First, regardless of their initial values, the $K$s of the individual
species appear to converge on an intermediate figure which puts all species
close to the phase boundary discussed in the last section.  This lends
support to the ideas of Kauffman and Johnsen that fitness is optimized at
this point (even though other arguments indicated that this might not be
the best choice for selfishly evolving species---see above).
Interestingly, the model also shows a power-law distribution of the sizes
of extinction events taking place; if we count up the number of species
becoming extinct at each time-step in the simulation and make a histogram
of these figures over the course of a long simulation, the result is of the
form shown in Figure~\fref{nkresults}.  The power-law has a measured
exponent of $\tau\simeq1$, which is not in good agreement with the figure
of $\tau\simeq2$ found in the fossil data (see Section~\sref{rates}), but
the mere existence of the power-law distribution is quite intriguing.
Kauffman and Neumann explain its appearance as the result of avalanches of
extinction which arise because the invasion of a niche by a new species
(with the resulting extinction of the niche's previous occupier) disturbs
the neighbouring species, perhaps making them susceptible to invasion by
further species.  Another possible mechanism arises from the uniformity of
genotypes which the invasion mechanism gives rise to.  As noted above, the
invasion of many niches by one particularly fit species tends to produce an
ecosystem with many similar species in it.  If a new species arises which
is able to compete successfully with these many similar species, then they
may all become extinct over a short period of time, resulting in an
extinction avalanche.

Why avalanches such as these should possess a power-law distribution is
not clear.  Kauffman and Neumann connect the phenomenon with the apparent
adaptation of the ecosystem to the phase boundary between the ordered and
chaotic regimes---the ``edge of chaos'' as Kauffman has called it.  A more
general explanation may come from the study of ``self-organized critical''
systems, which is the topic of the next section.

Kauffman and Neumann did not take the intermediate step of simulating a
system in which species are permitted to vary their values of $K$, but in
which there is no invasion mechanism.  Such a study would be useful for
clarifying the relative importance of the $K$-evolution and invasion
mechanisms.  Bak and Kauffman (unpublished, but discussed by Bak~(1996))
have carried out some simulations along these lines, but apparently found
no evidence for the evolution of the system to the critical point.
Bak~\etal~(1992) have argued on theoretical grounds that such evolution
should {\em not\/} occur in the maximally rugged case $K=N-1$, but the
argument does not extend to smaller values of $K$.  In the general case the
question has not been settled and deserves further study.

\section{The Bak--Sneppen model and variations}
\label{socmodels}
The models discussed in the last section are intriguing, but present a
number of problems.  In particular, most of the results about them come
from computer simulations, and little is known analytically about their
properties.  Results such as the power-law distribution of extinction sizes
and the evolution of the system to the ``edge of chaos'' are only as
accurate as the simulations in which they are observed.  Moreover, it is
not even clear what the mechanisms responsible for these results are,
beyond the rather general arguments we have already given.  In order to
address these shortcomings, Bak and Sneppen~(1993, Sneppen~\etal~1995,
Sneppen~1995, Bak~1996) have taken Kauffman's ideas, with some
modification, and used them to create a considerably simpler model of
large-scale coevolution which also shows a power-law distribution of
avalanche sizes and which is simple enough that its properties can, to some
extent, be understood analytically.  Although the model does not directly
address the question of extinction, a number of authors have interpreted
it, using arguments similar to those of Section~\sref{pseudoextinction}, as
a possible model for extinction by biotic causes.

The Bak--Sneppen model is one of a class of models that show
``self-organized criticality'', which means that regardless of the state in
which they start, they always tune themselves to a critical point of the
type discussed in Section~\sref{nkaval}, where power-law behaviour is seen.
We describe self-organized criticality in more detail in
Section~\sref{soc}.  First however, we describe the Bak--Sneppen model
itself.

\subsection{The Bak--Sneppen model}
\label{bs}
\begin{figure}[t]
\columnfigure{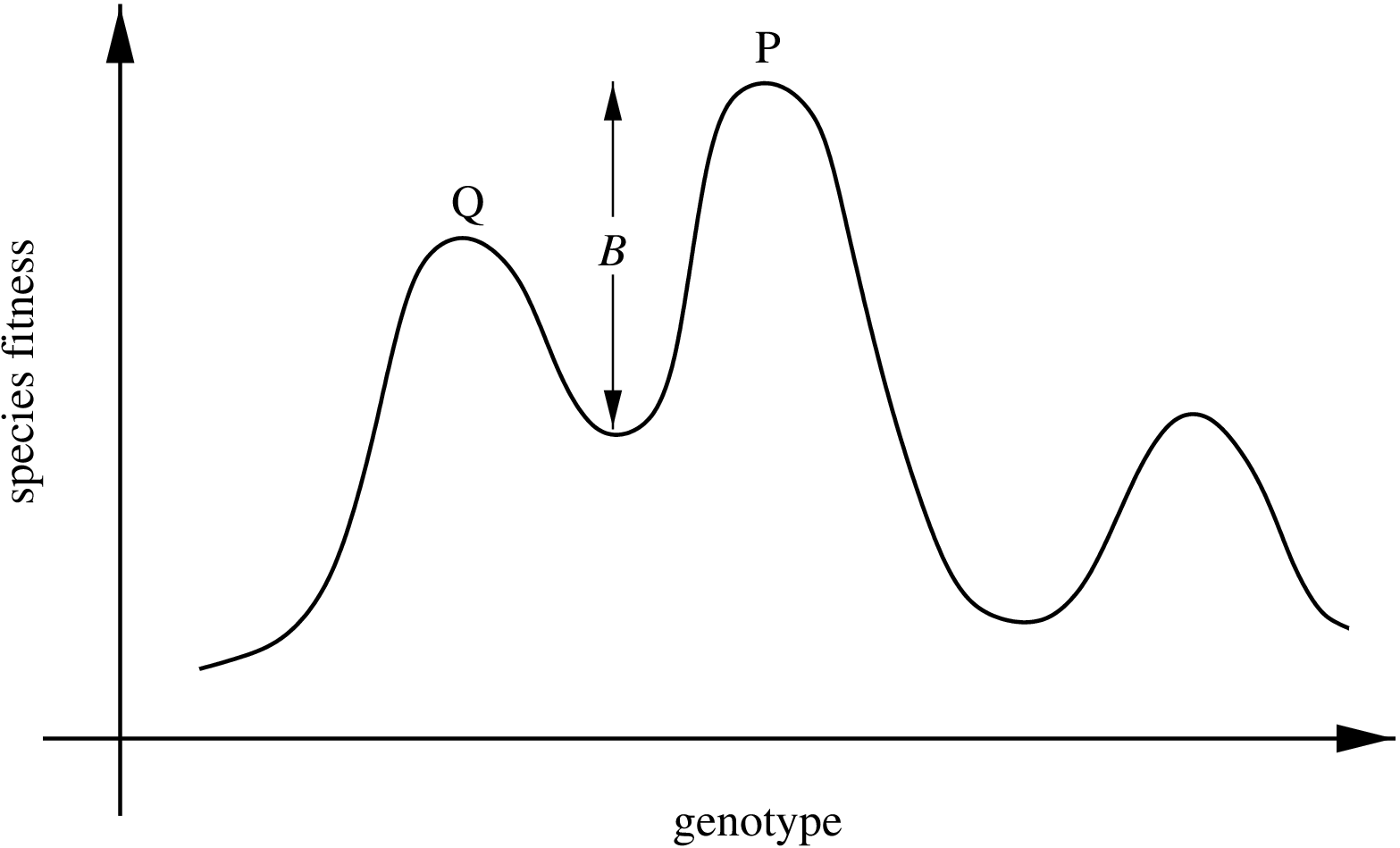}
\capt{In order to reach a new adaptive peak Q from an initial genotype P, a
  species must pass through an intervening fitness ``barrier'', or region
  of low fitness.  The height $B$ of this barrier is a measure of how
  difficult it is for the species to reach the new peak.}
\label{barrier}
\end{figure}

In the model of Bak and Sneppen there are no explicit fitness landscapes,
as there are in \NK\ models.  Instead the model attempts to mimic the
effects of landscapes in terms of ``fitness barriers''.  Consider
Figure~\fref{barrier}, which is a toy representation of a fitness landscape
in which there is only one dimension in the genotype (or phenotype) space.
If the mutation rate is low compared with the time-scale on which selection
takes place (as Kauffman assumed), then a population will spend most of its
time localized around a peak in the landscape (labelled~P in the figure).
In order to evolve to another, adjacent peak~(Q), we must pass through an
intervening ``valley'' of lower fitness.  This valley presents a barrier to
evolution because individuals with genotypes which fall in this region are
selected against in favour of fitter individuals closer to~P.  In their
model, Bak and Sneppen assumed that that the average time $t$ taken to
mutate across a fitness barrier of this type goes exponentially with the
height $B$ of the barrier:
\begin{equation}
t = t_0 \e^{B/T},
\label{bsassum}
\end{equation}
where $t_0$ and $T$ are constants.  The value of $t_0$ merely sets the time
scale, and is not important.  The parameter $T$ on the other hand depends
on the mutation rate in the population, and the assumption that mutation is
low implies that $T$ is small compared with the typical barrier heights $B$
in the landscape.  Equation~\eref{bsassum} was proposed by analogy with the
so-called Arrhenius law of statistical physics rather than by appealing to
any biological principles, and in the case of evolution on a rugged fitness
landscape it may well not be correct (see Section~\sref{bsshort}).
Nonetheless, as we will argue later, Equation~\eref{bsassum} may still be a
reasonable approximation to make.

Based on Equation~\eref{bsassum}, Bak and Sneppen then made a further
assumption.  If mutation rate (and hence $T$) is small, then the
time-scales $t$ for crossing slightly different barriers may be widely
separated.  In this case a species' behaviour is to a good approximation
determined by the lowest barrier which it has to cross to get to another
adaptive peak.  If we have many species, then each species $i$ will have
some lowest barrier to mutation $B_i$, and the first to mutate to a new
peak will be the one with the lowest value of $B_i$ (the ``lowest of the
low'', if you like).  The Bak--Sneppen model assumes this to be the case
and ignores all other barrier heights.

The dynamics of the model, which we now describe, have been justified in
different ways, some of them more reasonable than others.  Probably the
most consistent is that given by Bak~(private communication) which is as
follows.  In the model there are a fixed number $N$ of species.  Initially
each species $i$ is allotted a random number $0\le B_i<1$ to represent the
lowest barrier to mutation for that species.  The model then consists of
the repetition of two steps:
\begin{enumerate}
\item We assume that the species with the lowest barrier to mutation $B_i$
  is the first to mutate.  In doing so it crosses a fitness barrier and
  finds its way to a new adaptive peak.  From this new peak it will have
  some new lowest barrier for mutation.  We represent this process in the
  model by finding the species with the lowest barrier and assigning it a
  new value $0\le B_i<1$ at random.
\item We assume that each species is coupled to a number of neighbours.
Bak and Sneppen called this number $K$.  (The nomenclature is rather
confusing; the variables $N$ and $K$ in the Bak--Sneppen model correspond
to the variables $S$ and $S_i$ in the \NK\ model.)  When a species evolves,
it will affect the fitness landscapes of its neighbours, presumably
altering their barriers to mutation.  We represent this by also assigning
new random values $0\leq B_i<1$ for the $K$ neighbours.
\end{enumerate}
And that is all there is to the model.  The neighbours of a species can be
chosen in a variety of different ways, but the simplest is, as Kauffman and
Johnsen~(1991) also did, to put the species on a lattice and make the
nearest neighbours on the lattice neighbours in the ecological sense.  For
example, on a one dimensional lattice---a line---each species has two
neighbours and $K=2$.

\begin{figure*}
\sixfigure{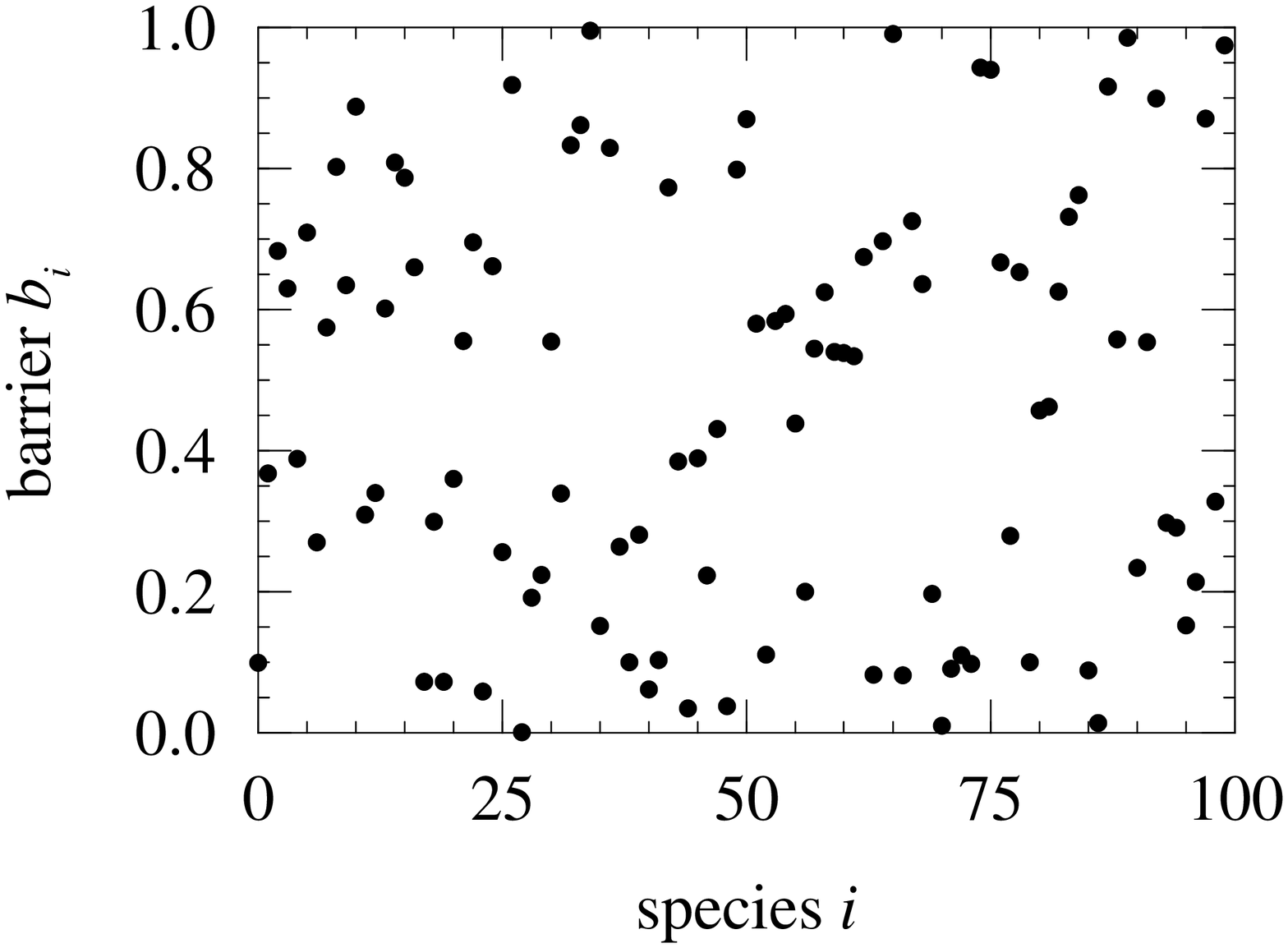}{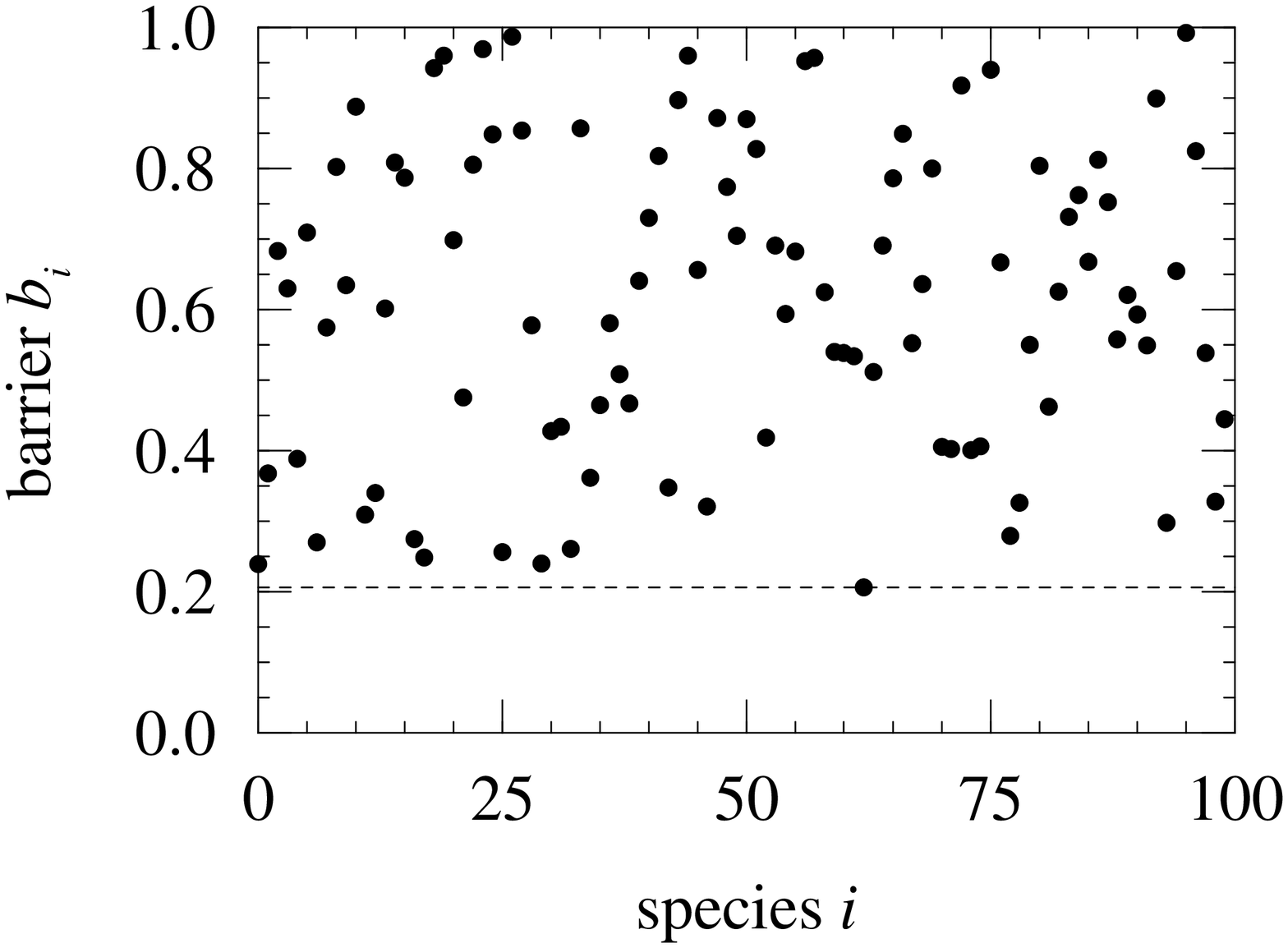}{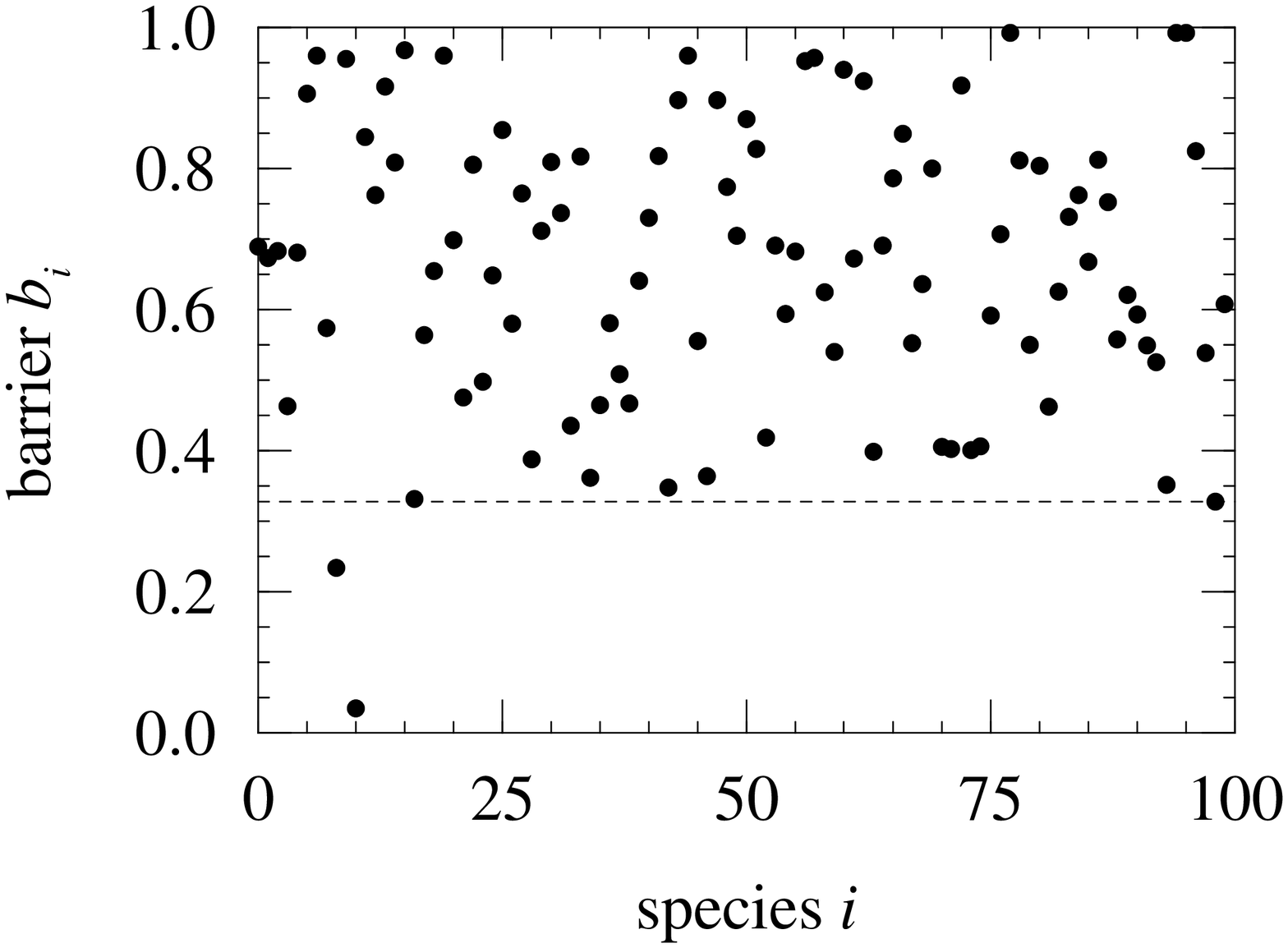}{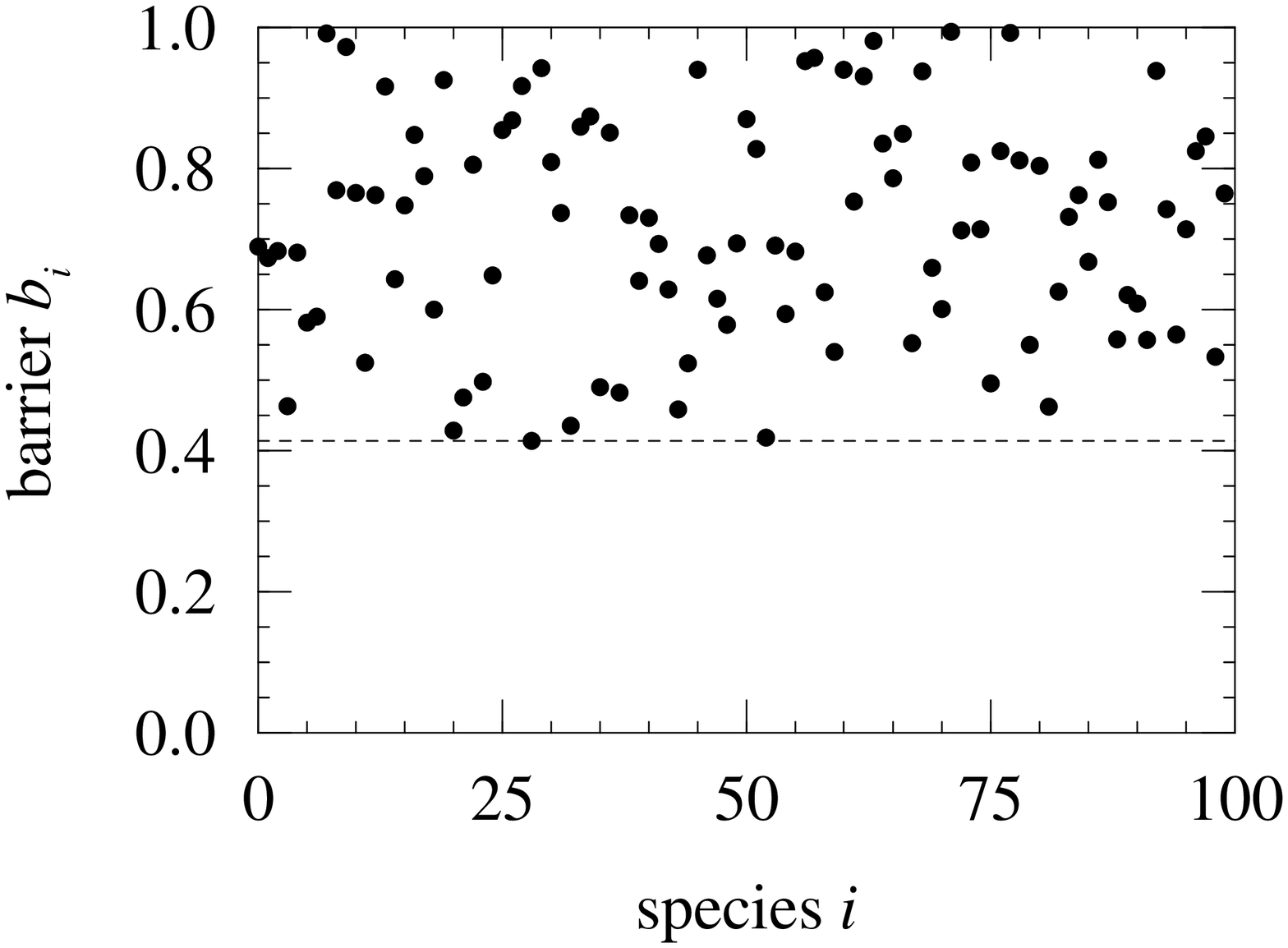}{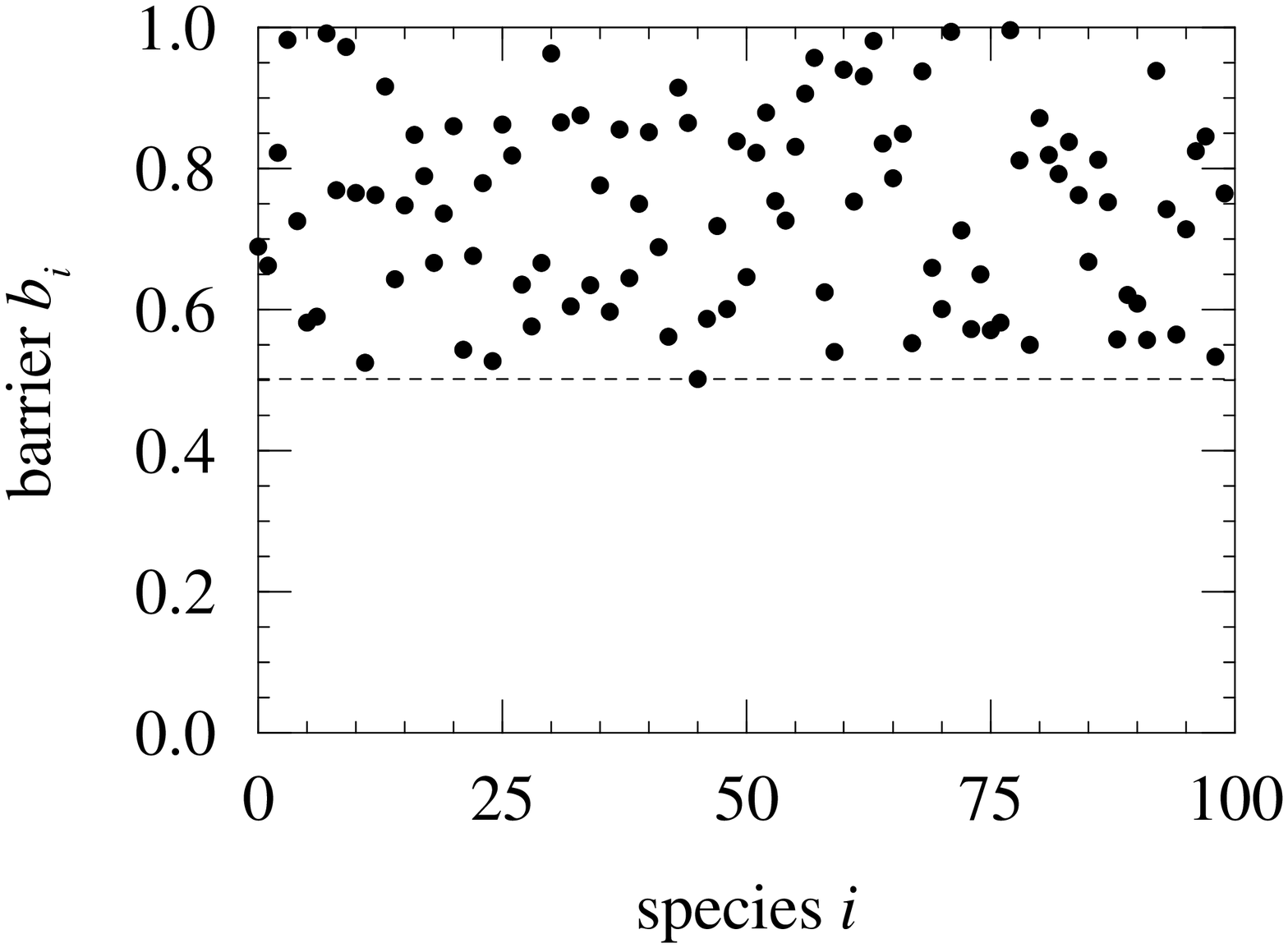}{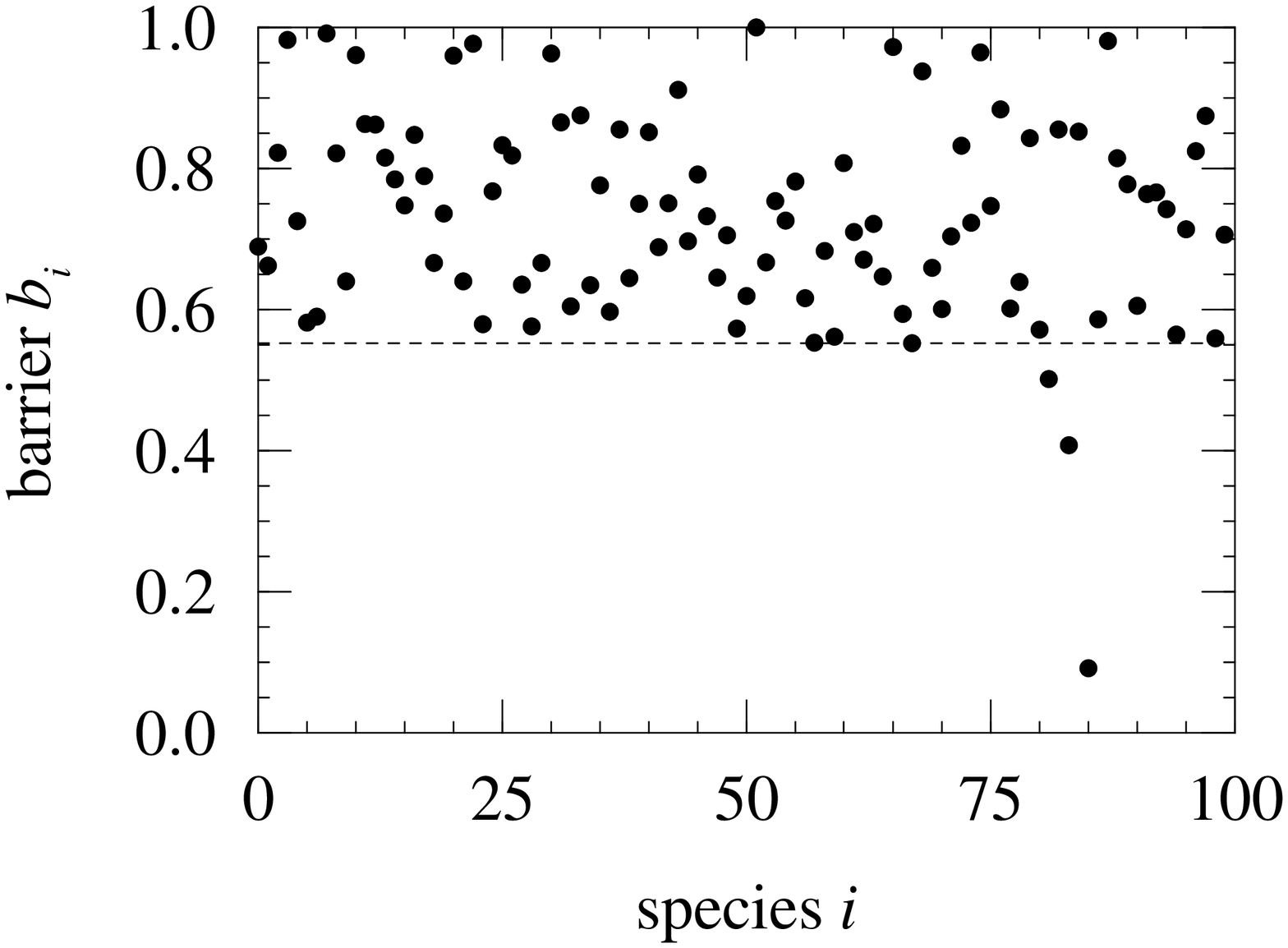}
\widecapt{The barrier values (dots) for a 100 species Bak--Sneppen model after
50, 100, 200, 400, 800 and 1600 steps of a simulation.  The dotted line in
each frame represents the approximate position of the upper edge of the
``gap'' described in the text.  In some frames a few species have barriers
below this level, indicating that they were taking part in an avalanche at
the moment when our snapshot of the system was taken.}
\label{snaps}
\end{figure*}

So what is special about this model?  Well, let us consider what happens as
we repeat the steps above many times.  Initially the barrier variables are
uniformly distributed over the interval between zero and one.  If $N$ is
large, the lowest barrier will be close to zero.  Suppose this lowest
barrier $B_i$ belongs to species $i$.  We replace it with a new random
value which is very likely to be higher than the old value.  We also
replace the barriers of the $K$ neighbours of $i$ with new random values.
Suppose we are working on a one-dimensional lattice, so that these
neighbours are species $i-1$ and $i+1$.  The new barriers we choose for
these two species are also very likely to be higher than $B_i$, although
not necessarily higher than the old values of $B_{i-1}$ and $B_{i+1}$.
Thus, the steps~(i) and~(ii) will on average raise the value of the lowest
barrier in the system, and will continue to do so as we repeat them again
and again.  This cannot continue forever however, since as the value of the
lowest barrier in the system increases, it becomes less and less likely
that it will be replaced with a new value which is higher.
Figure~\fref{snaps} shows what happens in practice.  The initial
distribution of barriers gets eaten away from the bottom at first,
resulting in a ``gap'' between zero and the height of the lowest barrier.
After a time however, the distribution comes to equilibrium with a value of
about $\frac23$ for the lowest barrier.  (The actual figure is measured to
be slightly over $\frac23$; the best available value at the time of writing
is $0.66702\pm0.00003$ (Paczuski, Maslov and Bak~1996).)

Now consider what happens when we make a move starting from a state which
has a gap like this at the bottom end of the barrier height distribution.
The species with the lowest barrier to mutation is right on the edge of the
gap.  We find this species and assign it and its $K$ neighbours new random
barrier values.  There is a chance that at least one of these new values
will lie in the gap, which necessarily makes it the lowest barrier in the
system.  Thus on the next step of the model, this species will be the one
to evolve.  We begin to see how avalanches appear in this model: there is a
heightened chance that the next species to evolve will be one of the
neighbours of the previous one.  In biological terms the evolution of one
species to a new adaptive peak changes the shape of the fitness landscapes
of neighbouring species, making them more likely to evolve too.  The
process continues, until, by chance, all new barrier values fall above the
gap.  In this case the next species to evolve will not, in general, be a
neighbour of one of the other species taking part in the avalanche, and for
this reason we declare it to be the first species in a new avalanche, the
old avalanche being finished.

As the size of the gap increases, the typical length of an avalanche also
increases, because the chances of a randomly chosen barrier falling in the
gap in the distribution become larger.  As we approach the equilibrium
value $B_c=0.667$ the mean avalanche size diverges, a typical sign of a
self-organized critical system.

\subsection{Self-organized criticality}
\label{soc}
So what exactly is self-organized criticality?  The phenomenon was first
studied by Bak, Tang and Wiesenfeld~(1987), who proposed what has now
become the standard example of a self-organized critical (SOC) model, the
self-organizing sand-pile.  Imagine a pile of sand which grows slowly as
individual grains of sand are added to it one by one at random positions.
As more sand is added, the height of the pile increases, and with it the
steepness of the pile's sides.  Avalanches started by single grains
increase in size with steepness until at some point the pile is so steep
that the avalanches become formally infinite in size, which is to say there
is bulk transport of sand down the pile.  This bulk transport in turn
reduces the steepness of the pile so that subsequent avalanches are
smaller.  The net result is that the pile ``self-organizes'' precisely to
the point at which the infinite avalanche takes place, but never becomes
any steeper than this.

A similar phenomenon takes place in the evolution model of Bak and Sneppen,
and indeed the name ``coevolutionary avalanche'' is derived from the
analogy between the two systems.  The size of the gap in the Bak--Sneppen
model plays the role of the steepness in the sandpile model.  Initially,
the gap increases as described above, and as it increases the avalanches
become larger and larger on average, until we reach the critical point at
which an infinite avalanche can occur.  At this point the rates at which
barriers are added and removed from the region below the gap exactly
balance, and the gap stops growing, holding the system at the critical
point thereafter.

It is interesting to compare the Bak--Sneppen model with the \NKCS\ model
discussed in Section~\sref{nkcoevo}.  Like the Bak--Sneppen model, the
\NKCS\ model also has a critical state in which power-law distributions of
avalanches occur, but it does not self-organize to that state.  It can be
critical, but not self-organized critical.  However the essence of both
models is that the evolution of one species distorts the shape of the
fitness landscape of another (represented by the barrier variables in the
Bak--Sneppen case), thus sometimes causing it to evolve too.  So what is
the difference between the two?  The crucial point seems to be that in the
Bak--Sneppen case the species which evolves is the one with the smallest
barrier to mutation.  This choice ensures that the system is always driven
towards criticality.

\begin{figure}[t]
\columnfigure{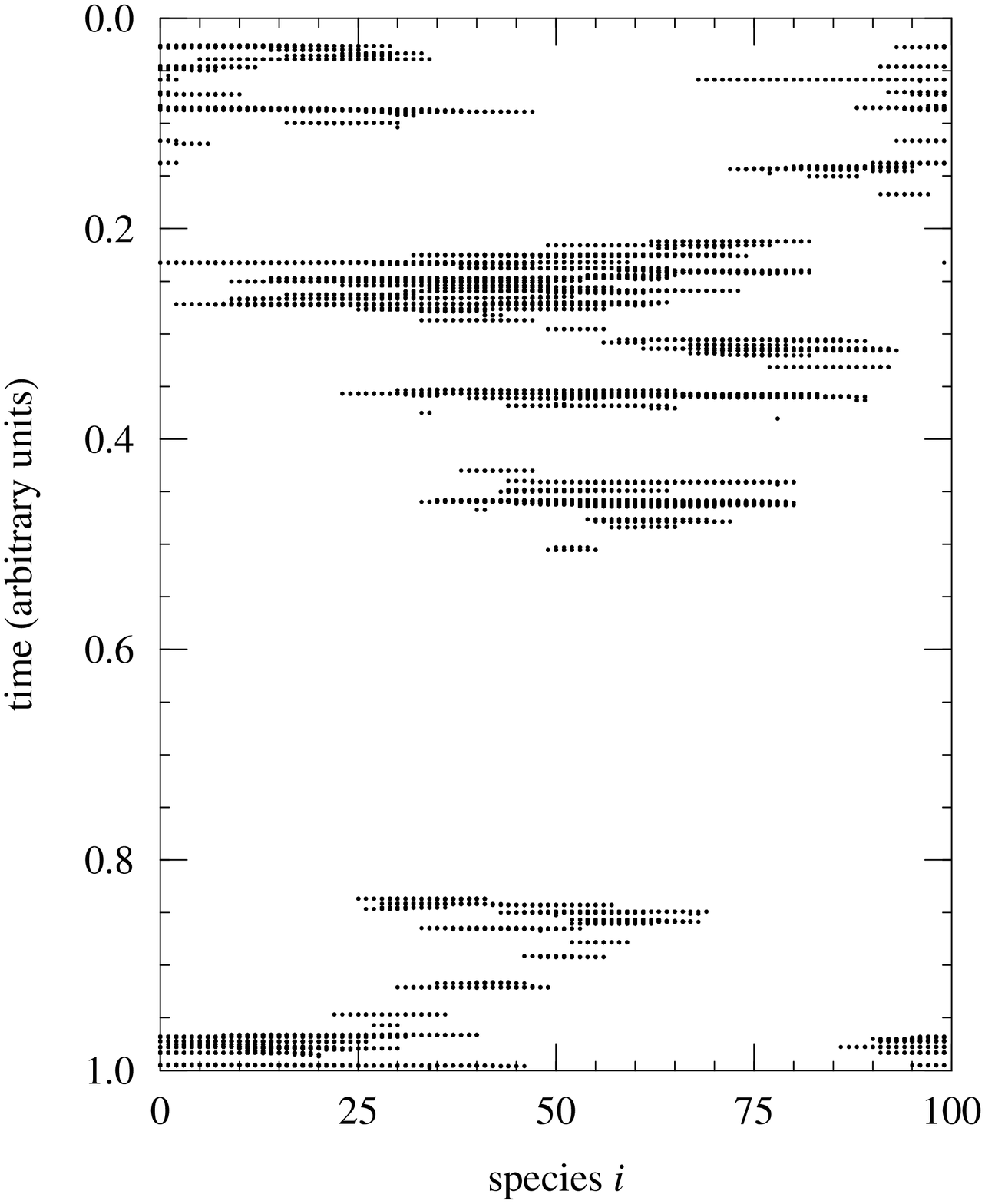}
\capt{A time-series of evolutionary activity in a simulation of the
  Bak--Sneppen model.  Each dot represents the action of choosing a new
  barrier value for one species.  Time in this figure runs down the page
  from top to bottom.}
\label{bsseries}
\end{figure}

At first sight, one apparent problem with the Bak--Sneppen model is that
the delineation of an ``avalanche'' seems somewhat arbitrary.  However the
avalanches are actually quite well separated in time because of the
exponential dependence of mutation timescale on barrier height given by
Equation~\eref{bsassum}.  As defined above, an avalanche is over when no
species remain with a barrier $B_i$ in the gap at the bottom of the barrier
height distribution, and the time until the next avalanche then depends on
the first barrier $B_i$ above the gap.  If the ``temperature'' parameter
$T$ is small, then the exponential in Equation~\eref{bsassum} makes this
inter-avalanche time much longer than typical duration of a single
avalanche.  If we make a plot of the activity of the Bak--Sneppen model as
a function of ``real'' time, (i.e.,~time measured in the increments
specified by Equation~\eref{bsassum}), the result looks like
Figure~\fref{bsseries}.  In this figure the avalanches in the system are
clearly visible and are well separated in time.

\begin{figure}[t]
\columnfigure{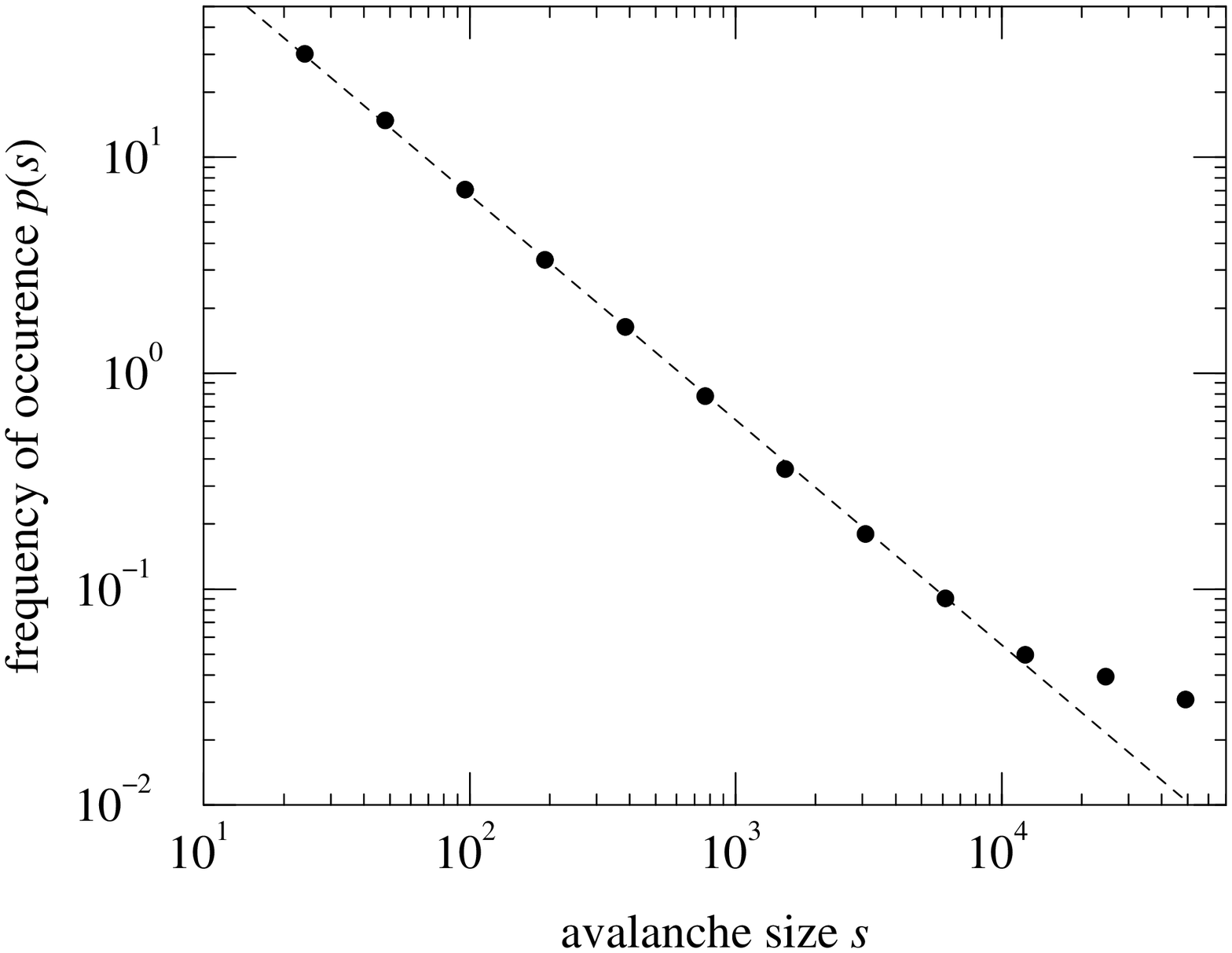}
\capt{Histogram of the sizes of avalanches taking place in a simulation of
an $N=100$ Bak--Sneppen model on a one-dimensional lattice.  The
distribution is very close to power-law over a large part of the range, and
the best-fit straight line (the dashed line above) gives a figure of
$\tau=1.04\pm0.01$ for the exponent.}
\label{bspower}
\end{figure}

One consequence of the divergence of the average avalanche size as the
Bak--Sneppen model reaches the critical point is that the distribution of
the sizes of coevolutionary avalanches becomes scale-free---the size scale
which normally describes it diverges and we are left with a distribution
which has no scale parameter.  The only (continuous) scale-free
distribution is the power law, Equation~\eref{powerlaw}, and, as
Figure~\fref{bspower} shows, the measured distribution is indeed a power
law.  Although the model makes no specific predictions about extinction,
its authors argued, as we have done in Section~\sref{pseudoextinction},
that large avalanches presumably give rise to large-scale pseudoextinction,
and may also cause true extinction via ecological interactions between
species.  They suggested that a power-law distribution of coevolutionary
avalanches might give rise in turn to a power-law distribution of
extinction events.  The exponent $\tau$ of the power law generated by the
Bak--Sneppen model lies strictly within the range $1\le\tau\le\frac32$ (Bak
and Sneppen~1993, Flyvbjerg~\etal~1993), and if the same exponent describes
the corresponding extinction distribution this makes the model incompatible
with the fossil data presented in Section~\sref{data}, which give
$\tau\simeq2$.  However, since the connection between the coevolutionary
avalanches and the extinction profile has not been made explicit, it is
possible that the extinction distribution could be governed by a different,
but related exponent which is closer to the measured value.

One of the elegant properties of SOC models, and critical systems in
general, is that exponents such as $\tau$ above are {\em universal}.  This
means that the value of the exponent is independent of the details of the
dynamics of the model, a point which has been emphasized by Bak~(1996).
Thus, although the Bak--Sneppen model is undoubtedly an extremely
simplified model of evolutionary processes, it may still be able to make
quantitative predictions about real ecosystems, because the model and the
real system share some universal properties.

\subsection{Time-scales for crossing barriers}
\label{bsshort}
Bak and Sneppen certainly make no claims that their model is intended to be
a realistic model of coevolution, and therefore it may seem unfair to level
detailed criticism at it.  Nonetheless, a number of authors have pointed
out shortcomings in the model, some of which have since been remedied by
extending the model in various ways.

Probably the biggest criticism which can be levelled at the model is that
the crucial Equation~\eref{bsassum} is not a good approximation to the
dynamics of species evolving on rugged landscapes.  Weisbuch~(1991) has
studied this question in detail.  He considers, as the models of Kauffman
and of Bak and Sneppen both also do, species evolving under the influence
of selection and mutation on a rugged landscape in the limit where the rate
of mutation is low compared with the timescale on which selection acts on
populations.  In this regime he demonstrates that the timescale $t$ for
mutation from one fitness peak across a barrier to another peak is given by
\begin{equation}
t = {1\over q P_0} \prod_i {F_0 - F_i\over q},
\label{bsweis}
\end{equation}
where $q$ is the rate of mutation per gene, $P_0$ is the size of the
population at the initial fitness peak, and $F_i$ are the fitnesses of the
mutant species at each genotype $i = 0, 1, 2,\ldots$ along the path in
genotype space taken by the evolving species.  The product over $i$ is
taken along this same path.  Clearly this expression does not vary
exponentially with the height of the fitness barrier separating the two
fitness peaks.  In fact, it goes approximately as a power law, with the
exponent depending on the number of steps in the evolutionary path taken by
the species.  If this is the case then the approximation implicit in
Equation~\eref{bsassum} breaks down and the dynamics of the Bak--Sneppen
model is incorrect.

This certainly appears to be a worrying problem, but there may be a
solution.  Bak~(1996) has suggested that the crucial point is that
Equation~\eref{bsweis} varies exponentially in the {\em number\/} of steps
along the path from one species to another, i.e.,~the number of genes which
must change to get us to a new genotype; in terms of the lengths of the
evolutionary paths taken through genotype space, the timescales for
mutation are exponentially distributed.  The assumption that the
``temperature'' parameter $T$ appearing in Equation~\eref{bsassum} is small
then corresponds to evolution which is dominated by short paths.  In other
words, mutations occur mostly between fitness peaks which are separated by
changes in only a small number of genes.  Whether this is in fact the case
historically is unclear, though it is certainly well known that mutational
mechanisms such as recombination which involve the simultaneous alteration
of large numbers of genes are also an important factor in biological
evolution.

\subsection[The exactly solvable multi-trait model]{The exactly solvable
multi-trait\\model}
\label{bsbp}
The intriguing contrast between the simplicity of the rules defining the
Bak--Sneppen model and the complexity of its behaviour has led an
extraordinary number of authors to publish analyses of its workings.  (See
Maslov~\etal~(1994), de Boer~\etal~(1995), Pang~(1997) and references
therein for a subset of these publications.)  In this review we will not
delve into these mathematical developments in any depth, since our primary
concern is extinction.  However, there are several extensions of the model
which {\em are\/} of interest to us.  The first one is the ``multi-trait''
model of Boettcher and Paczuski~(1996a, 1996b).  This model is a
generalization of the Bak--Sneppen model in which a record is kept of
several barrier heights for each species---barriers for mutation to
different fitness peaks.

In the model of Boettcher and Paczuski, each of the $N$ species has $M$
independent barrier heights.  These heights are initially chosen at random
in the interval $0\le B<1$.  On each step of the model we search through
all $MN$ barriers to find the one which is lowest.  We replace this one
with a new value, and we also change the value of one randomly chosen
barrier for each of the $K$ neighbouring species.  Notice that the other
$M-1$ barrier variables for each species are left untouched.  This seems a
little strange; presumably if a species is mutating to a new fitness peak,
all its barrier variables should change at once.  However, the primary aim
of Boettcher and Paczuski's model is not to mimic evolution more
faithfully.  The point is that their model is exactly solvable when
$M=\infty$, which allows us to demonstrate certain properties of the model
rigorously.

The exact solution is possible because when $M=\infty$ the dynamics of the
model separates into two distinct processes.  As long as there are barrier
variables whose values lie in the gap at the bottom of the barrier
distribution, then the procedure of finding the lowest barrier will always
choose a barrier in the gap.  However, the second step of choosing at
random one of the $M$ barriers belonging to each of $K$ neighbours will
{\em never\/} choose a barrier in the gap, since there are an infinite
number of barriers for each species, and only ever a finite number in the
gap.  This separation of the processes taking place allowed Boettcher and
Paczuski to write exact equations governing the dynamics of the system and
to show that the model does indeed possess true critical behaviour with a
power-law distribution of avalanches.

The Bak--Sneppen model is the $M=1$ limit of the multi-trait
generalization, and it would be very satisfying if it should turn out that
the analytic results of Boettcher and Paczuski could be extended to this
case, or indeed to any case of finite $M$.  Unfortunately, no such
extension has yet been found.

\subsection{Models incorporating speciation}
\label{bsspec}
One of the other criticisms levelled at the Bak--Sneppen model is that it
fails to incorporate speciation.  When a biological population gives rise
to a mutant individual which becomes the founder of a new species, the
original population does not always die out.  Fossil evidence indicates
that it is common for both species to coexist for some time after such a
speciation event.  This process is absent from the Bak--Sneppen model, and
in order to address this shortcoming Vandewalle and Ausloos~(1995,
Kramer~\etal~1996) suggested an extension of the model in which species
coexist on a phylogenetic tree structure, rather than on a lattice.  The
dynamics of their model is as follows.

\begin{figure}[t]
\sidefigure{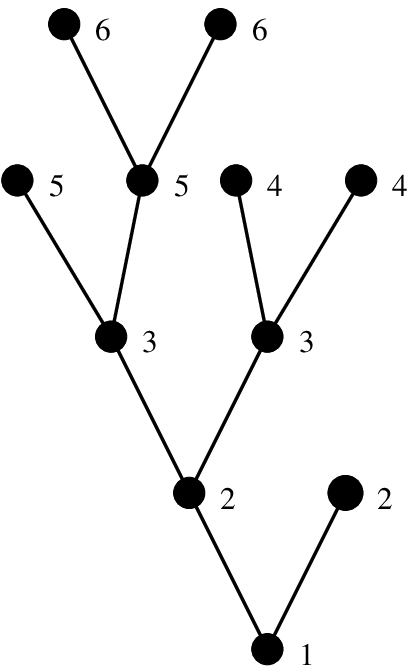}{An example of a phylogenetic tree generated by the
  model of Vandewalle and Ausloos~(1995).  The numbers indicate the
  order of growth of the tree.
\label{bstree}}
\end{figure}

Initially there is just a small number of species, perhaps only one, each
possessing a barrier to mutation $B_i$ whose value is chosen randomly in
the range between zero and one.  The species with the lowest barrier
mutates first, but now both the original species and the mutant are assumed
to survive, so that there is a branching of the tree leading to a pair of
coexisting species (Figure~\fref{bstree}).  One might imagine that the
original species should retain its barrier value, since this species is
assumed not to have changed.  However, if this were the case the model
would never develop a ``gap'' as the Bak--Sneppen model does and so never
self-organize to a critical point.  To avoid this, Vandewalle and Ausloos
specified that both species, the parent and the offspring should be
assigned new randomly-chosen barrier values after the speciation event.  We
might justify this by saying for example that the environment of the parent
species is altered by the presence of a closely-related (and possibly
competing) offspring species, thereby changing the shape of the parent's
fitness landscape.  Whatever the justification, the model gives rise to a
branching phylogenetic tree which contains a continuously increasing number
of species, by contrast with the other models we have examined so far, in
which the number was fixed.  As we pointed out in
Section~\sref{origination}, the number of species in the fossil record does
in fact increase slowly over time, which may be regarded as partial
justification for the present approach.

In addition to the speciation process, there is also a second process
taking place, similar to that of the Bak--Sneppen model: after finding the
species with the lowest barrier to mutation, the barrier variables $B_i$ of
all species within a distance $k$ of that species are also given new,
randomly-chosen values between zero and one.  Distances on the tree
structure are measured as the number of straight-line segments which one
must traverse in order to get from one species to another (see
Figure~\fref{bstree} again).  Notice that this means that the evolution of
one species to a new form is more likely to affect the fitness landscape of
other species which are closely related to it phylogenetically.  There is
some justification for this, since closely related species tend to exploit
similar resources and are therefore more likely to be in competition with
one another.  On the other hand predator-prey and parasitic interactions
are also very important in evolutionary dynamics, and these interactions
tend not to occur between closely related species.

Many of the basic predictions of the model of Vandewalle and Ausloos are
similar to those of the Bak--Sneppen model, indicating perhaps that Bak and
Sneppen were correct to ignore speciation events to begin with.  It is
found again that avalanches of coevolution take place, and that the system
organizes itself to a critical state in which the distribution of the sizes
of these avalanches follows a power law.  The measured exponent of this
power law is $\tau=1.49\pm0.01$ (Vandewalle and Ausloos~1997), which is
very close to the upper bound of $\frac32$ calculated by
Flyvbjerg~\etal~(1993) for the Bak--Sneppen model.  However, there are also
some interesting features which are new to this model.  In particular, it
is found that the phylogenetic trees produced by the model are
self-similar.  In Section~\sref{taxonomy} we discussed the work of
Burlando~(1990), which appears to indicate that the taxonomic trees of
living species are also self-similar.  Burlando made estimates of the
fractal (or Hausdorf) dimension $D_H$ of taxonomic trees for 44
previously-published catalogues of species taken from a wide range of taxa
and geographic areas, and found values ranging from $1.1$ to $2.1$ with a
mean of $1.6$.\footnote{In fact, $D_H$ is numerically equal to the exponent
$\beta$ for a plot such as that shown in Figure~\fref{willis} for the
appropriate group of species.} (The typical confidence interval for values
of $D_H$ was on the order of $\pm0.2$.)  These figures are in reasonable
agreement with the value of $D_H=1.89\pm0.03$ measured by Vandewalle and
Ausloos~(1997) for their model, suggesting that a mechanism of the kind
they describe could be responsible for the observed structure of taxonomic
trees.

The model as described does not explicitly include extinction, and
furthermore, since species are not replaced by their descendents as they
are in the Bak--Sneppen model, there is also no pseudoextinction.  However,
Vandewalle and Ausloos also discuss a variation on the model in which
extinction is explicitly introduced.  In this variation, they find the
species with the lowest barrier to mutation $B_i$ and then they randomly
choose either to have this species speciate with probability
$1-\exp(-B_i/r)$ or to have it become extinct with probability
$\exp(-B_i/r)$, where $r$ is a parameter which they choose.  Thus the
probability of extinction {\em decreases\/} with increasing height of the
barrier.  It is not at first clear how we are to understand this choice.
Indeed, it seems likely from reading the papers of Vandewalle~\etal\ that
there is some confusion between the barrier heights and the concept of
fitness; the authors argue that the species with higher {\em fitness\/}
should be less likely to become extinct, but then equate fitness with the
barrier variables $B_i$.  One way out of this problem may be to note that
on rugged landscapes with bounded fitness there is a positive correlation
between the heights of barriers and the fitness of species: the higher the
fitness the more likely it is that the lowest barrier to mutation will also
be high.

When $r=0$, this extinction model is equal to the first model described, in
which no extinction took place.  When $r$ is above some threshold value
$r_c$, which is measured to be approximately $0.48\pm0.01$ for $k=2$ (the
only case the authors investigated in detail), the extinction rate exceeds
the speciation rate and the tree ceases to grow after a short time.  In the
intervening range $0<r<r_c$ evolution and extinction processes compete and
the model shows interesting behaviour.  Again there is a power-law
distribution of coevolutionary avalanches, and a fractal tree structure
reminiscent of that seen in nature.  In addition there is now a power-law
distribution of extinction events, with the same exponent as the
coevolutionary avalanches, i.e.,~close to $\frac32$.  As with the
Bak--Sneppen model this is in disagreement with the figure of $2.0\pm0.2$
extracted from the fossil data.

Another variation of the Bak--Sneppen model which incorporates speciation
has been suggested by Head and Rodgers~(1997).  In this variation, they
keep track of the two lowest barriers to mutation for each species, rather
than just the single lowest.  The mutation of a species proceeds in the
same fashion as in the normal Bak--Sneppen model when one of these two
barriers is significantly lower than the other.  However, if the two
barriers are close together in value, then the species may split and evolve
in two different directions on the fitness landscape, resulting in
speciation.  How similar the barriers have to be in order for this to
happen is controlled by a parameter $\delta s$, such that speciation takes
place when
\begin{equation}
|B_1 - B_2| < \delta s,
\end{equation}
where $B_1$ and $B_2$ are the two barrier heights.  The model also
incorporates an extinction mechanism, which, strangely, is based on the
opposite assumption to the one made by Vandewalle and Ausloos.  In the
model of Head and Rodgers, extinction takes place when species have
particularly {\em high\/} barriers to mutation.  To be precise, a species
becomes extinct if its neighbour mutates (which would normally change its
fitness landscape and therefore its barrier variables) but both its
barriers are above some predetermined threshold value.  This extinction
criterion seems a little surprising at first: if, as we suggested above,
high barriers are positively correlated with high fitness, why should
species with high barriers become extinct?  The argument put forward by
Head and Rodgers is that species with high barriers to mutation find it
difficult to adapt to changes in their environment.  To quote from their
paper, ``A species with only very large barriers against mutation has
become so inflexible that it is no longer able to adapt and dies out''.  It
seems odd however, that this extinction process should take place precisely
in the species which are adjacent to others which are mutating.  In the
Bak--Sneppen model, these species have their barriers changed to new random
values as a result of the change in their fitness landscapes brought about
by the mutation of their neighbour.  Thus, even if they did indeed have
high barriers to mutation initially, their barriers would be changed when
their neighbour mutated, curing this problem and so one would expect that
these species would {\em not\/} become extinct.\footnote{A later paper on
  the model by Head and Rodgers (unpublished) has addressed this criticism
  to some extent.}

The model has other problems as well.  One issue is that, because of the
way the model is defined, it does not allow for the rescaling of time
according to Equation~\eref{bsassum}.  This means that evolution in the
model proceeds at a uniform rate, rather than in avalanches as in the
Bak--Sneppen model.  As a direct result of this, the distribution of the
sizes of extinction events in the model follows a Poisson distribution,
rather than the approximate power law seen in the fossil data
(Figure~\fref{extdist}).  The model does have the nice feature that the
number of species in the model tends to a natural equilibrium; there is a
balance between speciation and extinction events which causes the number of
species to stabilize.  This contrasts with the Bak--Sneppen model (and
indeed almost all the other models we discuss) in which the number of
species is artificially held constant, and also with the model of
Vandewalle and Ausloos, in which the number of species either shrinks to
zero, or grows indefinitely, depending on the value of the parameter $r$.
Head and Rodgers gave an approximate analytic explanation for their results
using a ``mean field'' technique similar to that employed by
Flyvbjerg~\etal~(1993) for the Bak--Sneppen model.  However, the question
of whether the number of species predicted by their model agrees with the
known taxon carrying capacity of real ecosystems has not been investigated.

\subsection[Model incorporating external stress]{Models incorporating
external\\stress}
\label{bsnoise}
Another criticism of the approach taken in Bak and Sneppen's work (and
indeed in the work of Kauffman discussed in Section~\sref{flmodels}) is
that real ecosystems are not closed dynamical systems, but are in reality
affected by many external factors, such as climate and geography.  Indeed,
as we discussed in Section~\sref{rates}, a number of the larger extinction
events visible in the fossil record have been tied quite convincingly to
particular exogenous events, so that any model ignoring these effects is
necessarily incomplete.  Newman and Roberts~(1995, Roberts and Newman~1996)
have proposed a variation on the Bak--Sneppen model which attempts to
combine the ideas of extinction via environmental stress and large-scale
coevolution.  The basic idea behind this model is that a large
coevolutionary avalanche will cause many species to move to new fitness
peaks, some of which may possess lower fitness than the peaks they previous
occupied.  Thus a large avalanche produces a number of new species which
have low fitness and therefore may be more susceptible to extinction as a
result of environmental stress.  This in fact is not a new idea.  Kauffman
for example has made this point clearly in his book {\sl The Origins of
  Order}~(Kauffman~1993): ``During coevolutionary avalanches, species fall
to lower fitness and hence are more likely to become extinct.  Thus the
distribution of avalanche sizes may bear on the distribution of extinction
events in the fossil record.''

Newman and Roberts incorporated this idea into their model as follows.  A
fixed number $N$ of species each possess a barrier $B_i$ to mutation, along
with another variable $F_i$ which measures their fitness at the current
adaptive peak.  On each step of the simulation the species with the lowest
barrier $B_i$ for mutation, and its $K$ neighbours, are selected, just as
in the Bak--Sneppen model.  The $B_i$ and $F_i$ variables of these $K+1$
species are all given new independent random values between zero and one,
representing the evolution of one species and the changed landscapes of its
neighbours.  Then, a positive random number $\eta$ is chosen which
represents the level of environmental stress at the current time, and all
species with $F_i<\eta$ are wiped out and replaced by new species with
randomly chosen $F_i$ and $B_i$.

The net result is that species with low fitness are rapidly removed from
the system.  However, when a large coevolutionary avalanche takes place,
many species receive new, randomly-chosen fitness values, some of which
will be low, and this process provides a ``source'' of low-fitness species
for extinction events.

\begin{figure}[t]
\columnfigure{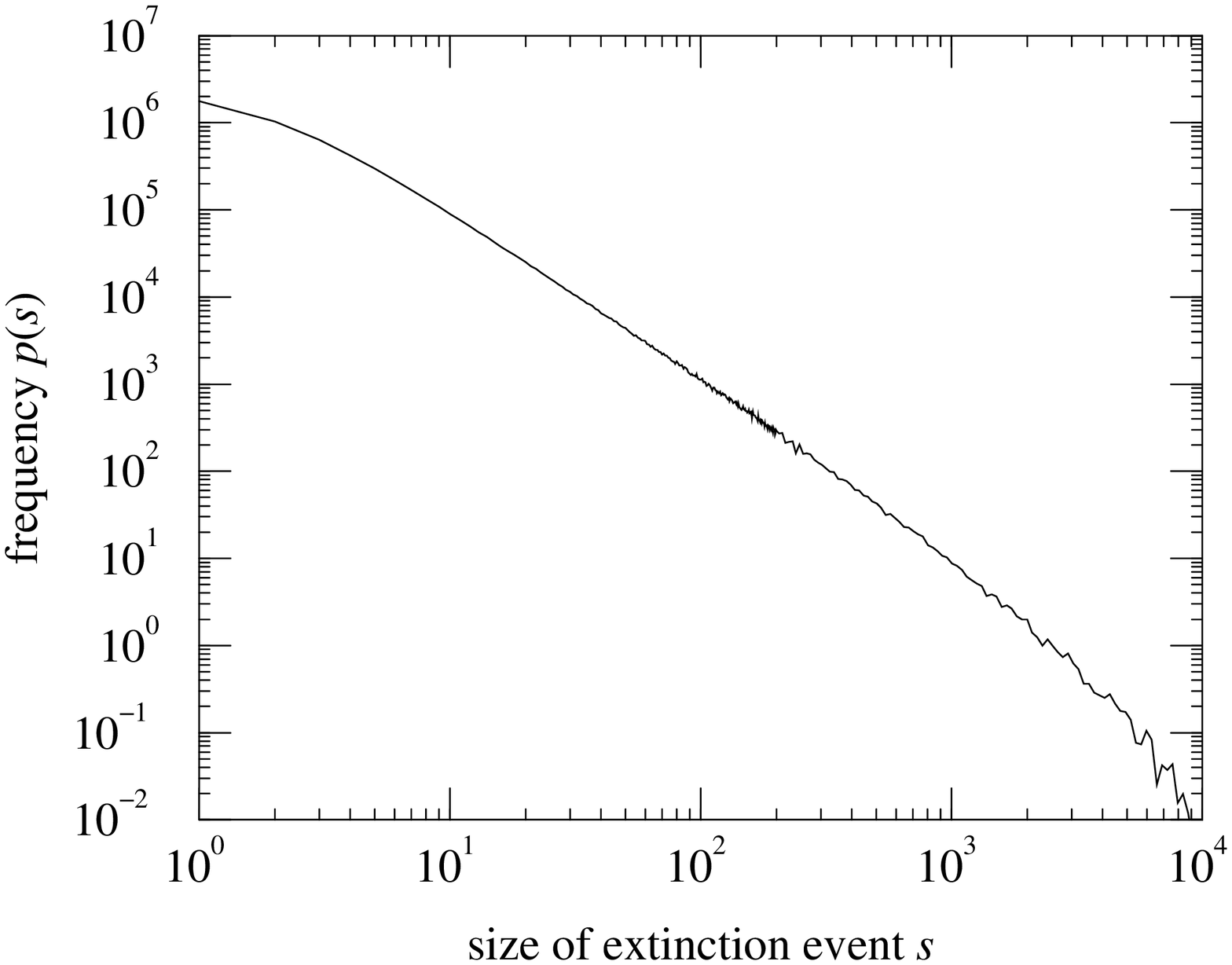}
\capt{The distribution of sizes of extinction events in a simulation of the
  model of Newman and Roberts~(1995) with $N=10000$ and $K=3$.  The
  measured exponent of the power law is $\tau=2.02\pm0.03$, which is in
  good agreement with the figure for the same quantity drawn from fossil
  data (see Section~\sref{rates}).}
\label{bsnrext}
\end{figure}

Interestingly, the distribution of extinction events in this model follows
a power law, apparently regardless of the distribution from which the
stress levels $\eta$ are chosen (Figure~\fref{bsnrext}).  Roberts and
Newman~(1996) offered an analytical explanation of this result within a
``mean field'' framework similar to the one used by Flyvbjerg~\etal~(1993)
for the original Bak--Sneppen model.  However, what is particularly
intriguing is that, even though the distribution of avalanche sizes in the
model still possesses an exponent in the region of $\frac32$ or less, the
{\em extinction\/} distribution is steeper, with a measured exponent of
$\tau=2.02\pm0.03$ in excellent agreement with the results derived from the
fossil data.

The model however has some disadvantages.  First, the source of the
power-law in the extinction distribution is almost certainly not a critical
process, even though the Bak--Sneppen model, from which this model is
derived, is critical.  In fact, the model of Newman and Roberts is just a
special case of the extinction model proposed later by Newman (see
Section~\sref{newman}), which does not contain any coevolutionary
avalanches at all.  In other words, the interesting behaviour of the
extinction distribution in this model is entirely independent of the
coevolutionary behaviour inherited from the Bak--Sneppen model.

A more serious problem with the model is the way in which the environmental
stress is imposed.  As we pointed out in Section~\sref{bs}, the time-steps
in the Bak--Sneppen model correspond to different durations of geological
time.  This means that there should be a greater chance of a large stress
hitting during time-steps which correspond to longer periods.  In the model
of Newman and Roberts however, this is not the case; the probability of
generating a given level of stress is the same in every time-step.  In the
model of stress-driven extinction discussed in Section~\sref{newman} this
shortcoming is rectified.

Another, very similar extension of the Bak--Sneppen model was introduced by
Schmoltzi and Schuster~(1995).  Their motivation was somewhat different
from that of Newman and Roberts---they were interested in introducing a
``real time scale'' into the model.  As they put it: ``The [Bak--Sneppen]
model does not describe evolution on a physical time scale, because an
update step {\em always\/} corresponds to a mutation of the species with
the smallest fitness and its neighbours.  This implies that we would
observe constant extinction intensity in morphological data and that there
will never be periods in which the system does not change.''  This is in
fact is only true if one ignores the rescaling of time implied by
Equation~\eref{bsassum}.  As Figure~\fref{bsseries} shows, there are very
clear periods in which the system does not change if one calculates the
time in the way Bak and Sneppen did.

The model of Schmoltzi and Schuster also incorporates an external stress
term, but in their case it is a local stress $\eta_i$, varying from species
to species.  Other than that however, their approach is very similar to
that of Newman and Roberts; species with fitness below $\eta_i$ are removed
from the system and replaced with new species, and all the variables
$\lbrace \eta_i \rbrace$ are chosen anew at each time step.  Their results
also are rather similar to those of Newman and Roberts, although their main
interest was to model neuronal dynamics in the brain, rather than
evolution, so that they concentrated on somewhat different measurements.
There is no mention of extinction, or of avalanche sizes, in their paper.

\section[Inter-species connection models]{Inter-species connection\\models}
\label{connection}
In the Bak--Sneppen model, there is no explicit notion of an interaction
strength between two different species.  It is true that if two species are
closer together on the lattice then there is a higher chance of their
participating in the same avalanche.  But beyond this there is no variation
in the magnitude of the influence of one species on another.  Real
ecosystems on the other hand have a wide range of possible interactions
between species, and as a result the extinction of one species can have a
wide variety of effects on other species.  These effects may be helpful or
harmful, as well as strong or weak, and there is in general no symmetry
between the effect of $A$ on $B$ and $B$ on $A$.  For example, if species
$A$ is prey for species $B$, then $A$'s demise would make $B$ less able to
survive, perhaps driving it also to extinction, whereas $B$'s demise would
aid $A$'s survival.  On the other hand, if $A$ and $B$ compete for a common
resource, then either's extinction would help the other.  Or if $A$ and $B$
are in a mutually supportive or symbiotic relationship, then each would be
hurt by the other's removal.

A number of authors have constructed models involving specific
species--species interactions, or ``connections''.  If species $i$ depends
on species $j$, then the extinction of $j$ may also lead to the extinction
of $i$, and possibly give rise to cascading avalanches of extinction.  Most
of these connection models neither introduce nor have need of a fitness
measure, barrier, viability or tolerance for the survival of individual
species; the extinction pressure on one species comes from the extinction
of other species.  Such a system still needs some underlying driving force
to keep its dynamics from stagnating, but this can be introduced by making
changes to the connections in the model, without requiring the introduction
of any extra parameters.

Since the interactions in these models are ecological in nature (taking
place at the individual level) rather than evolutionary (taking place at
the species level or the level of the fitness landscape), the
characteristic time-scale of the dynamics is quite short.  Extinctions
produced by ecological effects such as predation and invasion can take only
a single season, whereas those produced by evolutionary pressures are
assumed to take much longer, maybe thousands of years or more.

The models described in this section vary principally in their connection
topology, and in their mechanisms for replacing extinct species.  Sol\'e
and co-workers have studied models with no organized topology, each species
interacting with all others, or with a more-or-less random subset of them
(Sol\'e and Manrubia~1996, Sol\'e, Bascompte and Manrubia~1996,
Sol\'e~1996).  By contrast, the models of Amaral and Meyer~(1998) and
Abramson~(1997) involve very specific food-chain topologies.  The models of
Sol\'e~\etal\ keep a fixed total number of species, refilling empty niches
by invasion of surviving species.  Abramson's model also keeps the total
fixed, but fills empty niches with random new species, while Amaral and
Meyer use an invasion mechanism, but do not attempt to keep the total
number of species fixed.

\subsection{The Sol\'e--Manrubia model}
\label{sole}
Sol\'e and Manrubia~(1996, Sol\'e, Bascompte and Manrubia~1996,
Sol\'e~1996) have constructed a model that focuses on species--species
interactions through a ``connection matrix'' $\vJ$ whose elements give the
strength of coupling between each pair of species.  Specifically, the
matrix element $J_{ij}$ measures the influence of species $i$ on species
$j$, and $J_{ji}$ that of $j$ on $i$.  A positive value of $J_{ij}$ implies
that $i$'s continued existence helps $j$'s survival, whereas a negative
value implies that $j$ would be happy to see $i$ disappear.  The $J_{ij}$
values range between $-1$ and $1$, chosen initially at random.  In most of
their work, Sol\'e and Manrubia let every species interact with every other
species, so all $J_{ij}$s are non-zero, though some may be quite small.
Alternatively it is possible to define models in which the connections are
more restricted, for instance by placing all the species on a square
lattice and permitting each to interact only with its four neighbours
(Sol\'e~1996).

A species $i$ becomes extinct if its net support $\sum_j J_{ji}$ from
others drops below a certain threshold $\theta$.  The sum over $j$ here
is of course only over those species that (a)~are not extinct themselves,
and (b)~interact with $i$ (in the case of restricted connections).  Sol\'e
and Manrubia introduce a variable $S_i(t)$ to represent whether species $i$
is alive ($S_i=1$) or extinct ($S_i=0$) at time $t$, so the extinction
dynamics may be written
\begin{equation}
S_i(t+1) = \Theta\Bigl[\sum_j J_{ji}S_j(t) - \theta\Bigr],
\label{SMdynamics}
\end{equation}
where $\Theta(x)$ is the Heaviside step function, which is 1 for $x>0$ and
zero otherwise.  As this equation implies, time progresses in discrete
steps, with all updates occurring simultaneously at each step.  When
avalanches of causally connected extinctions occur, they are necessarily
spread over a sequence of successive time steps.

To complete the model, Sol\'e and Manrubia introduce two further features,
one to drive the system and one to replace extinct species.  The driving
force is simply a slow random mutation of the coupling strengths in the
connection matrix $\vJ$.  At each time step, for each species $i$, one of
the incoming connections $J_{ji}$ is chosen at random and given a new
random value in the interval between $-1$ and $1$.  This may cause one or
more species to become extinct though loss of positive support from other
species or through increase in the negative influences on it.  It is not
essential to think of these mutations as strictly biotic; external
environmental changes could also cause changes in the coupling between
species (and hence in species' viability).

The replacement of extinct species is another distinguishing feature of
Sol\'e and Manrubia's model.  All the niches that are left empty by
extinction are immediately refilled with copies of one of the surviving
species, chosen at random.  This is similar to the speciation processes
studied by Kauffman and Neumann in the variation of the \NKCS\ model
described in Section~\sref{nkcompete}, and in fact Sol\'e and Manrubia
refer to it as ``speciation''.  However, because the Sol\'e--Manrubia model
is a model of ecological rather than evolutionary processes, it is probably
better to think of the repopulation processes as being an invasion of empty
niches by survivor species, rather than a speciation event.  Speciation is
inherently an evolutionary process, and, as discussed above, takes place on
longer time-scales than the ecological effects which are the primary
concern of this model.

Invading species are copied to the empty slots along with all their
incoming and outgoing connections, except that a little noise is added to
these connections to introduce diversity.  Specifically, if species $k$ is
copied to fill a number of open niches $i$, then
\begin{equation}
J_{ij} = J_{kj} + \eta_{ij},\qquad J_{ji} = J_{jk} + \eta_{ji},
\label{SMclone}
\end{equation}
where $j$ ranges over the species with which each $i$ interacts, and the
$\eta$s are all chosen independently from a uniform random distribution in
the interval $(-\epsilon, \epsilon)$.

Because empty niches are immediately refilled, the $S_i(t)$ variables
introduced on the right hand side of Equation~\eref{SMdynamics} are
actually always $1$, and are therefore superfluous.  They do however make
the form of the dynamics formally very similar to that of spin-glasses in
physics (Fischer and Hertz~1991), and to that of Hopfield artificial neural
networks (Hertz~\etal~1991), and it is possible that these similarities
will lead to useful cross-fertilization between these areas of study.

\begin{figure}[t]
\columnfigure{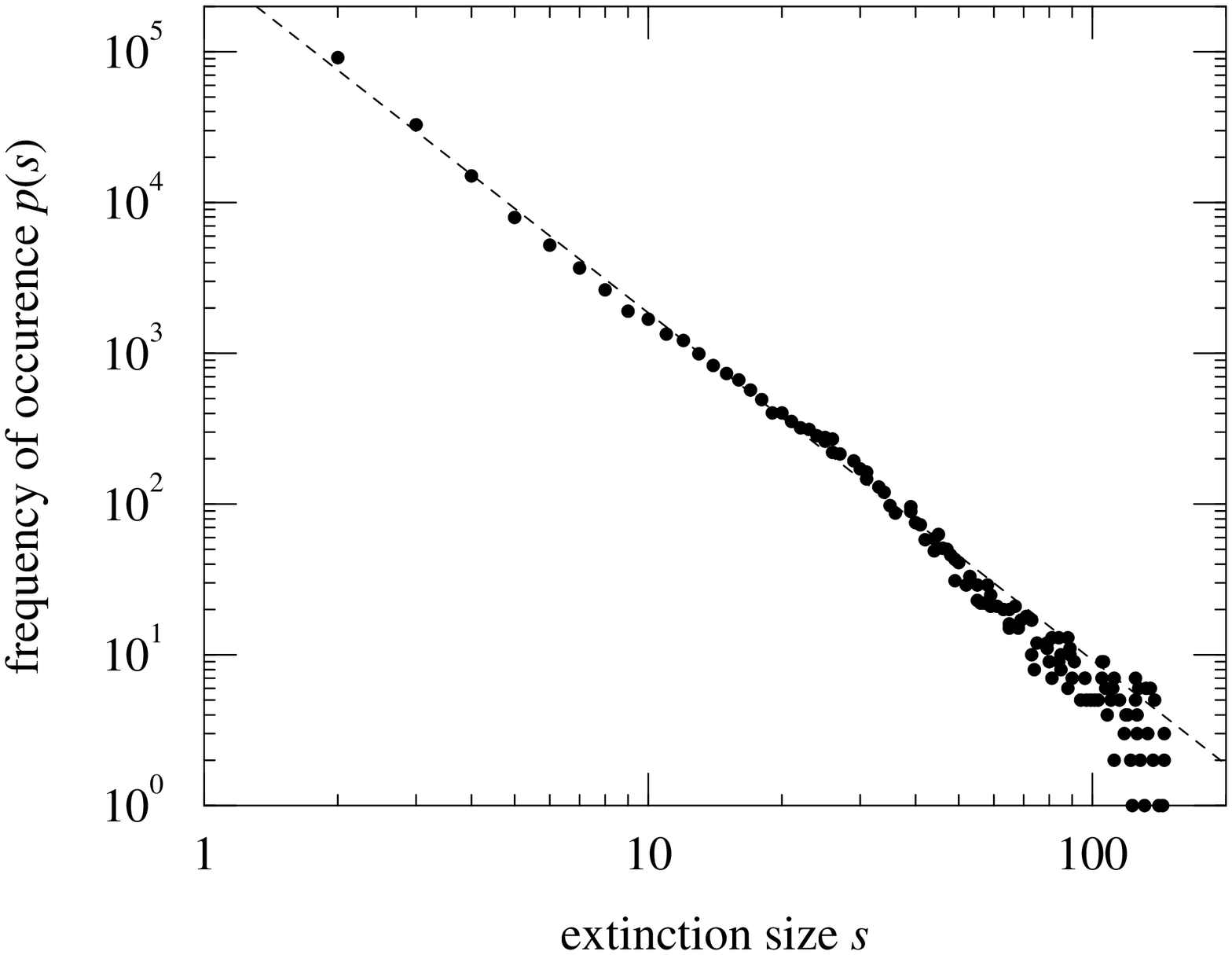}
\capt{The distribution of sizes of extinction events in a simulation of the
  model of Sol\'e and Manrubia~(1996) with $N=150$ species.  The
  distribution follows a power law with a measured exponent of
  $\tau=2.3\pm0.1$.}
\label{soleext}
\end{figure}

Sol\'e and Manrubia studied their model by simulation, generally using
$N=100$ to 150 species, $\theta = 0$, and $\epsilon=0.01$.  Starting from
an initial random state, they waited about $10\,000$ time steps for
transients to die down before taking data.  Extinction events in the model
were found to range widely in size $s$, including occasional large ``mass
extinction'' events that wiped out over 90\% of the population.  Such large
events were often followed by a long period with very little activity.  The
distribution $p(s)$ of extinction sizes was found to follow a power law, as
in Equation~\eref{powerlaw}, with $\tau = 2.3\pm 0.1$ (see
Figure~\fref{soleext}).  Later work by Sol\'e~\etal~(1996) using
$\epsilon=0.05$ gave $\tau = 2.05\pm0.06$, consistent with the value $\tau
= 2.0\pm 0.2$ from the fossil data (Section~\sref{rates}).

The diversified descendants of a parent species may be thought of as a
single genus, all sharing a common ancestor.  Since the number of offspring
of a parent species is proportional to the number of niches which need to
be filled following a extinction event, the distribution of genus sizes is
exactly the same as that of extinction sizes.  Thus Sol\'e and Manrubia
find an exponent in the vicinity of~2 for the taxonomic distribution as
well (see Equation~\eref{willislaw}), to be compared to $1.5\pm 0.1$ for
Willis's data (Figure~\fref{willis}) and to values between $1.1$ and $2.1$
for Burlando's analysis (Section~\sref{bsspec}).

The waiting time between two successive extinction events in the
Sol\'e--Manrubia model is also found to have a power-law distribution, with
exponent $3.0\pm 0.1$.  Thus events are correlated in time---a random
(Poisson) process would have an exponential distribution of waiting times.
The distribution of both species and genus lifetimes can in theory also be
measured in these simulations, although Sol\'e and Manrubia did not publish
any results for these quantities.  Further studies would be helpful here.
 
Sol\'e and Manrubia claim on the basis of their observed power laws that
their model is self-organized critical.  However, it turns out that this is
not the case (Sol\'e, private communication).  In fact, the model is an
example of an ordinary critical system which is tuned to criticality by
varying the parameter $\theta$, which is the threshold at which species
become extinct.  It is just coincidence that the value $\theta=0$ which
Sol\'e and Manrubia used in all of their simulations is precisely the
critical value of the model at which power laws are generated.  Away from
this value the distributions of the sizes of extinction events and of
waiting times are cut off exponentially at some finite scale, and therefore
do not follow a power law.  This then begs the question of whether there is
any reason why in a real ecosystem this threshold parameter should take
precisely the value which produces the power law distribution, rather than
any other value.  At present, no persuasive case has been made in favour of
$\theta=0$, and so the question remains open.


\subsection[Variations on the Sol\'e--Manrubia model]{Variations on the
Sol\'e--Manrubia\\model}
A number of variations on the basic model of Sol\'e and Manrubia are
mentioned briefly in the original paper (Sol\'e and Manrubia~1996).  The
authors tried relaxing the assumptions of total connectivity (letting some
pairs of species have no influence on each other), of $\theta=0$, and of
diversification (letting $\epsilon=0$).  They also tried letting each
$J_{ij}$ take only the values $+1$ or $-1$.  In all these cases they report
that they found the same behaviour with the same power-law exponents
(although as mentioned above, later results showed that in fact the
power-law behaviour is destroyed by making $\theta\ne0$).  This robustness
to changing assumptions is to be expected for critical phenomena, where
typically there occur large ``universality classes'' of similar behaviour
with identical exponents (see Section~\sref{soc}).

Sol\'e~(1996) presents a more significant extension of the model which does
change some of the exponents: he proposes a dynamical rule for the
connectivity itself.  At any time some pairs of sites $i,j$ are not
connected, so that in effect $J_{ij} = J_{ji} = 0$.  (Sol\'e introduces a
new connection variable to represent this, but that is not strictly
necessary.)  Initially the number of connections per site is chosen
randomly between 1 and $N-1$.  During the population of an empty niche $i$
by a species $k$, all but one of $k$'s non-zero connections are reproduced
with noise, as in Equation~\eref{SMclone}, but the last is discarded and
replaced entirely by a new random link from $i$ to a site to which $k$ is
{\em not\/} connected.

Sol\'e also replaces the mutation of $J_{ij}$, which provides the
fundamental random driving force in the Sol\'e--Manrubia model, by a rule
that removes one of the existing species at random at any step when no
extinction takes place.  Without this driving force the system would in
general become frozen.  The emptied niche is refilled by invasion as
always, but these ``random'' extinction events are not counted in the
statistical analysis of extinction.  (The waiting time would always be 1 if
they were counted.)  It is not clear whether this difference between the
models has a significant effect on the results.

The observed behaviour of this model is similar to that of the
Sol\'e--Manrubia model as far as extinction sizes are concerned; Sol\'e
reports an exponent $\tau = 2.02\pm0.03$ for the extinction size
distribution.  However the waiting-time distribution falls much more slowly
(so there are comparably more long waits), with an exponent $1.35\pm0.07$
compared to $3.0\pm0.1$ for the Sol\'e--Manrubia model.  The smaller
exponent seems more reasonable, though of course experimental waiting time
data is not available for comparison.  The number of connections itself
varies randomly through time, and a Fourier analysis shows a power spectrum
of the form $1/f^{\nu}$ with $\nu=0.99\pm0.08$.  Power spectra of this type
are another common feature of critical systems (Sol\'e~\etal~1997).

\subsection[Amaral and Meyer's food chain model]{Amaral and Meyer's
food chain\\model}
\label{amaral}
Whereas the Sol\'e--Manrubia model and its variants have a more or less
arbitrary connection topology between species, real ecosystems have very
specific sets of interdependencies.  An important part of the natural case
can be expressed in terms of food chains, specifying who eats whom.  Of
course food chains are not the only type of inter-species interaction, but
it is nevertheless of interest to consider models of extinction based on
food-chain dynamics.  Amaral and Meyer~(1998) and Abramson~(1997) have both
constructed and studied such models.

Amaral and Meyer~(1998) have proposed a model in which species are arranged
in $L$ trophic levels labelled $l=0,1,\ldots,L-1$.  Each level has $N$
niches, each of which may be occupied or unoccupied by a species.  A
species in level $l$ (except $l=0$) feeds on up to $k$ species in level
$l-1$; these are its prey.  If all of a species' prey become extinct, then
it too becomes extinct, so avalanches of extinction can occur.  This
process is driven by randomly selecting one species at level~0 at each
time-step and making it extinction extinction with probability $p$.  There
is no sense of fitness or of competition between species governing
extinction in this model.

To replace extinct species, Amaral and Meyer use a speciation mechanism.
At a rate $\mu$, each existing species tries to engender an offspring
species by picking a niche at random in its own level or in the level above
or below.  If that randomly selected niche is unoccupied, then the new
species is created and assigned $k$ preys at random from the existing
species on the level below.  The parameter $\mu$ needs to be large enough
that the average origination rate exceeds the extinction rate, or all
species will become extinct.  Note that, as pointed out earlier, speciation
is inherently an evolutionary process and typically takes place on longer
time-scales than extinction through ecological interactions, so there is
some question about whether it is appropriate in a model such as this.  As
with the Sol\'e--Manrubia model, it might be preferable to view the
repopulation of niches as an invasion process, rather than a speciation
one.

The model is initialized by populating the first level $l=0$ with some
number $N_0$ of species at random.  Assuming a large enough origination
rate, the population will then grow approximately exponentially until
limited by the number of available niches.

Amaral and Meyer presented results for a simulation of their model with
parameters $L=6$, $k=3$, $N=1000$, $N_0\approx50$, $p=0.01$ and $\mu=0.02$.
The statistics of extinction events are similar to those seen in many other
models.  The times series is highly intermittent, with occasional large
extinction events almost up to the maximum possible size $NL$.  The
distribution of extinction sizes $s$ fits a power law,
Equation~\eref{powerlaw}, with exponent $\tau = 1.97\pm 0.05$.  Origination
rates are also highly intermittent, and strongly correlated with extinction
events.\footnote{The authors report that they obtained similar results,
  with the same exponents, for larger values of $k$ too (Amaral, private
  communication).}

An advantage of this model is that the number of species is not fixed, and
its fluctuations can be studied and compared with empirical data.  Amaral
and Meyer compute a power spectrum for the number of species and find that
it fits a power law $p(f) \propto 1/f^{\nu}$ with $\nu=1.95\pm 0.05$.  The
authors argue that this reveals a ``fractal structure'' in the data, but it
is worth noting that a power-spectrum exponent of $\nu=2$ occurs for many
non-fractal processes, such as simple random walks, and a self-similar
structure only needs to be invoked if $\nu<2$.

Amaral and Meyer also compute a power spectrum for the extinction rate, for
comparison with the fossil data analysis of Sol\'e~\etal~(1997).  They find
a power law with $\nu\simeq1$ for short sequences, but then see a crossover
to $\nu\simeq0$ at longer times, suggesting that there is no long-time
correlation.

Drossel~(1999) has analysed the Amaral--Meyer model in some detail.  The
$k=1$ case is most amenable to analysis, because then the food chains are
simple independent trees, each rooted in a single species at level~0.  The
extinction size distribution is therefore equal to the tree size
distribution, which can be computed by master equation methods, leading to
$p(s) \propto s^{-2}$ (i.e., $\tau=2$) exactly in the limits $N\to\infty$,
$L\to\infty$.  Finite size effects (when $N$ or $L$ are not infinite) can
also be evaluated, leading to a cutoff in the power law at $s_{\rm max}
\sim N\log N$ if $L\gg\log N$ or $s_{\rm max} \sim \e^L$ if $L\ll\log N$.
These analytical results agree well with the simulation studies.

The analysis for $k>1$ is harder, but can be reduced in the case of large
enough $L$ and $N$ (with $L\ll\ln N$) to a recursion relation connecting
the lifetime distribution of species on successive levels.  This leads to
the conclusion that the lifetime distribution becomes invariant after the
first few levels, which in turn allows for a solution.  The result is again
a power-law extinction size distribution with $\tau=2$ and cutoff $s_{\rm
  max} \sim \e^L$.

Drossel also considers a variant of the Amaral--Meyer model in which a
species becomes extinct if {\em any\/} (instead of all) of its prey
disappear.  She shows that this too leads to a power law with $\tau = 2$,
although very large system sizes would be needed to make this observable in
simulation.  She also points out that other variations of the model (such
as making the speciation rate depend on the density of species in a layer)
do not give power laws at all, so one must be careful about attributing too
much universality to the ``critical'' nature of this model.

\subsection{Abramson's food chain model}
Abramson~(1997) has proposed a different food chain model in which each
species is explicitly represented as a population of individuals.  In this
way Abramson's model connects extinction to microevolution, rather than
being a purely macroevolutionary model.  There is not yet a consensus on
whether a theory of macroevolution can be built solely on microevolutionary
principles; see Stenseth~(1985) for a review.

Abramson considers only linear food chains, in which a series of species at
levels $i=1,2,\ldots,N$ each feed on the one below (except $i=1$) and are
fed on by the one above (except $i=N$).  If the population density at level
$i$ at time $t$ is designated by $n_i(t)$, then the changes in one time
step are given by
\begin{eqnarray}
n_i(t+1) - n_i(t) &=& k_in_{i-1}(t)n_i(t)[1-n_i(t)/c_i]\nonumber\\
                  & & - g_in_{i+1}(t)n_i(t).
\end{eqnarray}
Here $k_i$ and $g_i$ represent the predation and prey rates, and $c_i$ is
the carrying capacity of level $i$.  These equations are typical of
population ecology.  At the endpoints of the chain, boundary conditions may
be imposed by adjoining two fictitious species, $0$ and $N+1$ with $n_0 =
n_{N+1} = 1$.  For simplicity Abramson takes $c_i=1$ for all $i$, and sets
$g_i=k_{i+1}$.  The species are then parameterized simply by their $k_i$
and by their population size $n_i(t)$.  These are initially chosen randomly
in the interval $(0,1)$.

The population dynamics typically leads to some $n_i(t)$'s dropping
asymptotically to 0.  When they drop below a small threshold, Abramson
regards that species as extinct and replaces it with a new species with
randomly chosen $n_i$ and $k_i$, drawn from uniform distributions in the
interval between zero and one.  But an additional driving force is still
needed to prevent the dynamics from stagnating.  So with probability $p$ at
each time-step, Abramson also replaces one randomly chosen species, as if
it had become extinct.

The replacement of an extinct species by a new one with a larger population
size has in general a negative impact on the species below it in the food
chain.  Thus one extinction event can lead to an avalanche propagating down
the food chain.  Note that this is the precise opposite of the avalanches
in the Amaral--Meyer model, which propagate upwards due to loss of food
source.

Abramson studies the statistics of extinction events in simulations of his
model for values of $N$ from 50 to 1000.\footnote{Sol\'e~(private
  communication) has made the point that these values are unrealistically
  large for real food chains.  Real food chains typically have less than
  ten trophic levels.} He finds punctuated equilibrium in the extinction
event sizes, but the size distribution $p(s)$ does {\em not\/} fit a power
law.  It does show some scaling behaviour with $N$, namely $p(s) =
N^{\beta}f(sN^{\nu})$, where $\beta$ and $\nu$ are parameters and $f(x)$ is
a particular ``scaling function''.  Abramson attributes this form to the
system being in a ``critical state''.  The waiting time between successive
extinctions fits a power law over several decades of time, but the exponent
seems to vary with the system size.  Overall, this model does not have
strong claims for criticality and does not agree very well with the
extinction data.

\section{Environmental stress models}
\label{stress}
In Sections~\sref{flmodels} to~\sref{connection} we discussed several
models of extinction which make use of ideas drawn from the study of
critical phenomena.  The primary impetus for this approach was the
observation of apparent power-law distributions in a variety of statistics
drawn from the fossil record, as discussed in Section~\sref{data}; in other
branches of science such power laws are often indicators of critical
processes.  However, there are also a number of other mechanisms by which
power laws can arise, including random multiplicative processes (Montroll
and Shlesinger~1982, Sornette and Cont~1997), extremal random processes
(Sibani and Littlewood~1993) and random barrier-crossing dynamics
(Sneppen~1995).  Thus the existence of power-law distributions in the
fossil data is not on its own sufficient to demonstrate the presence of
critical phenomena in extinction processes.

Critical models also assume that extinction is caused primarily by biotic
effects such as competition and predation, an assumption which is in
disagreement with the fossil record.  As discussed in Section~\sref{rates},
all the plausible causes for specific prehistoric extinctions are abiotic
in nature.  Therefore an obvious question to ask is whether it is possible
to construct models in which extinction is caused by abiotic environmental
factors, rather than by critical fluctuations arising out of biotic
interactions, but which still give power-law distributions of the relevant
quantities.

Such models have been suggested by Newman~(1996, 1997) and by Manrubia and
Paczuski~(1998).  Interestingly, both of these models are the result of
attempts at simplifying models based on critical phenomena.  Newman's model
is a simplification of the model of Newman and Roberts (see
Section~\sref{bsnoise}), which included both biotic and abiotic effects;
the simplification arises from the realization that the biotic part can be
omitted without losing the power-law distributions.  Manrubia and
Paczuski's model was a simplification of the connection model of Sol\`e and
Manrubia (see Section~\sref{sole}), but in fact all direct species-species
interactions were dropped, leaving a model which one can regard as driven
only by abiotic effects.  We discuss these models in turn.

\subsection{Newman's model}
\label{newman}
The model proposed by Newman~(1996, 1997) has a fixed number $N$ of species
which in the simplest case are non-interacting.  Real species do interact
of course, but as we will see the predictions of the model are not greatly
changed if one introduces interactions, and the non-interacting version
makes a good starting point because of its extreme simplicity.  The absence
of interactions between species also means that critical fluctuations
cannot arise, so any power laws produced by the model are definitely of
non-critical origin.

As in the model of Newman and Roberts~(1995), the level of the
environmental stress is represented by a single number $\eta$, which is
chosen independently at random from some distribution $p_{\rm
  stress}(\eta)$ at each time-step.  Each species $i=1\ldots N$ possesses
some threshold tolerance for stress denoted $x_i$ which is high in species
which are well able to withstand stress and low in those which are not.
(See Jablonski~(1989) for a discussion of the selectivity of extinction
events in the fossil record.)  Extinction takes place via a simple rule: if
at any time-step the numerical value of the stress level exceeds a species'
tolerance for stress, $\eta>x_i$, then that species becomes extinct at that
time-step.  Thus large stresses (sea-level change, bolide impact) can give
rise to large mass extinction events, whilst lower levels of stress produce
less dramatic background extinctions.  Note that simultaneous extinction of
many species occurs in this model because the same large stress affects all
species, and not because of any avalanche or domino effects in the
ecosystem.

In order to maintain a constant number of species, the system is
repopulated after every time-step with as many new species as have just
become extinct.  The extinction thresholds $x_i$ for the new species can
either be inherited from surviving species, or can be chosen at random from
some distribution $p_{\rm thresh}(x)$.  To a large extent it appears that
the predictions of the model do not depend on which choice is made; here we
focus on the uniform case with $p_{\rm thresh}(x)$ a constant independent
of $x$ over some allowed range of $x$, usually $0\le x<1$.  In addition,
it is safe to assume that the initial values of the variables $x_i$ are
also chosen according to $p_{\rm thresh}(x)$, since in any case the effects
of the initial choices only persist as long as it takes to turn over all
the species in the ecosystem, which happens many times during a run of the
model (and indeed many times during the known fossil record).

There is one further element which needs to be added to the model in order
to make it work.  As described, the species in the system start off with
randomly chosen tolerances $x_i$ and, through the extinction mechanism
described above, those with the lowest tolerance are systematically removed
from the population and replaced by new species.  Thus, the number of
species with low thresholds for extinction decreases over time, in effect
creating a gap in the distribution, as in the Bak--Sneppen model.  As a
result the size of the extinction events taking place dwindles and
ultimately extinction ceases almost entirely, a behaviour which we know not
to be representative of a real ecosystem.  Newman suggests that the
solution to this problem comes from evolution.  In the intervals between
large stress events, species will evolve under other selection pressures,
and this will change the values of the variables $x_i$ in unpredictable
ways.  Adapting to any particular selection pressure might raise, lower, or
leave unchanged a species' tolerance to environmental stresses.
Mathematically this is represented by making random changes to the $x_i$,
either by changing them all slightly at each time-step, or by changing a
small fraction $f$ of them to totally new values drawn from $p_{\rm
  thresh}(x)$, and leaving the rest unchanged.  These two approaches can be
thought of as corresponding to gradualist and punctuationalist views of
evolution respectively, but it appears in practice that the model's
predictions are largely independent of which is chosen.  In his work Newman
focused on the punctuationalist approach, replacing a fraction $f$ of the
species by random new values.

This description fully defines Newman's model except for the specification
of $p_{\rm stress}(\eta)$ and $p_{\rm thresh}(x)$.  However it turns out
that we can, without loss of generality, choose $p_{\rm thresh}(x)$ to have
the simple form of a uniform distribution in the interval from 0 to~1,
since any other choice can be mapped onto this with the transformation
\begin{equation}
x \to x' = \int_{-\infty}^x p_{\rm thresh}(y)\>\d y.
\label{transform}
\end{equation}
The stress level must of course be transformed in the same way,
$\eta\to\eta'$, so that the condition $\eta'>x'_i$ corresponds precisely to
$\eta>x_i$.  This in turn requires a transformation
\begin{equation}
p_{\rm stress}(\eta') = p_{\rm stress}(\eta) {\d\eta\over\d\eta'}
                      = {p_{\rm stress}(\eta)\over p_{\rm thresh}(\eta)}
\end{equation}
for the stress distribution.

The choice of $p_{\rm stress}(\eta)$ remains a problem, since it is not
known what the appropriate distribution of stresses is in the real world.
For some particular sources of stress, such as meteor impacts, there are
reasonably good experimental results for the distribution (Morrison~1992,
Grieve and Shoemaker~1994), but overall we have very little knowledge about
stresses occurring either today or in the geologic past.  Newman therefore
tested the model with a wide variety of stress distributions and found
that, in a fashion reminiscent of the self-organized critical models, many
of its predictions are robust against variations in the form of $p_{\rm
  stress}(\eta)$, within certain limits.

\begin{figure}[t]
\columnfigure{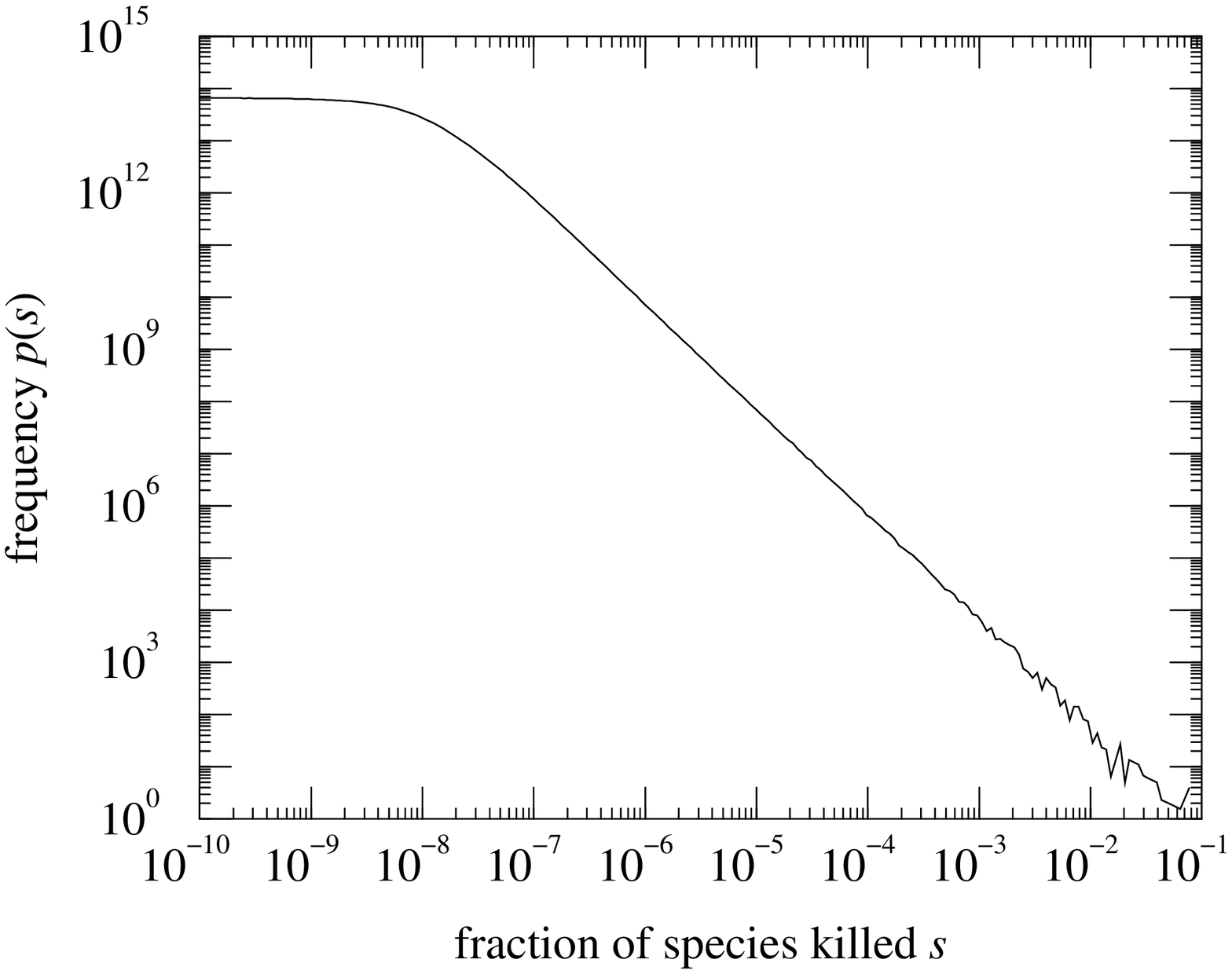}
\capt{Distribution of the sizes of extinction events taking place in the
  model of Newman~(1996).  The distribution is power-law in form with an
  exponent of $\tau=2.02\pm0.02$ except for extinctions of very small size,
  where it becomes flat.}
\label{sdist1}
\end{figure}

In Figure~\fref{sdist1} we show simulation results for the distribution
$p(s)$ of the sizes $s$ of extinction events in the model for one
particular choice of stress distribution, the Gaussian distribution:
\begin{equation}
p_{\rm stress}(\eta) \propto \exp\biggl[ -{\eta^2\over2\sigma^2} \biggr].
\label{normal}
\end{equation}
This is probably the commonest noise distribution occurring in natural
phenomena.  It arises as a result of the central limit theorem whenever a
number of different independent random effects combine additively to give
one overall stress level.  As the figure shows, the resulting distribution
of the sizes of extinction events in Newman's model follows a power law
closely over many decades.  The exponent of the power law is measured to be
$\tau=2.02\pm0.02$, which is in good agreement with the value of
$2.0\pm0.2$ found in the fossil data.  The only deviation from the
power-law form is for very small sizes $s$, in this case below about one
species in $10^8$, where the distribution flattens off and becomes
independent of $s$.  The point at which this happens is controlled
primarily by the value of the parameter $f$, which governs the rate of
evolution of species (Newman and Sneppen~1996).  No flat region is visible
in the fossil extinction distribution, Figure~\fref{extrates}, which
implies that the value of $f$ must be small---smaller than the smallest
fractional extinction which can be observed reliably in fossil data.
However, this is not a very stringent condition, since it is not possible
to measure extinctions smaller than a few per cent with any certainty.

\begin{figure}[t]
\columnfigure{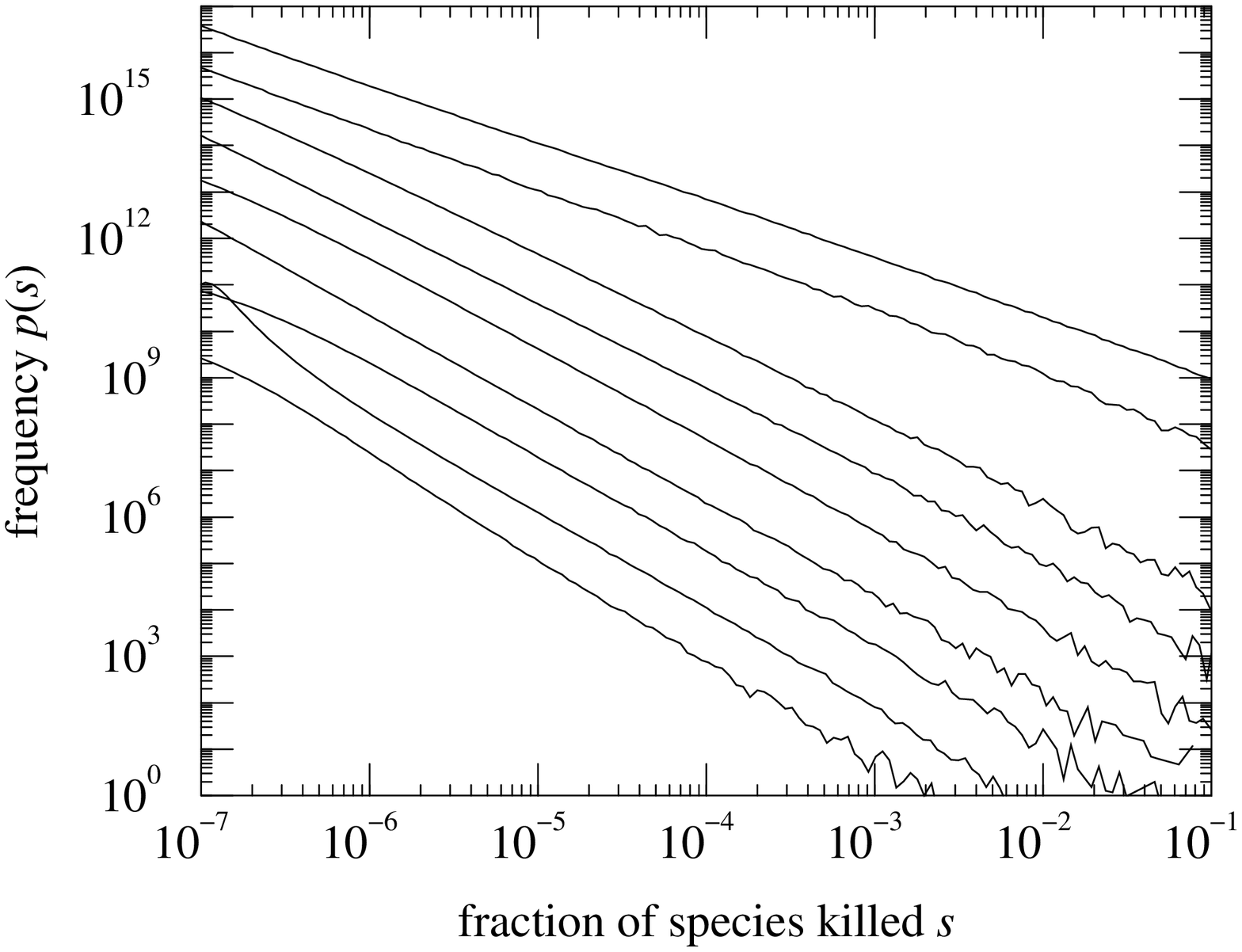}
\capt{Distribution of the sizes of extinction events for a variety of
  different stress distributions, including Gaussian, Lorentzian,
  Poissonian, exponential and stretched exponential.  In each case the
  distribution follows a power law closely over many decades.}
\label{sdist2}
\end{figure}

In Figure~\fref{sdist2} we show results for the extinction size distribution
for a wide variety of other distributions $p_{\rm stress}(\eta)$ of the
applied stress, including various different Gaussian forms, exponential and
Poissonian noise, power laws and stretched exponentials.  As the figure
shows, the distribution takes a power-law form in each case.  The exponent
of the power law varies slightly from one curve to another, but in all
cases it is fairly close to the value of $\tau\simeq2$ found in the fossil
record.
In fact, Sneppen and Newman~(1997) have shown analytically that for all
stress distributions $p_{\rm stress}(\eta)$ satisfying
\begin{equation}
\int_\eta^\infty p_{\rm stress}(x)\>\d x \sim p_{\rm stress}(\eta)^\alpha
\end{equation}
for large $\eta$ and some exponent $\alpha$, the distribution of extinction
sizes will take a power law form for large $s$.  This condition is exactly
true for exponential and power-law distributions of stress, and
approximately true for Gaussian and Poissonian distributions.  Since this
list covers almost all noise distributions which occur commonly in natural
systems, the predictions of the model should be reasonably robust,
regardless of the ultimate source of the stresses.

\begin{figure}[t]
\columnfigure{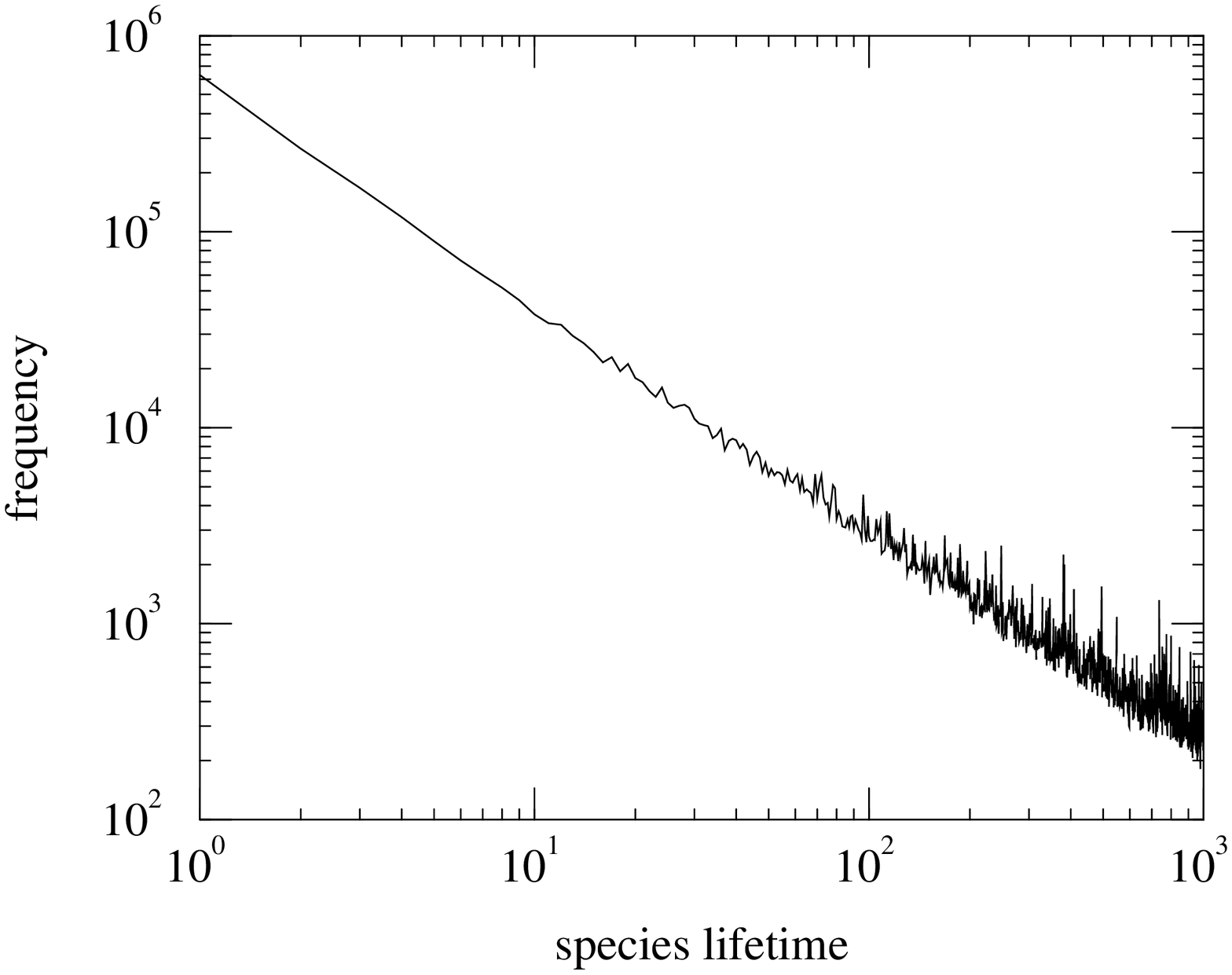}
\capt{The distribution of the lifetimes of species in the model of
  Newman~(1997).  The distribution follows a power law with an exponent in
  the vicinity of~1.}
\label{ldist}
\end{figure}

It is also straightforward to measure the lifetimes of species in
simulations of this model.  Figure~\fref{ldist} shows the distribution of
lifetimes measured in one particular run.  The distribution is power-law in
form as it is in the fossil data, with a measured exponent of
$1.03\pm0.05$.

Newman~(1997) has given a number of other predictions of his model.  In
particular, he has suggested how taxonomy can be incorporated into the
model to allow one to study the birth and death of genera and higher taxa,
in addition to species.  With this extension the model predicts a
distribution of genus lifetimes similar to that of species, with a
power-law form and exponent in the vicinity of one.  Note that although the
power-law form is seen also in the fossil data, an exponent of one is not
in agreement with the value of $1.7\pm0.3$ measured in the fossil lifetime
distribution (see Section~\sref{lifetimes}).  The model does however
correctly predict Willis's power-law distribution of the number of species
per genus (see Section~\sref{taxonomy}) with an exponent close to the
measured value of $\beta=\frac32$.

Another interesting prediction of the model is that of ``aftershock
extinctions''---strings of smaller extinctions arising in the aftermath of
a large mass extinction event (Sneppen and Newman~1997, Wilke~\etal~1998).
The mechanism behind these aftershock extinctions is that the repopulation
of ecospace after a large event tends to introduce an unusually high number
of species with low tolerance for stress.  (At other times such species are
rarely present because they are removed by the frequent small stresses
applied to the system.)  The rapid extinction of these unfit species
produces a high turnover of species for a short period after a mass
extinction, which we see as a series of smaller ``aftershocks''.  The model
makes the particular prediction that the intervals between these aftershock
extinctions should fall off with time as $t^{-1}$ following the initial
large event.  This behaviour is quite different from that of the critical
models of earlier sections, and therefore it could provide a way of
distinguishing in the fossil record between the two processes represented
by these models.  So far, however, no serious effort has been made to look
for aftershock extinctions in the fossil data, and indeed it is not even
clear that the available data are adequate for the task.  In addition,
later work by Wilke and Martinetz~(1997) calls into question whether one
can expect aftershocks to occur in real ecosystems.  (This point is
discussed further in Section~\sref{wilke}.)


\subsection{Shortcomings of the model}
\label{newshort}
Although Newman's model is simple and makes predictions which are in many
cases in good agreement with the fossil data, there are a number of
problems associated with it.

First, one could criticise the assumptions which go into the model.  For
example, the model assumes that species are entirely non-interacting, which
is clearly false.  In the version we have described here it also assumes a
``punctuated'' view of evolution in which species remain constant for long
periods and then change abruptly.  In addition, the way in which new
species are added to the model is questionable: new species are given a
tolerance $x_i$ for stress which is chosen purely at random, whereas in
reality new species are presumably descended from other earlier species and
therefore one might expect some correlation between the values of $x_i$ for
a species and its ancestors.

These criticisms lead to a number of generalizations of the model which
have been examined by Newman~(1997).  To investigate the effect of species
interactions, Newman looked at a variation of the model in which the
extinction of a species could give rise to the extinction of a neighbouring
species, in a way reminiscent of the avalanches of Kauffman's \NK\ model.
He placed the model on a lattice and added a step to the dynamics in which
the extinction of a species as a result of external stress caused the
knock-on extinction (and subsequent replacement) of all the species on
adjacent lattice sites.
In simulations of this version of the model, Newman found, inevitably,
spatial correlations between the species becoming extinct which are not
present in the original version.  Other than this however, it appears that
the model's predictions are largely unchanged.  The distributions of
extinction event sizes and taxon lifetimes for example are still power-law
in form and still possess approximately the same exponents.

Similarly it is possible to construct a version of the model in which
evolution proceeds in a ``gradualist'' fashion, with the values of the
variables $x_i$ performing a slow random walk rather than making punctuated
jumps to unrelated values.  And one can also create a version in which the
values of $x_i$ assumed by newly appearing species are inherited from
survivors, rather than chosen completely at random.  Again it appears that
these changes have little effect on the major predictions of the model,
although these results come primarily from simulations of the model; the
analytic results for the simplest version do not extend to the more
sophisticated models discussed here.

\subsection[The multi-trait version of the model]{The multi-trait version
of\\the model}
A more serious criticism of Newman's model is that it models different
types of stress using only a single parameter $\eta$.  Within this model
one can only say whether the stress level is high or low at a particular
time.  In the real world there are many different kinds of stress, such as
climatic stress, ecological stresses like competition and predation,
disease, bolide impact, changes in ocean chemistry and many more.  And
there is no guarantee that a period when one type of stress is high will
necessarily correspond to high stress of another type.  This clearly has an
impact on extinction profiles, since some species will be more susceptible
to stresses of a certain kind than others.  To give an example, it is
thought that large body mass was a contributing factor to extinction at the
Cretaceous--Tertiary boundary (Clemens~1986).  Thus the particular stress
which caused the K--T extinction, thought to be the result of a meteor
impact, should correspond to tolerance variables $x_i$ in our model which
are lower for large-bodied animals.  Another type of stress---sea-level
change, say---may have little or no correlation with body size.

To address this problem, Newman~(1997) has also looked at a variation of
his model in which there are a number $M$ of different kinds of stress.  In
this case each species also has a separate tolerance variable $x_i^{(k)}$
for each type of stress $k$ and becomes extinct if any one of the stress
levels exceeds the corresponding threshold.  As with the other variations
on the model, it appears that this ``multi-trait'' version reproduces the
important features of the simpler versions, including the power-law
distributions of the sizes of extinction events and of species lifetimes.
Sneppen and Newman~(1997) have explained this result with the following
argument.  To a first approximation, one can treat the probability of a
species becoming extinct in the multi-trait model as the probability that
the stress level exceeds the lowest of the thresholds for stress which that
species possesses.  In this case, the multi-trait model is identical to the
single-trait version but with a different choice for the distribution
$p_{\rm thresh}(x)$ from which the thresholds are drawn (one which reflects
the probability distribution of the lowest of $M$ random numbers).
However, as we argued earlier, the behaviour of the model is independent of
$p_{\rm thresh}(x)$ since we can map any distribution on the uniform one by
a simple integral transformation of $x$ (see Equation~\eref{transform}).

\subsection[The finite-growth version of the model]{The finite-growth
version of\\the model}
\label{wilke}
Another shortcoming of the model proposed by Newman is that the species
which become extinct are replaced instantly by an equal number of new
species.  In reality, fossil data indicate that the process of replacement
of species takes a significant amount of time, sometimes as much as a few
million years (Stanley~1990, Erwin~1996).  Wilke and Martinetz~(1997) have
proposed a generalization of the model which takes this into account.  In
this version, species which become extinct are replaced slowly according to
the logistic growth law
\begin{equation}
{\d N\over\d t} = gN(1-N/N_{\rm max}),
\end{equation}
where $N$ is the number of species as before, and $g$ and $N_{\rm max}$ are
constants.  Logistic growth appears to be a reasonable model for recovery
after large extinction events (Sepkoski~1991, Courtillot and
Gaudemer~1996).  When the growth parameter $g$ is infinite, we recover the
model proposed by Newman.  Wilke and Martinetz find, as one might expect,
that there is a transition in the behaviour of the system at a critical
value $g=g_c$ where the rate of repopulation of the system equals the
average rate of extinction.  They give an analytic treatment of the model
which shows how $g_c$ varies with the other parameters in the problem.  For
values of $g$ below $g_c$ life eventually dies out in the model, and it is
probably reasonable to assume that the Earth is not, for the moment at
least, in this regime.  For values of $g$ above $g_c$ it is found that the
power-law behaviour seen in the simplest versions of the model is retained.
The value of the extinction size exponent $\tau$ appears to decrease
slightly with increasing $g$, but is still in the vicinity of the value
$\tau\simeq2$ extracted from the fossil data.  Interestingly they also find
that the aftershock extinctions discussed in Section~\sref{newman} become
less well-defined for finite values of $g$, calling into question Newman's
contention that the existence of aftershocks in the fossil record could be
used as evidence in favour of his model.  This point is discussed further
by Wilke~\etal~(1998).

\subsection[The model of Manrubia and Paczuski]{The model of Manrubia and\\
Paczuski}
\label{mpmodel}
Another variation on the ideas contained in Newman's model has been
proposed by Manrubia and Paczuski~(1998).  Interestingly, although this
model is mathematically similar to the other models discussed in this
section, its inspiration is completely different.  In fact, it was
originally intended as a simplification of the connection model of Sol\'e
and Manrubia discussed in Section~\sref{sole}.

In Newman's model, there are a large number of species with essentially
constant fitness or tolerance to external stress, and those which fall
below some time-varying threshold level become extinct.  In the model of
Manrubia and Paczuski by contrast, the threshold at which species become
extinct is fixed and their fitness is varied over time.  In detail, the
model is as follows.

The model contains a fixed number $N$ of species, each with a fitness
$x_i$, or ``viability'' as Manrubia and Paczuski have called it.  This
viability measures how far a species is from becoming extinct, and might be
thought of as a measure of reproductive success.  All species are subject
to random coherent stresses, or ``shocks'', which additively increase or
decrease the viability of all species by the same amount $\eta$.  If at any
point the viability of a species falls below a certain threshold $x_0$,
that species becomes extinct and is replaced by speciation from one of the
surviving species.  In Newman's model there was also an ``evolution''
process which caused species with high viability to drift to lower values
over the course of time, preventing the system from stagnating when all
species with low viability had been removed.  The model of Manrubia and
Paczuski contains an equivalent mechanism, whereby the viabilities of all
species drift, in a stochastic fashion, toward lower values over the course
of time.  This also prevents stagnation of the dynamics.

\begin{figure}[t]
\columnfigure{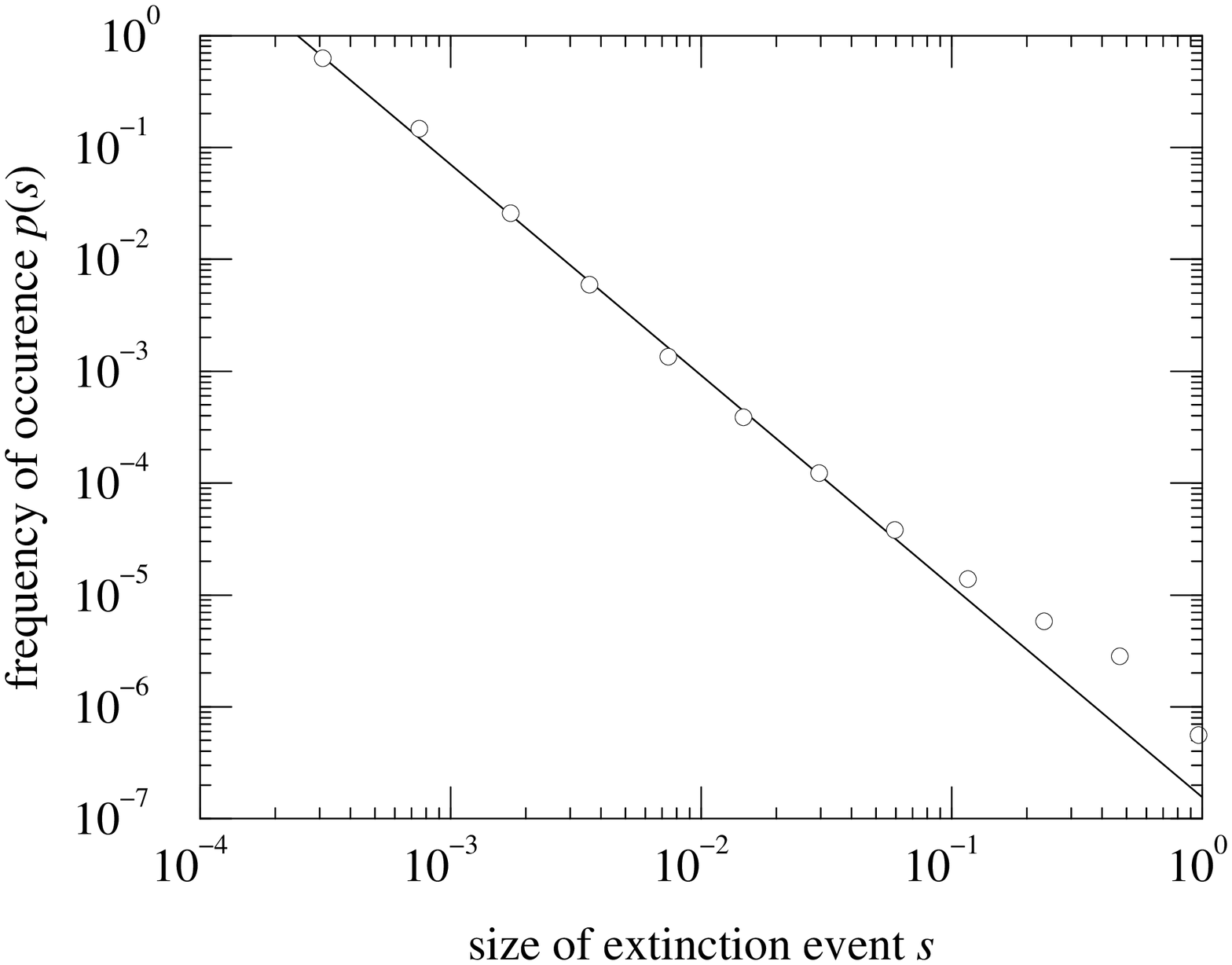}
\capt{The distribution of the sizes of extinction events in a simulation of
  the model of Manrubia and Paczuski, with $N=3200$ species (circles).  The
  best fit power law (solid line) has an exponent of $\tau=1.88\pm0.09$.
  After Manrubia and Paczuski~(1998).}
\label{manrubia}
\end{figure}

Although no one has shown whether the model of Manrubia and Paczuski can be
mapped exactly onto Newman's model, it is clear that the dynamics of the
two are closely similar, and therefore it is not surprising to learn that
the behaviour of the two models is also similar.  Figure~\fref{manrubia}
shows the distribution of the sizes $s$ of extinction events in a
simulation of the model with $N=3200$ species.  The distribution is close
to power-law in form with an exponent of $\tau=1.9$ similar to that of
Newman's model, and in agreement with the result $\tau\simeq2$ seen in the
fossil data.  The model also generates a power-law distribution in the
lifetimes of species and, as in Newman's model, a simple definition of
genus can be introduced and it can be shown that the distribution of number
of species per genus follows a power law as well.  The exponent of the
lifetime distribution turns out to be approximately~2, which is not far
from the value of $1.7\pm0.3$ found in the fossil data (see
Section~\sref{lifetimes}).\footnote{The exponent for the distribution of
  genus sizes is also~2 which is perhaps a shortcoming of this model;
  recall that Willis's value for flowering plants was $1.5$
  (Figure~\fref{willis}), and the comprehensive studies by Burlando~(1990,
  1993) gave an average value of $1.6$.}

What is interesting about this model however, is that its dynamics is
derived using a completely different argument from the one employed by
Newman.  The basic justification of the model goes like this.  We assume
first of all that it is possible to define a viability $x_i$ for species
$i$, which measures in some fashion how far a species is from the point of
extinction.  The point of extinction itself is represented by the threshold
value $x_0$.  The gradual downward drift of species' viability can be then
be accounted for as the result of mutation; the majority of mutations lower
the viability of the host.

Manrubia and Paczuski justify the coherent stresses in the system by
analogy with the model of Sol\'e and Manrubia~(1996) in which species feel
the ecological ``shock'' of the extinction of other nearby species.  In the
current model, the origin of the shocks is similarly taken to be the
extinction of other species in the system.  In other words it is the result
of biotic interaction, rather than exogenous environmental influences.
However, by representing these shocks as coherent effects which influence
all species simultaneously to the same degree, Manrubia and Paczuski have
removed from the dynamics the direct interaction between species which was
present in the original connection model.  Amongst other things, this
allows them to give an approximate analytic treatment of their model using
a time-averaged approximation similar to the one employed by Sneppen and
Newman~(1997) for Newman's model.

One further nice feature of the Manrubia--Paczuski model is that it is
particularly easy in this case to see how large extinction events arise.
Because species are replaced by speciation from others, the values of their
viabilities tend to cluster together: most species are copies, or near
copies, of other species in the system.  Such clusters of species tend all
to become extinct around the same time because they all feel the same
coherent shocks and are all driven below the extinction threshold together.
(A similar behaviour is seen in the Sol\'e--Manrubia model of
Section~\sref{sole}.)  This clustering and avalanche behaviour in the model
is reminiscent of the so-called ``phase-coherent'' models which have been
proposed as a mechanism for the synchronization of the flashing of
fireflies (Strogatz and Stewart~1993).  Although no one has yet made a
direct connection between these two classes of models, it is possible that
mathematical techniques similar to those employed with phase-coherent
models may prove profitable with models of type proposed by Manrubia and
Paczuski.

\section{Sibani's reset model}
\label{reset}
Sibani and co-workers have proposed a model of the extinction process,
which they call the ``reset model'' (Sibani~\etal~1995, 1998), which
differs from those discussed in the preceding sections in a fundamental
way; it allows for, and indeed relies upon, non-stationarity in the
extinction process.  That is, it acknowledges that the extinction record is
not uniform in time, as it is assumed to be (except for stochastic
variation) in the other models we have considered.  In fact, extinction
intensity has declined on average over time from the beginning of the
Phanerozoic until the Recent.  Within the model of Sibani~\etal, the
distributions of Section~\sref{data} are all the result of this decline,
and the challenge is then to explain the decline, rather than the
distributions themselves.

\subsection{Extinction rate decline}
\begin{figure}[t]
\columnfigure{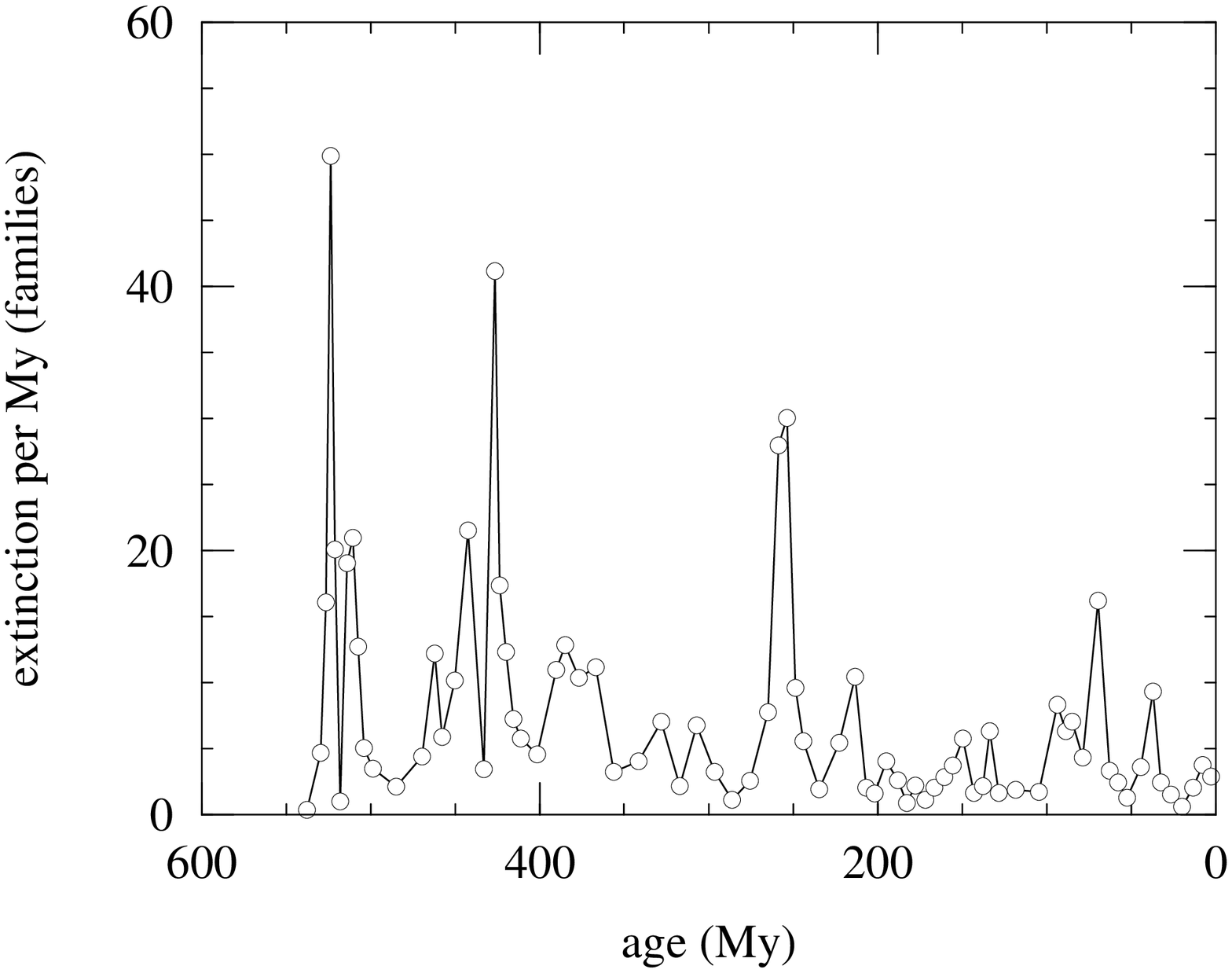}
\capt{The number of families of marine organisms becoming extinct per
  million years in each of the stages of the Phanerozoic.  The decline in
  average extinction rate is clearly visible in this plot.  The data are
  from the compilation by Sepkoski~(1992).}
\label{pertime}
\end{figure}

In Figure~\fref{diversity} we showed the number of known families as a
function of time over the last 600~My.  On the logarithmic scale of the
figure, this number appears to increase fairly steadily and although, as
we pointed out, some of this increase can be accounted for by the bias
known as the ``pull of the recent'', there is probably a real trend present
as well.  It is less clear that there is a similar trend in extinction
intensity.  The extinctions represented by the points in
Figure~\fref{extrates} certainly vary in intensity, but on average they
appear fairly constant.  Recall however, that Figure~\fref{extrates} shows
the number of families becoming extinct in each stage, and that the lengths
of the stages are not uniform.  In Figure~\fref{pertime} we show the
extinction intensity normalized by the lengths of the stages---the
extinction rate in families per million years---and on this figure it is
much clearer that there is an overall decline in extinction towards the
Recent.

\begin{figure}[t]
\columnfigure{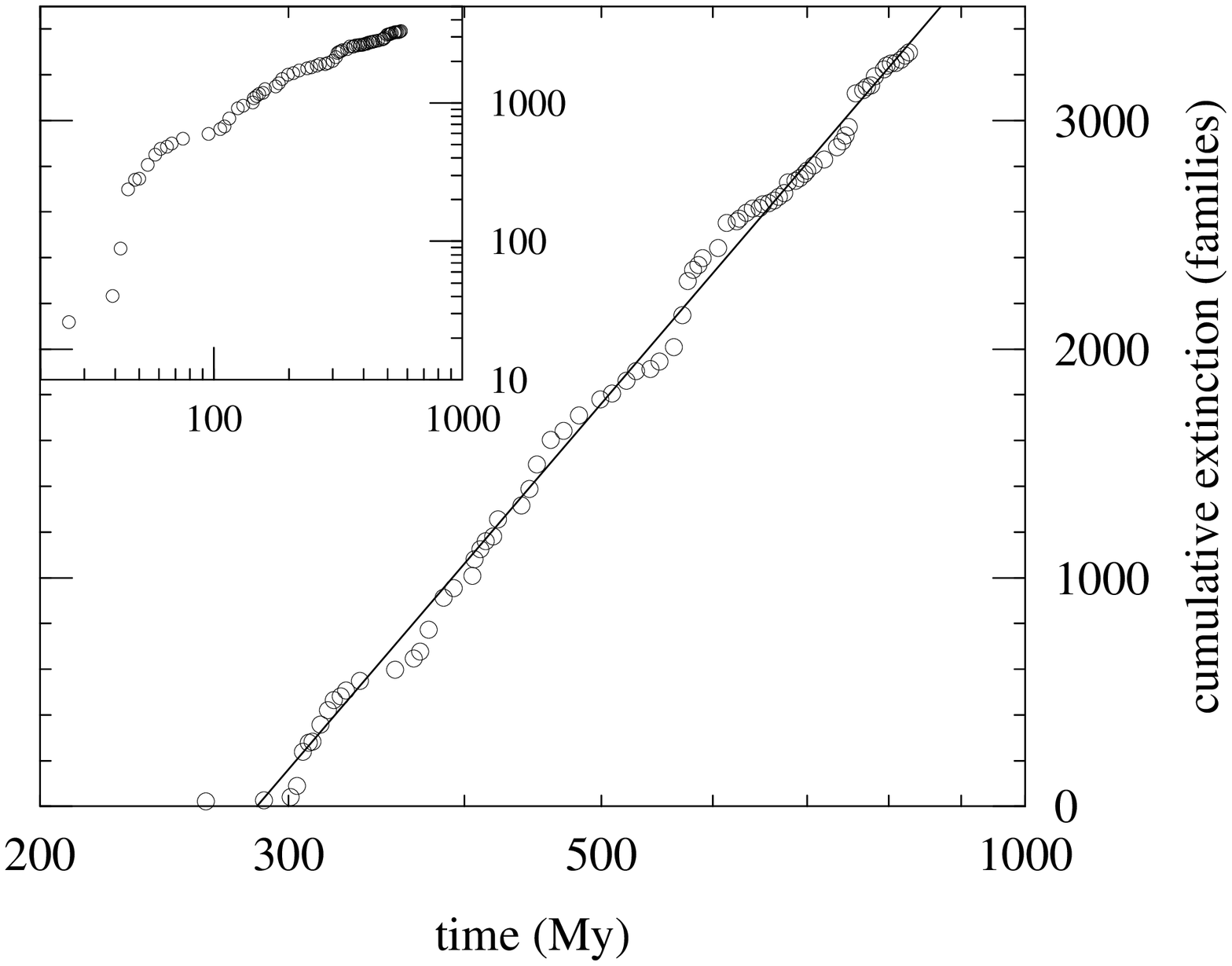}
\capt{Main figure: the cumulative extinction intensity as a function of
  time during the Phanerozoic on linear--log scales.  The straight line is
  the best logarithmic fit to the data.  Inset: the same data on log--log
  scales.  After Newman and Eble~(1999b).}
\label{accum}
\end{figure}

In order to quantify the decline in extinction rate, we consider the {\em
  cumulative\/} extinction intensity $c(t)$ as a function of time.  The
cumulative extinction at time $t$ is defined to be the number of taxa which
have become extinct up to that time.  In other words, if we denote the
extinction intensity at time $t$ by $x(t)$ then the cumulative extinction
intensity is
\begin{equation}
c(t) = \int_0^t x(t')\>\d t'.
\label{defsc}
\end{equation}
Figure~\fref{accum} shows this quantity for the marine families in
Sepkoski's database.  Clearly the plot has to be monotonically increasing.
Sibani~\etal\ suggested that it in fact has a power-law form, with an
exponent in the vicinity of~$0.6$.  Newman and Eble~(1999b) however have
pointed out that it more closely follows a logarithmic increase law---a
straight line on the linear--log scales of Figure~\fref{accum}.  (For
comparison we show the same data on log--log scales in the inset.  The
power-law form proposed by Sibani~\etal\ would appear as a straight line on
these scales.)  This implies that $c(t)$ can be written in the form
\begin{equation}
c(t) = A + B\log(t-t_0),
\label{accumulate}
\end{equation}
where $A$ and $B$ are constants and $t_0$ is the point of intercept of the
line in Figure~\fref{accum} with the horizontal axis.  (Note that $t_0$
lies before the beginning of the Cambrian.  If time is measured from $t=0$
at the start of the data set, which coincides roughly with the beginning of
the Cambrian, then the best fit of the form~\eref{accumulate} has
$t_0\simeq-260$~My.)

Combining Equations~\eref{defsc} and~\eref{accumulate} and differentiating
with respect to $t$ we get an expression for the extinction per unit time:
\begin{equation}
x(t) = {B\over t-t_0}.
\label{falloff}
\end{equation}
In other words the average extinction rate is falling off over time as a
power law with exponent $1$.  Sibani~\etal\ have pointed out that a
power-law decline in itself could be enough to explain the distribution of
the sizes of extinction events seen in Figure~\fref{extdist}.  For an
extinction profile of the form of Equation~\eref{falloff} the number of
time intervals in which we expect to see extinction events of a certain
size $s$ is given by
\begin{equation}
p(s) = {\d t\over\d x}\bigg|_{x=s} = -{B\over s^2}.
\label{diffform}
\end{equation}
In other words, the distribution of event sizes has precisely the power-law
form see in Figure~\fref{extlog}, with an exponent $\tau=2$ which is in
good agreement with the fossil data.  (If we use the power-law fit to the
cumulative extinction intensity suggested by Sibani~\etal, the exponent
works out at about $\tau=2.5$, which is outside the standard error on the
value measured in the fossil record---another reason for preferring the
logarithmic fit.)

There are problems with this argument.  The analysis assumes that the
extinction rate takes the idealized form of Equation~\eref{falloff},
whereas in fact this equation represents only the average behaviour of the
real data.  In reality, there is a great deal of fluctuation about this
form.  For example, Equation~\eref{falloff} implies that all the large
extinction events happened in the earliest part of the fossil record,
whereas in fact this is not true.  The two largest events of all time (the
late-Permian and end-Cretaceous events) happened in the second half of the
Phanerozoic.  Clearly then this analysis cannot tell the entire story.

A more serious problem is that this theory is really just ``passing the
buck''.  It doesn't tell us how, in biological terms, the observed
extinction size distribution comes about.  All it does is tell us that one
distribution arises because of another.  The extinction size distribution
may be a result of the fall-off in the average extinction rate, but where
does the fall-off come from?

The origin of the decline in the extinction rate has been a topic of debate
for many years.  It has been suggested that the decline may be a sampling
bias in the data, arising perhaps from variation in the quality of the
fossil record through geologic time (Pease~1992) or from changes in
taxonomic structure (Flessa and Jablonski~1985).  As with the increase in
diversity discussed in Section~\sref{origination}, however, many believe
that these biases are not enough to account entirely for the observed
extinction decline.  Raup and Sepkoski~(1982) have suggested instead that
the decline could be the result of a slow evolutionary increase in the mean
fitness of species, fitter species becoming extinct less easily than their
less fit ancestors.  This appears to be a plausible suggestion, but it has
a number of problems.  With respect to what are we measuring fitness in
this case?  Do we mean fitness relative to other species?  Surely not,
since if all species are increasing in fitness at roughly the same rate,
then their fitness relative to one another will remain approximately
constant.  (This is another aspect of van~Valen's ``Red Queen hypothesis'',
which we mentioned in Section~\sref{early}.)  Do we then mean fitness with
respect to the environment, and if so, how is such a fitness defined?  The
reset model attempts to address these questions and quantify the theory of
increasing species fitness.

\subsection{The reset model}
The basic idea of the reset model is that species are evolving on
high-dimensional rugged fitness landscapes of the kind considered
previously in Section~\sref{flmodels}.  Suppose a species is evolving on
such a landscape by mutations which take it from one local peak to another
at approximately regular intervals of time.  (This contrasts with the
picture proposed by Bak and Sneppen~(1993)---see Section~\sref{bs}---in
which the time between evolutionary jumps is not constant, but depends on a
barrier variable which measures how difficult a certain jump is.)  If the
species moves to a new peak where the fitness is higher than the fitness at
the previous peak, then the new strain will replace the old one.  If the
dimensionality of the landscape is sufficiently high then the chance of a
species retracing its steps and encountering the same peak twice is small
and can be neglected.  In this case, the process of sampling the fitness at
successive peaks is equivalent to drawing a series of independent random
fitness values from some fixed distribution, and keeping a record of the
highest one encountered so far.  Each time the current highest value is
replaced by a new one, an evolutionary event has taken place in the model
and such events correspond to pseudoextinction of the ancestral species.
Sibani~\etal\ refer to this process as a ``resetting'' of the fitness of
the species (hence the name ``reset model''), and to the entire dynamics of
the model as a ``record dynamics''.

The record dynamics is simple enough to permit the calculation of
distributions of a number of quantities of interest.  First of all,
Sibani~\etal\ showed that the total number of evolution/extinction events
happening between an initial time $t_0$ and a later time $t$ goes as
$\log(t-t_0)$ on average, regardless of the distribution from which the
random numbers are drawn.  This of course is precisely the form seen in the
fossil data, Equation~\eref{accumulate}, and immediately implies that the
number of events per unit time falls off as $1/(t-t_0)$.  Then the
arguments leading up to Equation~\eref{diffform} tell us that we should
expect a distribution of sizes of extinction events with an exponent
$\tau=2$, as in the fossil data.

We can also calculate the distribution of the lifetimes of species.
Assuming that the lifetime of a species is the interval between the
evolutionary event which creates it and the next event, in which it
disappears, it turns out that the reset model implies a distribution of
lifetimes which is power-law in form with an exponent $\alpha=1$, again
independent of the distribution of the random numbers used.  This is some
way from the value $\alpha=1.7\pm0.3$ observed in the fossil data
(Section~\sref{lifetimes}), but no more so than for most of the other
models discussed previously.

\begin{table*}
\begin{center}
\begin{tabular}{|l|c|c|c|}
\hline
 & \multicolumn{3}{c|}{exponent of distribution} \\
\cline{2-4}
 & extinction size & taxon lifetime & species per genus \\
 & $\tau$          & $\alpha$       & $\beta$ \\
\hline
\hline
fossil data            & $2.0\pm0.2$      & $1.7\pm0.3$   & $1.5\pm0.1$   \\
\hline
\NKCS\                 & $\simeq1$        & --            & --            \\
Bak and Sneppen        & $1$ to $\frac32$ & $1$           & --            \\
Vandewalle and Ausloos & $1.49\pm0.01$    & --            & $1.89\pm0.03$ \\
Newman and Roberts     & $2.02\pm0.03$    & --            & --            \\
Sol\'e and Manrubia    & $2.05\pm0.06$    & --            & $2.05\pm0.06$ \\
Amaral and Meyer       & $1.97\pm0.05$    & --            & --            \\
Newman                 & $2.02\pm0.02$    & $1.03\pm0.05$ & $1.6\pm0.1$   \\
Manrubia and Paczuski  & $1.9\pm0.1$      & $\simeq2$     & $\simeq2$     \\
Sibani~\etal           & $2$              & $1$           & --            \\
\hline
\end{tabular}
\end{center}
\null\vspace{0.4cm}
\tcapt{Exponents of various distributions as measured in the fossil record,
and in some of the models described in this review.}
\end{table*}

\subsection{Extinction mechanisms}
The model described so far contains only a pseudoextinction mechanism;
there is no true extinction taking place, a situation which we know not to
be representative of the fossil record.  Sibani~\etal\ suggested an
extension of their model to incorporate a true extinction mechanism based
on competition between species.  In this version of the model each species
interacts with a number of neighbouring species.  Sibani~\etal\ placed the
species on a lattice and allowed each one to interact with its nearest
neighbours on the lattice.  (Other choices would also be possible, such as
the random neighbours of the \NK\ and Sol\'e--Manrubia models, for
instance.)  If a species increases its fitness to some new value through an
evolutionary event, then any neighbouring species with fitness lower than
this new value becomes extinct.  The justification for this extinction
mechanism is that neighbouring species are in direct competition with one
another and therefore the fitter species tends to wipe out the less fit one
by competitive exclusion.  As in most of the other models we have
considered, the number of species in the model is maintained at a constant
level by repopulating empty niches with new species whose fitnesses are, in
this case, chosen at random.  Curiously, Sibani~\etal\ did not calculate
the distribution of the sizes of extinction events in this version of the
model, although they did show that the new version has a steeper species
lifetime distribution; it is still a power law but has an exponent of
$\alpha=2$, a value somewhat closer to the $\alpha=1.7\pm0.3$ seen in the
fossil data.

\section{Conclusions}
\label{conclusions}
In this paper we have reviewed a large number of recent quantitative models
aimed at explaining a variety of large-scale trends seen in the fossil
record.  These trends include the occurrence of mass extinctions, the
distribution of the sizes of extinction events, the distribution of the
lifetimes of taxa, the distribution of the numbers of species per genus,
and the apparent decline in the average extinction rate.  None of the
models presented match all the fossil data perfectly, but all of them offer
some suggestion of possible mechanisms which may be important to the
processes of extinction and origination.  In this section we conclude our
review by briefly running over the properties and predictions of each of
the models once more.  Much of the interest in these models has focussed on
their ability (or lack of ability) to predict the observed values of
exponents governing distributions of a number of quantities.  In Table~1 we
summarize the values of these exponents for each of the models.

Most of the models we have described attempt to provide possible
explanations for a few specific observations.  (1)~The fossil record
appears to have a power-law (i.e.,~scale-free) distribution of the sizes of
extinction events, with an exponent close to~$2$ (Section~\sref{rates}).
(2)~The distribution of the lifetimes of genera also appears to follow a
power law, with exponent about $1.7$ (Section~\sref{lifetimes}).  (3)~The
number of species per genus appears to follow a power law with exponent
about $1.5$ (Section~\sref{taxonomy}).

One of the first models to attempt an explanation of these observations was
the \NK\ model of Kauffman and co-workers.  In this model extinction is
driven by coevolutionary avalanches.  When tuned to the critical point
between chaotic and frozen regimes, the model displays a power-law
distribution of avalanche sizes with an exponent of about $1$.  It has been
suggested that this could in turn lead to a power-law distribution of the
sizes of extinction events, although the value of $1$ for the exponent is
not in agreement with the value $2$ measured in the fossil extinction
record.  It is not clear by what mechanism the extinction would be produced
in this model.

Building on Kauffman's ideas, Bak and Sneppen proposed a simpler model
which not only produces coevolutionary avalanches, but also self-organizes
to its own critical point, thereby automatically producing a power-law
distribution of avalanche sizes, regardless of other parameters in the
system.  Again the exponent of the distribution is in the vicinity of one,
which is not in agreement with the fossil record.  Many extensions of the
Bak--Sneppen model have been proposed.  We have described the multi-trait
model of Boettcher and Paczuski which is less realistic but has the
advantage of being exactly solvable, the model of Vandewalle and Ausloos
which incorporates speciation effects and phylogenetic trees, the model of
Head and Rodgers which also proposes a speciation mechanism, and the model
of Newman and Roberts which introduces true extinction via environmental
stress.

A different, but still biotic, extinction mechanism has been investigated
by Sol\'e and Manrubia, who proposed a ``connection'' model based on ideas
of ecological competition.  It is not clear whether ecological effects have
made an important contribution to the extinction we see in the fossil
record, although the current consensus appears to be that they have not.
The Sol\'e--Manrubia model, like Kauffman's \NK\ model, is a true critical
model, which only produces power-law distributions when tuned to its
critical point.  Unlike Kauffman's model however, the model of Sol\'e and
Manrubia produces the correct value for the extinction size distribution
when tuned to this point.  We have also described two other models of
extinction through ecological interaction: the food chain models of Amaral
and Meyer and of Abramson.

A third distinct extinction mechanism is extinction through environmental
stress, which has been investigated in modelling work by Newman.  In
Newman's model, species with low tolerance for stress become extinct during
periods of high stress, and no species interactions are included at all.
The model gives a value of $2$ for the extinction size distribution, the
same as that seen in the fossil record.  Wilke and Martinetz have proposed
a more realistic version of the same model in which recovery after mass
extinctions takes place gradually, rather than instantaneously.  Another
related model is that of Manrubia and Paczuski in which extinction is also
caused by coherent ``shocks'' to the ecosystem, although the biological
justification for these shocks is different from that given by Newman.
Their model also generates a power-law distribution of extinction sizes
with exponent~2.

Finally, we have looked at the ``reset model'' of Sibani~\etal, which
proposes that the distribution of sizes of extinction events is a result of
declining extinction intensity during the Phanerozoic.  The decline is in
turn explained as a result of increasing average fitness of species as they
evolve.

Clearly there are a large number of competing models here, and simply
studying quantities such as the distribution of the sizes of extinction
events is not going to allow us to distinguish between them.  In
particular, the question of whether the dominant mechanisms of extinction
are biotic or abiotic is interesting and thus far undecided.  However, the
models we have give us a good feeling for what mechanisms might be
important for generating these distributions.  A sensible next step would
be to look for signatures, in the fossil record or elsewhere, which might
allow us to distinguish between these different mechanisms.

\section*{Acknowledgements}
The authors would like to thank Per Bak, Stefan Boettcher, Gunther Eble,
Doug Erwin, Wim Hordijk, Stuart Kauffman, Tim Keitt, Erik van Nimwegen,
Andreas Pedersen, David Raup, Jack Sepkoski, Paolo Sibani, Kim Sneppen and
Ricard Sol\'e for useful discussions.  Special thanks are due also to Chris
Adami, Gunther Eble, Doug Erwin and Jack Sepkoski for providing data used
in a number of the figures.  This work was supported by the Santa Fe
Institute and DARPA under grant number ONR N00014--95--1--0975.

\clearpage

\addcontentsline{toc}{section}{References}
\markboth{References}{References}

\def\refer#1#2#3#4#5#6{\item{\frenchspacing\sc#1}\hspace{4pt}
                       #2.\hspace{8pt}#3 {\frenchspacing#4} {\bf#5}, #6.}
\def\bookref#1#2#3#4{\item{\frenchspacing\sc#1}\hspace{4pt}
                     #2.\hspace{8pt}{\it#3}  #4.}

\section*{References}
\baselineskip=15pt

\begin{list}{}{\leftmargin=2em \itemindent=-\leftmargin%
\itemsep=3pt \parsep=0pt \small}

\refer{Abramson, G.}{1997}{Ecological model of extinctions.}{\it Phys. Rev.
  E\/}{55}{785--788}

\refer{Adami, C.}{1995}{Self-organized criticality in living systems.}{\it
  Phys. Letts. A\/}{203}{29--32}

\refer{Alvarez, L. W.}{1983}{Experimental evidence that an asteroid impact
  led to the extinction of many species 65 million years ago.}{\it
  Proc. Natl. Acad. Sci.}{80}{627--642}

\refer{Alvarez, L. W.}{1987}{Mass extinctions caused by large bolide
  impacts.}{\it Physics Today\/}{40}{24--33}

\refer{Alvarez, L. W., Alvarez, W., Asara, F. \& Michel,
  H. V.}{1980}{Extraterrestrial cause for the Cretaceous--Tertiary
  extinction.}{\it Science\/}{208}{1095--1108}

\refer{Amaral, L. A. N. \& Meyer, M.}{1999}{Environmental changes,
  coextinction, and patterns in the fossil record.}{\it
  Phys. Rev. Lett.}{82}{652--655}

\bookref{Bak, P.}{1996}{How Nature Works: The Science of Self-Organized
  Criticality.}{Copernicus (New York)}

\refer{Bak. P., Flyvbjerg, H. \& Lautrup, B.}{1992}{Coevolution in a rugged
  fitness landscape.}{\it Phys. Rev. A\/}{46}{6724--6730}

\refer{Bak, P. \& Sneppen, K.}{1993}{Punctuated equilibrium and
  criticality in a simple model of
  evolution.}{\it Phys. Rev. Lett.}{71}{4083--4086}

\refer{Bak, P., Tang, C. \& Wiesenfeld, K.}{1987}{Self-organized
  criticality: An explanation of $1/f$ noise.}{\it
  Phys. Rev. Lett.}{59}{381--384}

\refer{Benton, M. J.}{1987}{Progress and competition in
  macroevolution.}{\it Biol. Rev.}{62}{305--338}

\item {\sc Benton, M. J.}\ \ 1991.\ \ Extinction, biotic replacements and clade
  interactions.  In {\it The Unity of Evolutionary Biology,} Dudley,
  E. C. (ed.), Dioscorides (Portland).

\bookref{Benton, M. J.}{1993}{The Fossil Record 2.}{Chapman and Hall
  (London)}

\refer{Benton, M. J.}{1995}{Diversification and extinction in the
  history of life.}{\it Science\/}{268}{52--58}

\bookref{Binney, J. J., Dowrick, N. J., Fisher, A. J. \& Newman, M. E.
  J.}{1992}{The Theory of Critical Phenomena.}{Oxford University Press
  (Oxford)}

\refer{Boettcher, S. \& Paczuski, M.}{1996}{Exact results for
  spatiotemporal correlation in a self-organized critical model of
  punctuated equilibrium.}{\it Phys. Rev. Lett.}{76}{348--351}

\refer{Bourgeois, T., Clemens, W. A., Spicer, R. A., Ager, T. A., Carter,
  L. D. \& Sliter, W. V.}{1988}{A tsunami deposit at the
  Cretaceous--Tertiary boundary in Texas.}{\it Science\/}{241}{567--571}

\refer{Bowring, S. A., Grotzinger, J. P., Isachsen, C. E., Knoll, A. H.,
  Pele\-chaty, S. M. \& Kolosov, P.}{1993}{Calibrating rates of early
  Cambrian evolution.}{\it Science\/}{261}{1293--1298}

\refer{Burlando, B.}{1990}{The fractal dimension of taxonomic systems.}{\it
  J. Theor. Biol.}{146}{99--114}

\refer{Burlando, B.}{1993}{The fractal geometry of evolution.}{\it J.
  Theor. Biol.}{163}{161--172}

\refer{Chiappe, L. M.}{1995}{The first 85 million years of avian
  evolution.}{\it Nature\/}{378}{349--355}

\item {\sc Clemens, W. A.}\ \ 1986.\ \ Evolution of the vertebrate fauna
  during the Cretaceous--Tertiary transition.  In {\it Dynamics of
    Extinction,} Elliott, D. K. (ed.), Wiley (New York).

\refer{Courtillot, V., Feraud, G., Malushi, H., Vandamme, D., Moreau,
  M. G. \& Besse, J.}{1988}{Deccan flood basalts and the Cretaceous/Tertiary
  boundary.}{\it Nature\/}{333}{843--846}

\refer{Courtillot, V. \& Gaudemer, Y.}{1996}{Effects of mass extinctions on
  biodiversity.}{\it Nature\/}{381}{146--148}

\refer{Davis, M., Hut, P. \& Muller, R. A.}{1984}{Extinction of species by
  periodic comet showers.}{\it Nature\/}{308}{715--717}

\refer{de Boer, J., Jackson, A. D. \& Wettig, T.}{1995}{Criticality in
  simple models of evolution.}{\it Phys. Rev. E\/}{51}{1059--1073}

\refer{Derrida, B.}{1980}{Random energy model: The limit of a family of
    disordered models.}{\it Phys. Rev. Lett.}{45}{79--82}

\refer{Derrida, B.}{1981}{Random energy model: An exactly solvable model of
  disordered systems.}{\it Phys. Rev. B\/}{24}{2613--2626}

\refer{Drossel, B.}{1999}{Extinction events and species lifetimes
  in a simple ecological model.}{\it Phys. Rev. Lett.}{81}{5011--5014}

\refer{Duncan, R. A. \& Pyle, D. G.}{1988}{Rapid eruption of the Deccan
  basalts at the Cretaceous/Tertiary boundary.}{\it
  Nature\/}{333}{841--843}

\item {\sc Eble, G. J.}\ \ 1998\ \ The role of development in
  evolutionary radiations.  In {\it Biodiversity Dynamics: Turnover of
    Populations, Taxa and Communities,} M.~L.~McKinney (ed.), Columbia
  University Press (New York).
  
\refer{Eble, G. J.}{1999}{Originations: Land and sea compared.}{\it
  Geobios\/}{32}{223--234}
  
\refer{Ellis, J. \& Schramm, D. M.}{1995}{Could a nearby supernova
  explosion have caused a mass extinction?}{\it
  Proc. Natl. Acad. Sci.}{92}{235--238}

\item {\sc Erwin, D. H.}\ \ 1996.\ \ Understanding biotic recoveries.  In
  {\it Evolutionary Paleobiology,} Jablonski, D., Erwin, D. \& Lipps,
  I. (eds.), University of Chicago Press (Chicago).
  
\refer{Flessa, K. W. \& Jablonski, D.}{1983}{Extinction is here to
  stay.}{\it Paleobiology\/}{9}{315--321}

\refer{Flessa, K. W. \& Jablonski, D.}{1985}{Declining Phanerozoic
  background extinction rates: Effect of taxonomic structure?}{\it
  Nature\/}{313}{216--218}

\refer{Flyvbjerg, H., Sneppen, K. \& Bak, P.}{1993}{Mean field theory for
  a simple model of evolution.}{\it Phys. Rev. Lett.}{71}{4087--4090}

\refer{Fox, W. T.}{1987}{Harmonic analysis of periodic extinctions.}{\it
  Paleobiology\/}{13}{257--271}

\refer{Gauthier, J. A.}{1986}{Saurischian monophyly and the origin of
  birds.}{\it Mem. Calif. Acad. Sci.}{8}{1--47}

\refer{Gilinsky, N. L. \& Bambach, R. K.}{1987}{Asymmetrical patterns
  of origination and extinction in higher taxa.}{\it
  Paleobiology\/}{13}{427--445}

\bookref{Glen, W.}{1994}{The Mass Extinction Debates.}{Stanford
  University Press (Stanford)}

\item {\sc Grieve, R. A. F. \& Shoemaker, E. M.}\ \ 1994.\ \ The record of
  past impacts on Earth.  In {\it Hazards Due to Comets and Asteroids,}
  Gehrels, T. (ed.), University of Arizona Press (Tucson).

\bookref{Grimmett, G. R. \& Stirzaker, D. R.}{1992}{Probability and Random
  Processes, 2nd Edition.}{Oxford University Press (Oxford)}

\refer{Hallam, A.}{1989}{The case for sea-level change as a dominant
  causal factor in mass extinction of marine
  invertebrates.}{\it Phil. Trans. R. Soc. B\/}{325}{437--455}

\refer{Hallock, P.}{1986}{Why are large foraminifera large?}{\it
  Paleobiology\/}{11}{195--208}

\bookref{Harland, W. B., Armstrong, R., Cox, V. A., Craig, L. E.,
  Smith, A. G. \& Smith, D. G.}{1990}{A Geologic Time Scale
  1989.}{Cambridge University Press (Cambridge)}

\refer{Head, D. A. \& Rodgers, G. J.}{1997}{Speciation and extinction in a
  simple model of evolution.}{\it Phys. Rev. E\/}{55}{3312--3319}

\bookref{Hertz, J. A., Krogh, A. S. \& Palmer, R. G.}{1991}{Introduction to
  the Theory of Neural Computation.}{Addison-Wesley (Reading)}

\bookref{Hoffman, A. A. \& Parsons, P. A.}{1991}{Evolutionary
  Genetics and Environmental Stress.}{Oxford University Press (Oxford)}

\refer{Hut, P., Alvarez, W., Elder, W. P., Hansen, T., Kauffman, E. G.,
  Killer, G., Shoemaker, E. M. \& Weissman, P. R.}{1987}{Comet showers as a
  cause of mass extinctions.}{\it Nature\/}{329}{118--125}

\item {\sc Jablonski, D.}\ \ 1985.\ \ Marine regressions and mass
  extinctions: a test using the modern biota.  In {\it Phanerozoic
  diversity patters,} Valentine, J. W. (ed.), Princeton University Press
  (Princeton).
  
\refer{Jablonski, D.}{1986}{Background and mass extinctions: The
  alternation of macroevolutionary regimes.}{\it
  Science\/}{231}{129--133}

\refer{Jablonski, D.}{1989}{The biology of mass extinction: A
  palaeontological view.}{\it Phil. Trans. R. Soc. B\/}{325}{357--368}

\refer{Jablonski, D.}{1991}{Extinctions: A paleontological
  perspective.}{\it Science\/}{253}{754--757}

\refer{Jablonski, D.}{1993}{The tropics as a source of evolutionary
  novelty through geological time.}{\it Nature\/}{364}{142--144}

\item {\sc Jablonski, D. \& Bottjer, D. J.}\ \ 1990a\ \ The ecology
  of evolutionary innovation: the fossil record.  In {\it Evolutionary
    Innovations,} M.~Nitecki, (ed.), University of Chicago Press
  (Chicago).
  
\item {\sc Jablonski, D. \& Bottjer, D. J.}\ \ 1990b\ \ The origin and
  diversification of major groups: Environmental patterns and
  macroevolutionary lags.  In {\it Major Evolutionary Radiations,}
  P.~D.~Taylor \& G.~P.~Larwood, (eds.), Oxford University Press (Oxford).
  
\item {\sc Jablonski, D. \& Bottjer, D. J.}\ \ 1990c\ \ 
  Onshore-offshore trends in marine invertebrate evolution.  In {\it
    Causes of Evolution: A Paleontological Perspective,} R.~M.~Ross
  \& W.~D.~Allmon, (eds.), University of Chicago Press (Chicago).
  
\bookref{Kauffman, S. A.}{1993}{Origins of Order: Self-Organization and
  Selection in Evolution.}{Oxford University Press (Oxford)}

\bookref{Kauffman, S. A.}{1995}{At Home in the Universe.}{Oxford University
  Press (Oxford)}

\refer{Kauffman, S. A. \& Johnsen, S.}{1991}{Coevolution to the edge of
  chaos: Coupled fitness landscapes, poised states, and coevolutionary
  avalanches.}{\it J. Theor. Biol.}{149}{467--505}

\refer{Kauffman, S. A. \& Levin, S.}{1987}{Towards a general theory of
  adaptive walks on rugged landscapes.}{\it J. Theor. Biol.}{128}{11--45}

\bookref{Kauffman, S. A. \& Perelson, A. S.}{1990}{Molecular Evolution on
  Rugged Landscapes: Proteins, RNA, and the Immune
  Response.}{Addison--Wesley (Reading)}

\refer{Kauffman, S. A. \& Weinberger, E. W.}{1989}{The \NK\ model of rugged
  fitness landscapes and its application to maturation of the immune
  response.}{\it J. Theor. Biol.}{141}{211--245}


\refer{Kramer, M., Vandewalle, N. \& Ausloos, M.}{1996}{Speciations and
  extinction in a self-organizing critical model of tree-like
  evolution.}{\it J. Phys. I France\/}{6}{599--606}

\bookref{Langton, C. G.}{1995}{Artificial Life: An Overview.}{MIT Press
  (Cambridge)}

\refer{Loper, D. E., McCartney, K. \& Buzyna, G.}{1988}{A model of
  correlated periodicity in magnetic-field reversals, climate and mass
  extinctions.}{\it J. Geol.}{96}{1--15}

\bookref{Lyell, C.}{1832}{Principles of Geology, Vol. 2.}{Murray (London)}

\refer{Macken, C. A. \& Perelson, A. S.}{1989}{Protein evolution on rugged
  landscapes.}{\it Proc. Natl. Acad. Sci.}{86}{6191--6195}

\refer{Manrubia, S. C. \& Paczuski, M.}{1998}{A simple model of large scale
  organization in evolution.}{\it Int. J. Mod. Phys. C\/}{9}{1025--1032}

\refer{Maslov, S., Paczuski, M. \& Bak, P.}{1994}{Avalanches and $1/f$
  noise in evolution and growth models.}{\it
  Phys. Rev. Lett.}{73}{2162--2165}

\refer{May, R. M.}{1990}{How many species?}{\it
  Phil. Trans. R. Soc. B\/}{330}{293--304}

\refer{Maynard Smith, J.}{1989}{The causes of extinction.}{\it
  Phil. Trans. R. Soc. B\/}{325}{241--252}

\refer{Maynard Smith, J. and Price, G. R.}{1973}{The logic of animal
  conflict.}{\it Nature\/}{246}{15--18}

\refer{McLaren, D. J.}{1988}{Detection and significance of mass
  killings.}{\it Historical Biology\/}{2}{5--15}

\item {\sc McNamara, K. J.}\ \ 1990.\ \ Echinoids.  In {\it Evolutionary
      Trends,} McNamara, K. J. (ed.), Belhaven Press (London).

\bookref{Mitchell, M.}{1996}{An Introduction to Genetic Algorithms.}{MIT
  Press (Cambridge)}

\refer{Montroll, E. W. \& Shlesinger, M. F.}{1982}{On $1/f$ noise and
  other distributions with long tails.}{\it
  Proc. Natl. Acad. Sci.}{79}{3380--3383}

\bookref{Morrison, D.}{1992}{The Spaceguard Survey: Report of the NASA
  International Near-Earth Object Detection Workshop.}{Jet Propulsion
  Laboratory (Pasadena)}

\refer{Newman, M. E. J.}{1996}{Self-organized criticality, evolution
  and the fossil extinction record.}{\it
  Proc. R. Soc. London B\/}{263}{1605--1610}

\refer{Newman, M. E. J.}{1997}{A model of mass extinction.}{\it
  J. Theor. Biol.}{189}{235--252}

\refer{Newman, M. E. J. \& Eble, G. J.}{1999a}{Power spectra of extinction
  in the fossil record.}{\it Proc. R. Soc. London B\/}{266}{1267--1270}

\item {\sc Newman, M. E. J. \& Eble, G. J.}\ \ 1999b.\ \ Decline in
  extinction rates and scale invariance in the fossil record.  {\it
    Paleobiology,} in press.

\refer{Newman, M. E. J., Fraser, S. M., Sneppen, K. \& Tozier,
  W. A.}{1997}{Comment on ``Self-organized criticality in living
  systems''.}{\it Phys. Lett. A\/}{228}{201--203}

\refer{Newman, M. E. J. \& Roberts, B. W.}{1995}{Mass extinction:
  Evolution and the effects of external influences on unfit species.}{\it
  Proc. R. Soc. London B\/}{260}{31--37}

\refer{Newman, M. E. J. \& Sibani, P.}{1999}{Extinction, diversity and
  survivorship of taxa in the fossil record.}{\it Proc. R. Soc. London
  B\/}{266}{1593--1600}

\refer{Newman, M. E. J. \& Sneppen, K.}{1996}{Avalanches, scaling and
  coherent noise.}{\it Phys. Rev. E\/}{54}{6226--6231}

\refer{Paczuski, M., Maslov, S. \& Bak, P.}{1996}{Avalanche dynamics in
  evolution, growth, and depinning models.}{\it Phys. Rev.
  E\/}{53}{414--443}

\refer{Pang, N. N.}{1997}{The Bak--Sneppen model: A self-organized critical
  model of biological evolution.}{\it Int. J. Mod. Phys.
  B\/}{11}{1411--1444}

\refer{Parsons, P. A.}{1993}{Stress, extinctions and evolutionary
  change: From living organisms to fossils.}{\it Biol. Rev.}{68}{313--333}

\refer{Patterson, C. \& Smith, A. B.}{1987}{Is the periodicity of
  extinctions a taxonomic artifact?}{\it Nature\/}{330}{248--251}

\refer{Patterson, C. \& Smith, A. B.}{1989}{Periodicity in extinction: the
  role of the systematics.}{\it Ecology\/}{70}{802--811}

\refer{Patterson, R. T. \& Fowler, A. D.}{1996}{Evidence of self
  organization in planktic foraminiferal evolution: Implications for
  interconnectedness of paleoecosystems.}{\it Geology\/}{24}{215--218}

\refer{Pease, C. M.}{1992}{On the declining extinction and origination rates
  of fossil taxa.}{\it Paleobiology\/}{18}{89--92}

\refer{Plotnick, R. E. \& McKinney, M. L.}{1993}{Ecosystem
  organization and extinction dynamics.}{\it Palaios\/}{8}{202--212}

\refer{Rampino, M. R. \& Stothers, R. B.}{1984}{Terrestrial mass
  extinctions, cometary impacts and the sun's motion perpendicular to the
  galactic plane.}{\it Nature\/}{308}{709--712}

\refer{Raup, D. M.}{1979a}{Biases in the fossil record of species and
  genera.}{\it Bulletin of the Carnegie Museum of Natural
  History\/}{\bf13}{85--91}

\refer{Raup, D. M.}{1979b}{Size of the Permo-Triassic bottleneck and
  its evolutionary implications.}{\it Science\/}{206}{217--218}

\refer{Raup, D. M.}{1985}{Magnetic reversals and mass extinctions.}{\it
  Nature}{314}{341--343}

\refer{Raup, D. M.}{1986}{Biological extinction in Earth history.}{\it
  Science\/}{231}{1528--1533}

\bookref{Raup, D. M.}{1991a}{Extinction: Bad Genes or Bad Luck?}{Norton
  (New York)}

\refer{Raup, D. M.}{1991b}{A kill curve for Phanerozoic marine
  species.}{\it Paleobiology\/}{17}{37--48}

\refer{Raup, D. M.}{1992}{Large-body impact and extinction in the
  Phanerozoic.}{\it Paleobiology\/}{18}{80--88}

\item {\sc Raup, D. M.}\ \ 1996.\ \ Extinction models.  In {\it Evolutionary
  Paleobiology,} Jablonski,~D., Erwin, D. H. \& Lipps, J. H., (eds.),
  University of Chicago Press (Chicago).

\refer{Raup, D. M. \& Boyajian, G. E.}{1988}{Patterns of generic
  extinction in the fossil record.}{Paleobiology\/}{14}{109--125}

\refer{Raup, D. M. \& Sepkoski, J. J., Jr.}{1982}{Mass extinctions in the
marine fossil record}{\it Science\/}{215}{1501--1503}

\refer{Raup, D. M. \& Sepkoski, J. J., Jr.}{1984}{Periodicity of
  extinctions in the geologic past.}{\it
  Proc. Natl. Acad. Sci.}{81}{801--805}

\refer{Raup, D. M. \& Sepkoski, J. J., Jr.}{1986}{Periodic extinctions of
  families and genera.}{\it Science\/}{231}{833--836}

\refer{Raup, D. M. \& Sepkoski, J. J., Jr.}{1988}{Testing for periodicity
  of extinction.}{\it Science\/}{241}{94--96}

\refer{Ray, T. S.}{1994a}{An evolutionary approach to synthetic
  biology.}{\it Artificial Life\/}{1}{179--209}

\refer{Ray, T. S.}{1994b}{Evolution, complexity, entropy and artificial
  reality.}{\it Physica D\/}{75}{239--263}

\refer{Roberts, B. W. \& Newman, M. E. J.}{1996}{A model for evolution and
  extinction.}{\it J. Theor. Biol.}{180}{39--54}

\bookref{Rosenzweig, M. L.}{1995}{Species Diversity in Space and
  Time.}{Cambridge University Press (Cambridge)}

\refer{Roy, K.}{1996}{The roles of mass extinction and biotic interaction
in large-scale replacements.}{\it Paleobiology\/}{22}{436--452}

\refer{Schmoltzi, K. \& Schuster, H. G.}{1995}{Introducing a real time scale
  into the Bak--Sneppen model.}{\it Phys. Rev. E\/}{52}{5273--5280}

\refer{Sepkoski, J. J., Jr.}{1988}{Perodicity of extinction and the problem
  of catastophism in the history of life.}{\it
  J. Geo. Soc. London\/}{146}{7--19}

\item {\sc Sepkoski, J. J., Jr.}\ \ 1990.\ \ The taxonomic structure of
  periodic extinction.  In {\it Global Catastrophes in Earth History,}
  Sharpton, V. L. \& Ward, P. D. (eds.), {\it Geological Society of
  America Special Paper\/} {\bf247}, 33--44.

\item {\sc Sepkoski, J. J., Jr.}\ \ 1991\ \ Diversity in the Phanerozoic
  oceans: A partisan review.  In {\it The Unity of Evolutionary Biology,}
  Dudley, E. C. (ed.), Dioscorides (Portland).

\item {\sc Sepkoski, J. J., Jr.}\ \ 1993.\ \ A compendium of fossil marine
  animal families, 2nd edition.  {\it Milwaukee Public Museum
    Contributions in Biology and Geology\/} {\bf83}.

\item {\sc Sepkoski, J. J., Jr.}\ \ 1996\ \ Patterns of Phanerozoic
  extinction: A perspective from global databases.  In {\it Global
  events and event stratigraphy,} O.~H.~Walliser, (ed.),
  Springer-Verlag (Berlin).
  
\refer{Sepkoski, J. J., Jr.}{1998}{Rates of speciation in the fossil
  record.}{\it Phil. Trans. R. Soc. B\/}{353}{315--326}

\refer{Sepkoski, J. J., Jr. \& Kendrick, D. C.}{1993}{Numerical experiments
  with model monophyletic and paraphyletic taxa.}{\it
  Paleobiology\/}{19}{168--184}

\refer{Sibani, P. \& Littlewood, P.}{1993}{Slow dynamics from noise
  adaptation.}{\it Phys. Rev. Lett.}{71}{1482--1485}
  
\refer{Sibani, P., Schmidt, M. R. and Alstr\o{}m, P.}{1995}{Fitness
  optimization and decay of extinction rate through biological
  evolution.}{\sl Phys. Rev. Lett.}{75}{2055--2058}
  
\refer{Sibani, P., Schmidt, M. R. and Alstr\o{}m, P.}{1998}{Evolution
  and extinction dynamics in rugged fitness landscapes.}{\sl Int. J.
  Mod. Phys. B\/}{12}{361--391}
  
\item {\sc Signor, P. W. \& Lipps, J. H.}\ \ 1982.\ \ Sampling bias,
  gradual extinction patterns, and catastrophes in the fossil record.  In
  {\it Geological Implications of Impacts of Large Asteroids and Comets on
  the Earth,} Silver, L. T. \& Schultz, P. H. (eds.), {\it Geological
  Society of America Special Paper\/} {\bf190}, 291--296.
  
\refer{Simpson, G. G.}{1952}{How many species?}{\it Evolution\/}{6}{342--342}

\refer{Sneppen, K.}{1995}{Extremal dynamics and punctuated
  co-evolution.}{\it Physica A\/}{221}{168--179}

\refer{Sneppen, K., Bak, P., Flyvbjerg, H. \& Jensen,
  M. H.}{1995}{Evolution as a self-organized critical
  phenomenon.}{\it Proc. Natl. Acad. Sci.}{92}{5209--5213}

\refer{Sneppen, K. \& Newman, M. E. J.}{1997}{Coherent noise, scale
  invariance and intermittency in large systems.}{\it Physica
  D\/}{110}{209--222}

\refer{Sol\'e, R. V.}{1996}{On macroevolution, extinctions and critical
  phenomena.}{\it Complexity}{1}{40--44}

\refer{Sol\'e, R. V. \& Bascompte, J.}{1996}{Are critical phenomena
  relevant to large-scale evolution?}{\it Proc. R. Soc. London
  B\/}{263}{161--168}

\refer{Sol\'e, R. V., Bascompte, J., \& Manrubia, S. C.}{1996}{Extinction:
  Bad genes or weak chaos?}{\it Proc. R. Soc. London
  B\/}{263}{1407--1413}

\refer{Sol\'e, R. V. \& Manrubia, S. C.}{1996}{Extinction and
  self-organized criticality in a model of large-scale evolution.}{\it
  Phys. Rev. E\/}{54}{R42--R45}

\refer{Sol\'e, R. V., Manrubia, S. C., Benton, M. \& Bak,
  P.}{1997}{Self-similarity of extinction statistics in the fossil
  record.}{\it Nature\/}{388}{764--767}

\refer{Sornette, D. \& Cont, R.}{1997}{Convergent multiplicative processes
  repelled from zero: power laws and truncated power laws.}{\it J. Phys. I
  France\/}{7}{431--444}

\item {\sc Stanley, S. M.}\ \ 1984.\ \ Marine mass extinction: A dominant
  role for temperature.  In {\it Extinctions,} Nitecki, M. H. (ed.),
  University of Chicago Press (Chicago).

\refer{Stanley, S. M.}{1988}{Paleozoic mass extinctions: Shared patterns
  suggest global cooling as a common cause.}{\it
  Am. J. Sci.}{288}{334--352}

\refer{Stanley, S. M.}{1990}{Delayed recovery and the spacing of major
  extinctions.}{\it Paleobiology\/}{16}{401--414}

\item {\sc Stenseth, N. C.}\ \ 1985.\ \ Darwinian evolution in ecosystems:
  The red queen view.  In {\sl Evolution,} Cambridge University Press
  (Cambridge).
  
\refer{Strogatz, S. H. \& Stewart, I.}{1993}{Coupled oscillators and
    biological synchronization.}{\it Scientific American\/}{269}{102--109}

\refer{Van~Valen, L.}{1973}{A new evolutionary law.}{\it Evol.\ 
  Theory\/}{1}{1--30}

\refer{Vandewalle, N. \& Ausloos, M.}{1995}{The robustness of
  self-organized criticality against extinctions in a tree-like model of
  evolution.}{Europhys. Lett.}{32}{613--618}

\refer{Vandewalle, N. \& Ausloos, M.}{1997}{Different universality classes
  for self-organized critical models driven by extremal
  dynamics.}{\it Europhys. Lett.}{37}{1--6}

\bookref{Vermeij, G. J.}{1987}{\it Evolution as Escalation.}{Princeton
  University Press (Princeton)}

\bookref{Weisbuch, G.}{1991}{Complex Systems Dynamics.}{Addison-Wesley
  (Reading)}

\refer{Whitmire, D. P. \& Jackson, A. A.}{1984}{Are periodic mass
  extinctions driven by a distant solar companion?}{\it
  Nature\/}{308}{713--715}

\refer{Wilde, P. \& Berry, W. B. N.}{1984}{Destabilization of the oceanic
  density structure and its significance to marine extinction events.}{\it
  Palaeogeog. Palaeoclimatol. Palaeoecol.}{48}{142--162}

\refer{Wilke, C., Altmeyer, S. \& Martinetz, T.}{1998}{Aftershocks in
  coherent-noise models.}{\it Physica D\/}{120}{401--417}

\refer{Wilke, C. \& Martinetz, T.}{1997}{Simple model of evolution with
  variable system size.}{\it Phys. Rev. E\/}{56}{7128--7131}

\refer{Williams, C. B.}{1944}{Some applications of the logarithmic series
  and the index of diversity to ecological problems.}{\it J.
  Ecol.}{32}{1--44}

\bookref{Willis, J. C.}{1922}{Age and Area.}{Cambridge University Press
  (Cambridge)}

\bookref{Zipf, G. K.}{1949}{Human Behavior and the Principle of Least
Effort.}{Addison--Wesley (Reading)}

\end{list}

\end{document}